\RequirePackage{fix-cm}

\documentclass[twocolumn,epjc3]{svjour3}  

\usepackage{hyperref}       % hyperlinks
\usepackage{url}            % simple URL typesetting
\usepackage[LGRgreek]{mathastext}
\usepackage{graphicx}
\usepackage{xspace}

\usepackage{graphicx}
\usepackage{epsfig}
\usepackage{url}
\usepackage[cal=esstix]{mathalfa}
\usepackage{amsmath}
\usepackage{amssymb} 
\usepackage{float, placeins}
\usepackage{subcaption}
\usepackage{dblfloatfix}%,fixltx2e}
\usepackage{xfrac}
\usepackage{xspace}
\usepackage{hyperref}
\usepackage[binary-units=true]{siunitx}
\usepackage{accents}
\usepackage[super]{nth}
\usepackage{algorithm}
\usepackage{algpseudocode}
\usepackage{bm} % bold math symbols
\DeclareMathOperator{\Lagr}{L}%\mathcal{L}}

\newcommand{\adam}{\textsc{Adam}\xspace}

\newcommand{\Root}{\textsc{Root}\xspace}

\newcommand{\geant}{\textsc{Geant}\xspace}

\newcommand{\lumin}{\textsc{Lumin}\xspace}
\newcommand{\pytorch}{\textsc{PyTorch}\xspace}

\newcommand{\sfCaption}[2]{(\protect\subref*{#1}) #2}

\newcommand{\sfSmall}{0.48}

\newcommand{\sfMid}{0.8}

\newcommand{\tev}{\tera\electronvolt}
\newcommand{\gev}{\giga\electronvolt}
\newcommand{\mev}{\mega\electronvolt}
\newcommand{\mm}{\milli\meter}

\journalname{Eur. Phys. J. C}

\begin{document}

\title{Calorimetric Measurement of Multi-TeV Muons via Deep Regression} %[MODE-21-002]} 
\titlerunning{Calorimetric Measurement of Multi-TeV Muons via Deep Regression}

\author{Jan Kieseler\thanksref{e1,addr1} \and Giles C. Strong\thanksref{e2,addr2,addr3} \and Filippo Chiandotto\thanksref{e3,addr2} \and Tommaso Dorigo\thanksref{e4,addr3} \and Lukas Layer\thanksref{e5,addr4,addr3} }

\thankstext{e1}{e-mail: jan.kieseler@cern.ch}
\thankstext{e2}{e-mail: giles.chatham.strong@cern.ch}
\thankstext{e3}{e-mail: filippo.chiandotto@studenti.unipd.it}
\thankstext{e4}{e-mail: dorigo@pd.infn.it}
\thankstext{e5}{e-mail: lukas.layer@cern.ch}

\institute{CERN \label{addr1}
\and
Università di Padova \label{addr2}
\and
INFN, Sezione di Padova \label{addr3}
\and
Università di Napoli ``Federico II" \label{addr4}
}

\date{Received: 6 August 2021 / Accepted: 4 January 2022 / Published online: 27 January 2022 \\
Eur. Phys. J. C (2022) 82:79 \url{https://doi.org/10.1140/epjc/s10052-022-09993-5} \\
\copyright The Author(s) 2022}

\maketitle
\begin{abstract}
The performance demands of future particle-physics experiments investigating the high-energy frontier pose a number of new challenges, forcing us to find improved solutions for the detection, identification, and measurement of final-state particles in subnuclear collisions. One such challenge is the precise measurement of muon momentum at very high energy, where an estimate of the curvature provided by conceivable magnetic fields in realistic detectors proves insufficient for achieving good momentum resolution when detecting, {\em e.g.}, a narrow, high mass resonance decaying to a muon pair.

In this work we study the feasibility of an entirely new avenue for the measurement of the energy of muons based on their radiative losses in a dense, finely segmented calorimeter. This is made possible by exploiting spatial information of the clusters of energy from radiated photons in a regression task. The use of a task-specific deep learning architecture based on convolutional layers allows us to treat the problem as one akin to image reconstruction, where images are constituted by the pattern of energy released in successive layers of the calorimeter. A measurement of muon energy with better than \SI{20}{\%} relative resolution is shown to be achievable for ultra-\si{\tev} muons.

\end{abstract}

%%%%%%%%%%%%%%%%%%%%%%%%%%%%%%%%%%%%%%%%%%%%%
\section{Introduction \label{s:introduction}}

Muons have been used as clean probes of new phenomena in particle physics ever since their discovery in cosmic showers~\cite{muon_disc1,muon_disc2}. Their detection and measurement enabled many groundbreaking discoveries, from those of heavy quarks~\cite{richter,lederman,topdisc} and weak bosons~\cite{rubbia} to that of the Higgs boson~\cite{higgsatlas,higgscms} through its decay into weak bosons; most recently, a first evidence for $H \to \mu \mu$ decays has also been reported by CMS~\cite{hmmcms}, highlighting the importance of muons for searches as well as measurements of standard model parameters. 
The uniqueness of muons is due to their 
intrinsic physical properties, which produce a distinctive phenomenology of interactions with matter. Endowed with a mass 200 times higher than that of the electron, the muon loses little energy by electromagnetic radiation as it traverses dense media; it behaves as a minimum ionizing particle in a wide range of energies, where it is easily distinguishable from long-lived light hadrons such as charged pions and kaons. 

In continuity with their glorious past, muons will remain valuable probes of new physics phenomena in future searches at high-energy colliders. A number of heavy particles predicted by new-physics models are accessible preferentially, and in some cases exclusively, by the detection of their decay to final states that include electrons or muons; in particular, the reconstruction of the resonant shape of dileptonic decays of new $Z'$ gauge bosons resulting from the addition of an extra U(1) group or higher symmetry structures to the Standard Model~\cite{zprime1,zprime2} constitutes a compelling reason for seeking the best possible energy resolution for electrons and muons of high energy.
\begin{center}
\begin{figure}[h!]
\centerline{\includegraphics[width=0.48\textwidth]{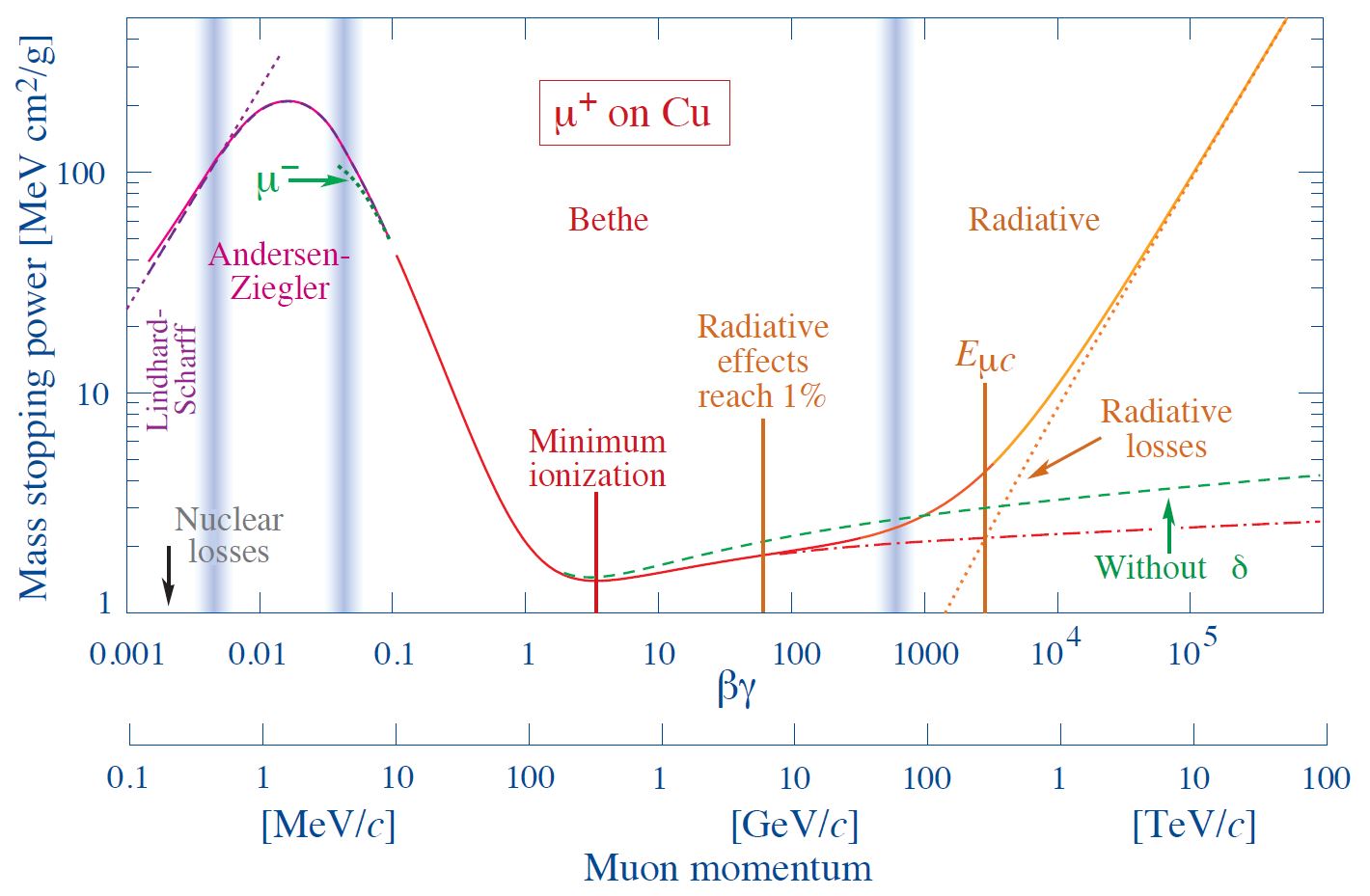}}
\caption{Mass stopping power for muons in the \SI{0.1}{\mev} to  \SI{100}{\tev} range, in \si{\mev\centi\meter^2\per\gram}. The rise in radiative loss becomes important above \SI{100}{\gev}. 
Image reproduced with permission from Ref.~\cite{pdg}.
}
\label{f:muonEloss}
\end{figure}
\end{center}

Unfortunately, the very features that make muons special and easily distinguishable from backgrounds also hinder the precise measurement of their energy in the ultra-relativistic regime. While the energy of electrons is effectively inferred from the electromagnetic showers they initiate in dense calorimeters, muon energy estimates rely solely on the determination of the curvature of their trajectory in a magnetic field. If we consider the ATLAS and CMS detectors as a reference, we observe that the relative resolution of muon transverse momentum achieved in those state-of-the-art instruments at \SI{1}{\tev} ranges from 8 to 20\% in ATLAS, and from 6 to 17\% in CMS~\cite{atlasmuon,cmsmuon}, depending on detection and reconstruction details; by comparison, for electrons of the same energy the resolution ranges from 0.5 to 1.0\% in ATLAS, and from 1 to 2\% in CMS~\cite{atlaselec,cmselec}. 
Clearly, for non-minimum-ionizing particles, calorimetric measurements win over curvature determinations at high energy, due to the different scaling properties of the respective resolution functions: relative uncertainty of curvature-driven estimates grows linearly with energy, while the one of calorimetric estimates decreases with $\sqrt{E}$.

However, ultra-relativistic muons do not behave as minimum-ionizing particles; rather, they show a rise in their radiative energy loss~\cite{pdg} above roughly \SI{100}{\gev} (see \autoref{f:muonEloss}). The effect is clear, although undeniably very small in absolute terms; for example, a \SI{1}{\tev} muon is expected to lose a mere \SI{2.3}{\gev} in traversing the $\SI{25.8}{X_0}$ of the CMS electromagnetic calorimeter~\cite{CMSDetPaper}. For that reason, patterns of radiative losses have never been exploited to estimate muon energy in collider detectors\footnote {To our knowledge, a measurement strategy has been demonstrated only in the IceCube experiment~\cite{Abbasi_2013,Aartsen_2014}, where the energy of muons of interest is still higher than that investigated in this work, and the thickness in radiation lengths of the traversed detector material is over an order of magnitude larger than that of present-day collider detectors. Moreover, attempts to aid the muon reconstruction with the energy sum in the calorimeters have been made by the ATLAS Collaboration~\cite{Nikolopoulos:2007zz,theatlascollaboration2009expected}}. It is the purpose of this work to show how low-energy photons radiated by TeV-energy muons and detected in a sufficiently thick and fine-grained calorimeter may be successfully exploited to estimate muon energy even in collider detector applications. Crucially, we will also demonstrate how the input of such a measurement is not only the magnitude, but also the pattern of the detected energy depositions in the calorimeter cells.

%%%% 
The spatial patterns of calorimeter deposits are a well known and heavily exploited feature for object identification purposes, {\em e.g.} to distinguish electromagnetic showers from hadronic showers by comparing the depth profile of the energy deposits~\cite{CMSElecID,atlaselec}. Recently, in the context of proposals for calorimeters endowed with fine grained lateral and longitudinal segmentation, it has been shown that this granularity not only improves the identification purity, but also allows for an accurate determination of the energy of hadronic showers, by identifying individual patterns of their electromagnetic and hadronic sub-components~\cite{Belayneh_2020,Adloff:2012gv,Chefdeville:2015,Israeli:2018byq,HGCAL-TDR,atlas_pions}. In parallel, machine learning techniques have proven to be very powerful for reconstructing individual showers~\cite{Neubuser:2705432,neubuser2021optimising,HGCAL-TDR} as well as multiple, even overlapping showers while at the same time being adaptable to the particularities of the involved detector geometries~\cite{Qasim_2019,ju2020graph,qasim2021multiparticle}. Also pattern recognition applications for quick identification of pointing and non-pointing showers at trigger level have been proposed~\cite{Alimena_2020,Iiyama_2021}. 
Following the success of such applications, we chose a deep learning approach to the problem, based on convolutional neural networks and loosely inspired by the techniques used for reconstructing hadronic showers in~\cite{Neubuser:2705432,neubuser2021optimising}.

The plan of this document is as follows. In \autoref{s:detector} we describe the idealised calorimeter we have employed for this study. In \autoref{s:regressor} we discuss the architecture of the convolutional neural network we used for the regression of muon energy from the measured energy deposits. In \autoref{s:results} we detail our results. We offer some concluding remarks in \autoref{s:conclusion}. In Appendix~\ref{s:features} we describe the high-level features we constructed from energetic and spatial information of each muon interaction event; these features are used as additional input for the regression task. In Appendix~\ref{sec:ablation} we offer an extensive ablation study of the model architecture and loss, the training schedule, and other technical aspects of our approach. Finally, in \autoref{sec:requirements} we describe the hardware and time requirements of both the study and the regressor.

A public version of the research code is available from Ref.~\cite{muon_reg_repo}. The pre-processed datasets are available from Ref.~\cite{muon_reg_dataset}, and are designed to be used directly with the code-base.

%%%%%%%%%%%%%%%%%%%%%%%%%%%%%%%%%%%%%%%%%%%%%%%%%%%%%%%%%%%%
\section{Detector geometry and simulation}\label{s:detector}

%%%%%%%%%%%%%%%%%%%%%%%%%%%%%%
\subsection{Detector geometry}

\noindent
Since our goal in this work is to show the feasibility of muon-energy estimation from energy deposits in a calorimeter, we strip the problem of any complication from factors that are ancillary to the task. For that reason, we consider a homogeneous lead tungstate cuboid calorimeter with a total depth in $z$ of $\SI{2032}{\mm}$ and a spatial extent of $\SI{120}{\mm}$ in $x$ and $y$. The calorimeter is segmented into 50 layers in $z$, each with a thickness of $\SI{39.6}{\mm}$; this corresponds to 4.5 radiation lengths. Such a longitudinal segmentation allows for electromagnetic showers to be well resolvable. Each layer is further segmented in $x$ and $y$ in $32 \times 32$ cells, with a size of $\SI{3.73}{\mm} \times \SI{3.73}{\mm}$. This results in \num{51200} channels in total. 

We assume that the calorimeter is embedded in a uniform 2-Tesla magnetic field, provided by an external solenoid or dipole magnet. The chosen magnet strength equals that of the ATLAS detector, and is in the range of what future collider detectors will likely be endowed with. We note that the magnetic bending of muon tracks inside the calorimeter volume is very small in the energy range of our interest (\SI{1}{\tev} and above), and its effect on the regression task is negligible there\footnote{A \SI{1}{\tev} muon traversing a uniform 2-Tesla field for \SI{2.032}{\meter} withstands a transverse displacement of \SI{1.24}{\mm} from its original trajectory, which is less than a third of a calorimeter cell.}. In the studies reported {\em infra} we will both compare the curvature-based momentum estimate provided by an ATLAS-like detector to the radiative losses-driven one, and combine the two to show their complementarity.

\begin{figure}[ht]
    \begin{center}
        \begin{subfigure}[t]{\sfSmall\textwidth}
            \begin{center}
               \includegraphics[width=\textwidth]{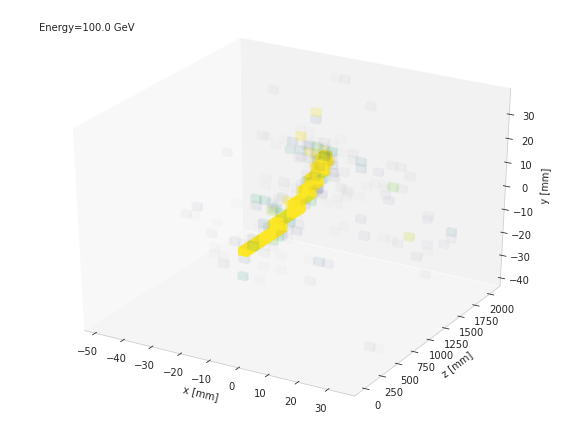}
                \caption{A low-energy \SI{100}{\gev} muon, in which few off-track deposits are produced, and the magnetic bending is clearly visible.}
            \end{center}
        \end{subfigure}
        \begin{subfigure}[t]{\sfSmall\textwidth}
            \begin{center}
               \includegraphics[width=\textwidth]{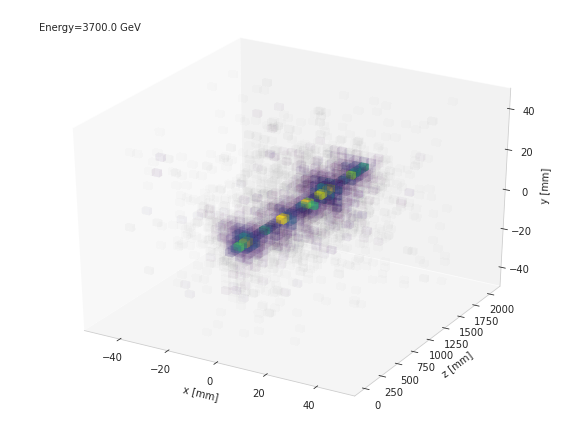}
                \caption{A high-energy \SI{3700}{\gev} muon, which produces many off-track deposits and is less affected by the magnetic bending.}
            \end{center}
        \end{subfigure}
        \caption{Pair of examples of muons entering the simulated calorimeter in the $z$ direction. The colour palette indicates the relative logarithmic energy of deposits, moving from blue, through green, to yellow in increasing energy.}
        \label{f:geometry}
    \end{center}
\end{figure}

%%%%%%%%%%%%%%%%%%%%%%%%%%%%%%%%
\subsection {Data generation}

\noindent
We generate unpolarised muons of both charges with a momentum $P = P_z$ in the $z$ direction, of magnitude ranging between $\SI{50}{\gev}$ and $\SI{8}{\tev}$. This interval extends beyond the conceivable momentum range of muons produced by a future high-energy electron-positron collider such as CepC or FCC-ee~\cite{fcc-ee}, and it therefore enables an unbiased study of the measurement of that quantity in an experimentally interesting scenario. 

The generated initial muon position in the $z$ coordinate is set to $z=-\SI{50}{\mm}$ with respect to the calorimeter front face; its $x$ and $y$ coordinates are randomly chosen within $|x|\leq\SI{20}{\mm}$ and $|y|\leq\SI{20}{\mm}$. The momentum components in $x$ and $y$ direction are set to zero. As mentioned {\em supra}, to compare the curvature-based and calorimetric measurement we assume that the calorimeter is immersed in a constant $B=2T$ magnetic field, oriented along the positive $y$ direction.
The detector geometry and the radiation pattern of a muon entering the calorimeter are shown in \autoref{f:geometry}. Even at a relatively low energy of $\SI{100}{\gev}$, the produced pattern of radiation deposits is clearly visible and we can also see that the multiplicity of deposits grows with the muon energy. The interaction of the muons with the detector material is simulated with \geant~4~\cite{GEANT4_0,GEANT4_1} using the FTFP\_BERT physics list.  

For the training and validation tasks of the regression problem a total of \num{886716} muons are generated, sampled from a Uniform distribution in the \SIrange{0.05}{8}{\tev} range. Additional muon samples, for a total of \num{429750} muons, are generated at fixed values of muon energy (E=100, 500, 900, 1300, 1700, 2100, 2500, 2900, 3300, 3700, \SI{4100}{\gev}) in order to verify the posterior distributions in additional tests discussed {\em infra}, \autoref{s:results}. Such a discrete-energy dataset allows us to compute precisely the resolution of the trained regressor at specific muon energies, rather than having to bin muons according to their energy. 

%\FloatBarrier
%%%%%%%%%%%%%%%%%%%%%%%%%%%%%%%%%%%%%%%%%%%%%%%%%%%%
\section{The CNN regression task}\label{s:regressor}

\begin{figure*}[ht]
        \begin{center}
            \begin{subfigure}[t]{\sfSmall\textwidth}
                \begin{center}
                   \includegraphics[width=\textwidth]{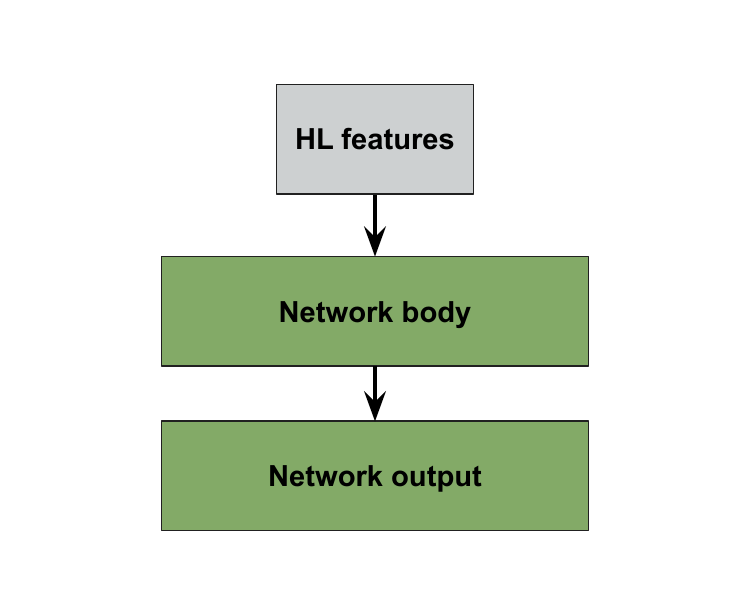}
                    \caption{High-level model, in which flat, continuous high-level features are fed directly to the network body.}
                    \label{fig:models:hl}
                \end{center}
            \end{subfigure}
            \begin{subfigure}[t]{\sfSmall\textwidth}
                \begin{center}
                   \includegraphics[width=\textwidth]{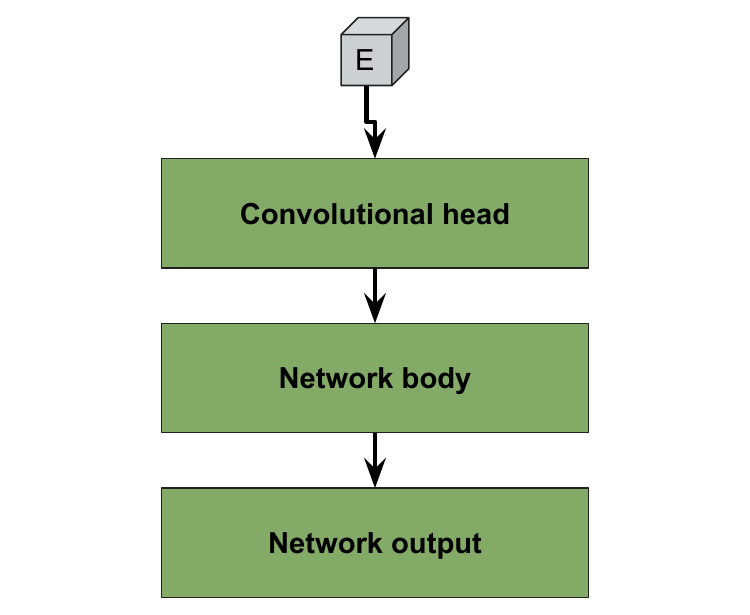}
                    \caption{Convolutional model, in which the 3D grid of raw energy deposits is fed through the convolutional head.}
                    \label{fig:model:conv}
                \end{center}
            \end{subfigure}
            \begin{subfigure}[t]{\sfSmall\textwidth}
                \begin{center}
                   \includegraphics[width=\textwidth]{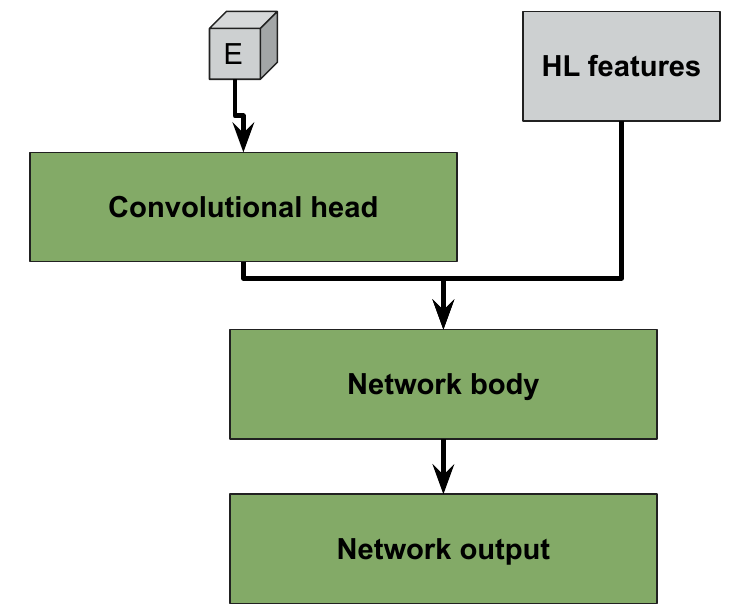}
                    \caption{The hybrid model, in which both the 3D grid of raw energy deposits and the high-level features are used.}
                    \label{fig:blocks:hybrid}
                \end{center}
            \end{subfigure}
            \caption{Diagrams illustrating the three types of models used.}
            \label{fig:models}
        \end{center}
    \end{figure*}

    Three regressor architectures are considered: regressors that only use continuous input-features (such as the energy sum and other high-level features) pass their inputs through a set of fully-connected layers (referred to as the network \textit{body}), ending with a single-neuron output; when the 3D grid of energy deposits is considered, the body is prepended with a series of 3D convolutional layers (referred to as the \textit{head}), which act to reduce the size of the grid, whilst learning high-level features of the data, prior to passing the outs to the body; the main model used is a hybrid model combining both approaches, in which the energy deposits are passed through the head, and the pre-computed high-level features are passed directly to the body. Layout diagrams for these three models are illustrated in \autoref{fig:models}, and a technical description of component is included in the following subsection. Models are implemented and trained using \pytorch~\cite{pytorch} wrapped by \lumin~\cite{lumin} - a high-level API which includes implementations of the advanced training techniques and architecture components we make use of in the regressor.
    
    %%%%%%%%%%%%%%%%%%%%%%%%%%%%%%%%%%%%%%%%%%%%%%%%%%%%
    \subsection{Architecture components}\label{sec:arch}
        %%%%%%%%%%%%%%%%%%%%%%%%%%%%%%%%%%
        \subsubsection{Convolutional head}
            The head architecture is inspired by domain knowledge and is based on the fact that the sum of the energy deposits is related to the energy of the traversing muon, however accurate correspondence requires that the deposits receive small corrections based on the distribution of surrounding deposits. The convolutional architecture draws on both the \textsc{DenseNet}~\cite{densenet} and \textsc{ResNet}~\cite{resnet} architectures, and is arranged in blocks of several layers. Within each block, new channels are computed based on incoming channels (which include the energy deposits) using a pair of 3D convolutional layers. The channels computed by the convolutional layers are weighted by a squeeze-excitation (SE) block~\cite{senet}. The convolutional plus SE path is by-passable via a residual sum to an identity path. At the output of the block, the channel corresponding to the energy deposits is concatenated (channel-wise) to the output of the addition of the convolutional layers and the identity path \footnote{By convention, the energy channel is kept as the zeroth channel.}. In this way, convolutional layers always have direct access to the energy deposits, allowing their outputs to act as the ``small corrections" required. 
            
\begin{figure*}[ht]
                \begin{center}  
                    \begin{subfigure}[t]{\sfSmall\textwidth}
                        \begin{center} \includegraphics[width=\textwidth]{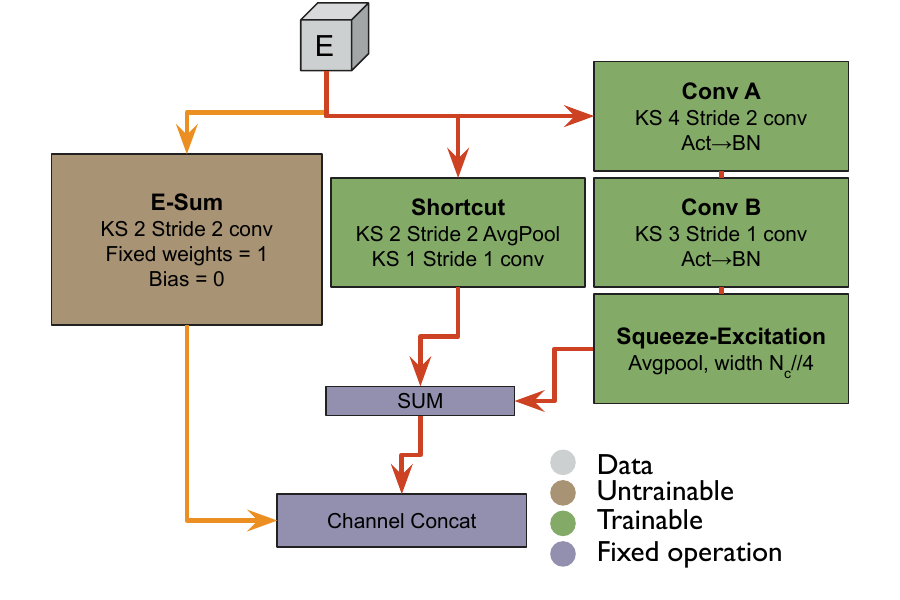}
                            \caption{Initial block taking the grid of reconstructed energy, E, as input and downsampling. The E-Sum layer uses a kernel size equal to its stride and so reduces the size of the grid by two without missing any deposits. The identity and convolutional paths also use stride two filters in their first layers, in order to produce representations of the same dimensionality. Since downsampling is required, the identity path is a convolutional shortcut, consisting of a stride two average pooling, followed by a stride one convolution. ``Act" = activation function, ``BN" = batch normalisation, ``AvgPool" = average-pooling layer, and ``conv" = convolutional layer.}
                            \label{fig:blocks:initial}
                        \end{center}
                    \end{subfigure}
                    \begin{subfigure}[t]{\sfSmall\textwidth}
                        \begin{center} \includegraphics[width=\textwidth]{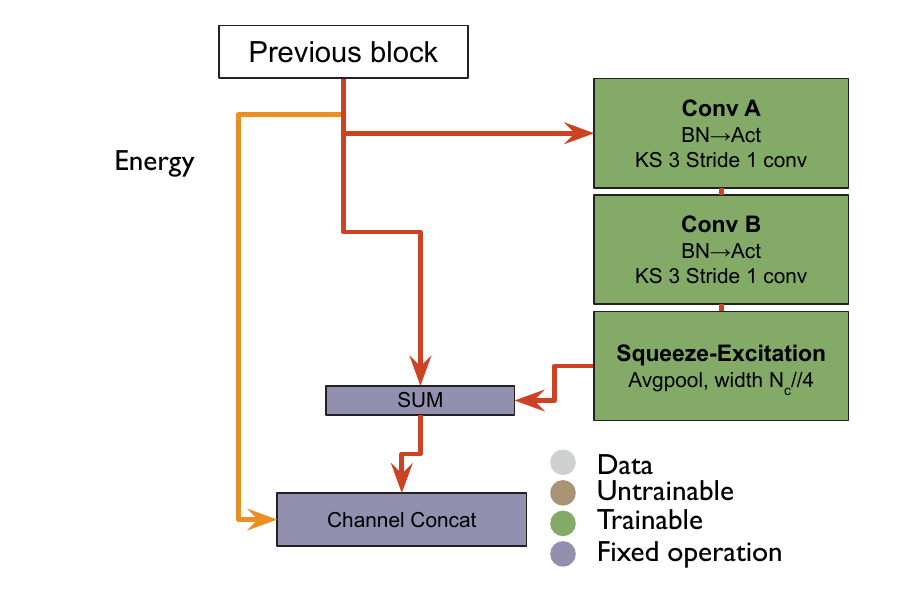}
                            \caption{A subsequent block in which no downsampling occurs. The Energy path refers only to the zeroth channel of the output of the previous block. The identity and convolutional blocks act on all channels. The identity path here is straightforward, as no downsampling or changes in number of channels is required. Note that the convolutional path now has the batch normalisation layers (BN) before the activation and convolutional layers, a la pre-activation \textsc{ResNet}.}
                            \label{fig:blocks:subsequent}
                        \end{center}
                    \end{subfigure}
                    \begin{subfigure}[t]{\sfSmall\textwidth}
                        \begin{center} \includegraphics[width=\textwidth]{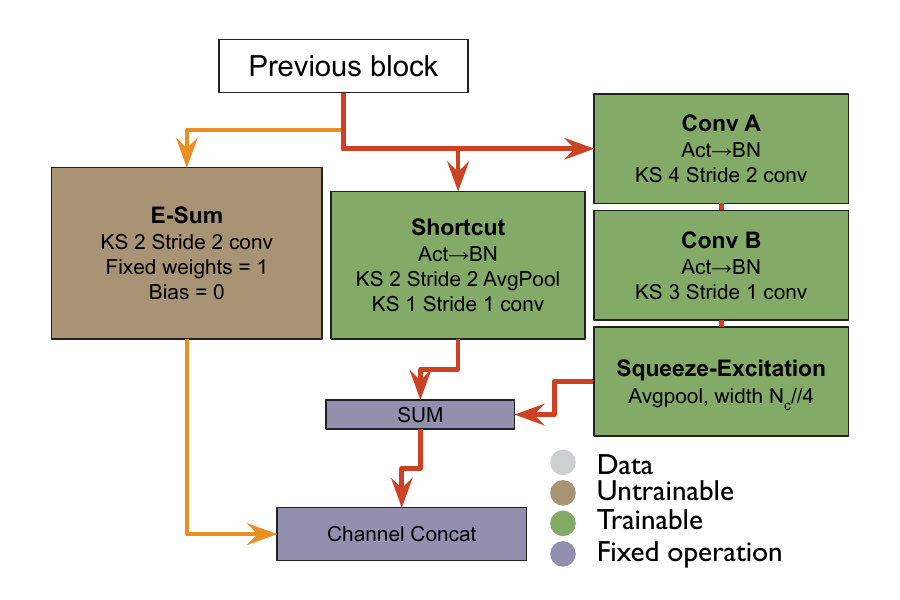}
                            \caption{A subsequent block in which downsampling does occur. This is effectively a pre-activation version of the block illustrated in Fig.~\sfCaption{fig:blocks:initial}, except that E-Sum layer acts only on the zeroth channel of the output of the previous block and the identity path includes activation and batch normalisation layers. The identity and convolutional blocks act on all channels.}
                            \label{fig:blocks:downsample}
                        \end{center}
                    \end{subfigure}
                    \caption{Diagrams illustrating the three types of blocks used to construct the convolutional heads.}
                    \label{fig:blocks}
                \end{center}
            \end{figure*}

\begin{figure}[ht]
                \begin{center}
                    \includegraphics[width=\sfSmall\textwidth]{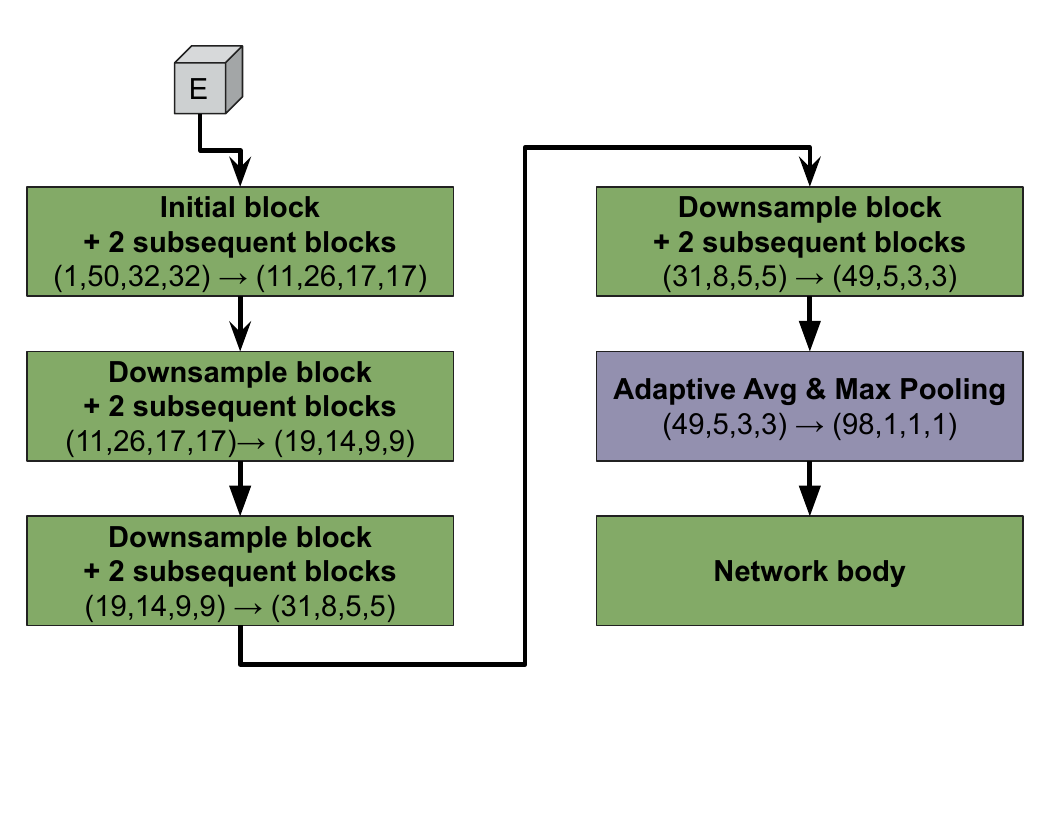}
                    \caption{Block layout for the convolutional head. Tensor dimensions are indicated in the form $(\text{channel},z,x,y)$. The convention is to increase the number of channels to eight in the first downsample, and then increase the number of channels at each downsample by a factor of \num{1.5}. The number of channels increases by one in each block due to the energy concatenation. Prior to being fed into the network body, the tensor is pooled by computing the maximum and mean of each channel. The data-batch dimension is not shown, for simplicity.}
                    \label{fig:conv_head}
                \end{center}
            \end{figure}

            The architecture becomes slightly more complicated when the energy is downsampled; in such cases, convolutional shortcuts~\cite{preact_resnet} are used on the identity path, and fixed, unit-weighted convolutional layers with strides equal to their kernel size are applied to the energy deposits. These fixed kernels act to sum up the energy deposited within each sub-cube of the detector, and are referred to here as the ``E-sum layers". This approach is strongly inspired by~\cite{Neubuser:2705432,neubuser2021optimising}.
            Additionally, for blocks after the very first one, a \textit{pre-activation} layout~\cite{preact_resnet} is adopted with regards to the placement of batch normalisation layers. Figure~\ref{fig:blocks} illustrates and discusses the general configurations of the three types of blocks used.
            Sets of these convolutional blocks are used to construct the full convolutional head. In all cases, the grid is downsampled four times, each time with a reduction by a factor of two. However, non-downsampling blocks (\autoref{fig:blocks:subsequent}) may be inserted in between the downsampling blocks in order to build deeper networks. Figure~\ref{fig:conv_head} illustrates the layout of the full convolutional head.
            
            \paragraph{Technical specification}
                In the convolutional layers, the kernel sizes of all convolutional and average-pooling layers are set to three, with the exception of the first convolution in downsampling and initial blocks, which use a kernel size of four, to match the stride and padding of the E-sum layer. Zero-padding of size one is used (two when the kernel size is four). Swish activation-functions~\cite{swish} are used with $\beta=1$ (Swish-1). Weights are initialised using the Kaiming rule~\cite{He}, with the exception of the E-Sum layers, which are initialised with ones. No biases are used.
                
                The squeeze-excitation blocks feed the channel means into a fully connected layer of width $\max\left(2,N_c//4\right)$ ($N_c$ = number of channels, $//$ indicates integer division, Kaiming weight initialisation and zero bias initialisation) and a Swish-1 activation, followed by a fully connected layer of width $N_c$ (Glorot~\cite{Glorot} weight initialisation and zero bias initialisation) and a sigmoid activation. This provides a set of multiplicative weights per channel which are used to rescale each channel prior to the residual sum.
                
                Due to the sparse nature of the data, we found in necessary to use \textit{running} batch-normalisation~\cite{running_bn}. This modifies the batch normalisation layers to apply the same transformation during both training and inference, {\em i.e.} during training, the batch statistics are used to update the running averages of the transformation, and then the averaged transformation is applied to the batch (normally only batch-wise statistics are used to transform the training batches, causing potential differences between training and inference computations). Additionally, running averages of the sums and squared sums of the incoming data are tracked, rather than the mean and standard deviation, allowing the true standard deviation to be computed on the fly (normally the average standard deviation is used). Together, these changes with respect to a default batch normalisation implementation provide greater stability during training, and enabled generalisation to unseen data. All batch normalisation layers use a momentum of 0.1, meaning that the running average of statistic $\theta$ is tracked according to $\bar{\theta}\leftarrow0.9\bar{\theta}+0.1\theta_{\text{batch}}$.
            
        %%%%%%%%%%%%%%%%%%%%%%%%%%%%    
        \subsubsection{Network body and output}
            The body of the network is relatively simple, and consists of three fully connected layers, each with 80 neurons. Weights are initialised using the Kaiming rule, and biases are initialised with zeros. Swish-1 activation functions are placed after every layer. No batch normalisation is used.

            The output layer of the network consists of a single neuron. Weights are initialised using the Glorot rule, and the bias is initialised to zero. No activation function is used.

    %%%%%%%%%%%%%%%%%%%%%%%%%%%%%%%%%%%%%%%%%            
    \subsection{Training}\label{sec:training}
        %%%%%%%%%%%%%%%%%%%%%%%%%%%%%%%%%%%%%%%%%%%%%%
        \subsubsection{Data}\label{s:data_description}
            Models are trained on simulated data for the full considered range of muon true energy, \SIrange{50}{8000}{\gev}. The 3D grid of raw energy deposits does not undergo any preprocessing, nor do the target energies. When used, the measured energy extracted from the curvature fit (V[24], see {\em infra}, Appendix~\ref{s:features}) is clamped between 0 and \SI{10}{\tev}~\footnote{The computation code provides a signed energy according to the direction of curvature as dictated by the charge of the muon, but the sign is dropped before using the feature.}. All high-level features are then standardised and normalised by mean-subtraction and division by standard deviation.
            
            The full training dataset consists of \num{886716} muons. This is split into 36 \textit{folds} of \num{24631} muons; the zeroth fold is used to provide a hold-out validation dataset on which model performance is compared. During training a further fold is used to provide monitoring validation to evaluate the general performance of the network and catch the point of highest performance.
            
            Prior to using the discrete-energy testing-data to compute the resolution, the continuous-energy validation dataset is finely binned in true energy, allowing us to compute an approximation of the resolution at the central energy of the bin (computed as the median true-energy of muons in the bin).
        
        %%%%%%%%%%%%%%%%%%%%
        \subsubsection{Loss}
            Models are trained to minimise a Huberised~\cite{huber} version of the mean fractional squared error (MFSE):
            \begin{equation}
                \Lagr\!\left(y,\hat{y}\right)=\frac{1}{N}\sum_{n=1}^N\frac{\left(y_n-\hat{y_n}\right)^2}{y_n},
                \label{eq:MFSE}
            \end{equation}
            where $y$ is the true muon-energy, $\hat{y}$ is the predicted energy, and $N$ is the batch size. The form of this loss function reflects the expectation of a linear scaling of the variance of the energy measurement with true energy, as is normally the case for calorimeter showers when the energy resolution is dominated by the stochastic term. In this study, the batch size used for training the models is 256.
            
            \paragraph{Huber loss}\label{s:huber_loss}
                To prevent non-Gaussian tails of the regressed muon energy distribution from dominating the loss estimate, element-wise losses are first computed as the squared error, $\left(y_n-\hat{y_n}\right)^2$, and high-loss predictions above a threshold are modified such that they correspond to a linear extrapolation of the loss at the threshold:
                \begin{equation}
                    \Lagr_{\mathrm{Huber},i} = t+\left(2\sqrt{t}\left(\left|y_i-\hat{y_i}\right|-\sqrt{t}\right)\right),
                \end{equation}
                where $i$ are indices of the data-points with a squared-error loss greater than the threshold $t$. This Huberised element-wise loss is then divided by the true energy to obtain the fractional error, which is then multiplied by element-wise weights (discussed below) and averaged over the data points in the batch.
                
                Since the loss values vary significantly across the true-energy spectrum, data points are grouped into five equally sized bins, each of which has its own threshold used to define the transition to the absolute error. The transition point used for a given bin is the \nth{68} percentile of the distribution of squared-error losses in that bin (allowing the threshold to always be relevant to the current scale of the loss as training progresses). However, since for a batch size of 256 one expects only 51 points per bin, the threshold can vary significantly from one batch to another. To provide greater stability, the bin-wise thresholds are actually running averages of the past \nth{68} percentiles, again with a momentum of 0.1, {\em i.e.} for bin $j$, the threshold is tracked as $t_j\leftarrow0.9t_j+0.1\Lagr_{\mathrm{SE},j,\nth{68}}$, where $\Lagr_{\mathrm{SE},j,\nth{68}}$ is the \nth{68} percentile of the squared errors in bin $j$.
            
            \paragraph{Data weighting}
                Models are trained on muons of true energy in the \SIrange{50}{8000}{\gev} range, but will only be evaluated in the range \SIrange{100}{4000}{\gev} in order to avoid biases due to edge effects; effectively the regressor can learn that no targets exist outside of the data range, and so it is more efficient to only predict well within the data-range. This leads to an overestimation of low-energy muons, and an underestimation of high-energy muons. By training on an extended range and then evaluating on the intended range, these edge-effects can be mitigated. Yet we still want the network to focus on the intended range; rather than generating data with a pre-defined PDF in true energy, we use a uniform PDF and down-weight data with true muon energy outside the range of interest.
                
\begin{figure}[ht]
            		\begin{center}
            			\includegraphics[width=\sfSmall\textwidth]{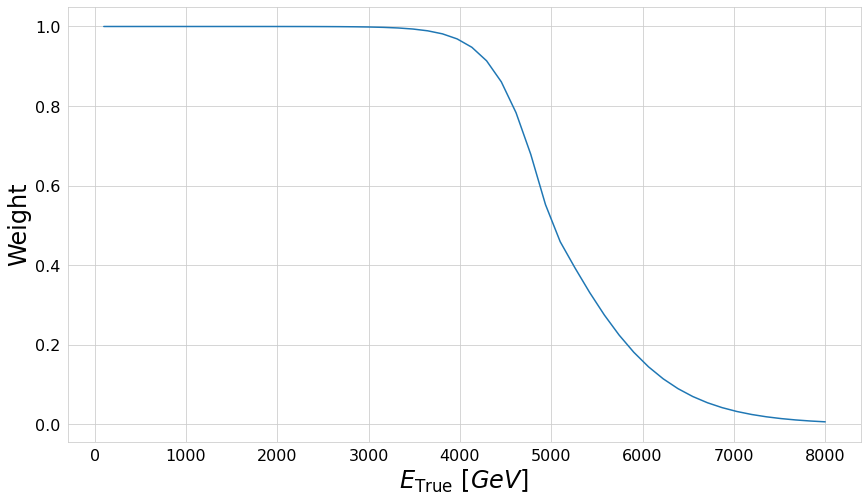}
            			\caption{Data weight as a function of true muon energy.}
            			\label{fig:downweighting}
            		\end{center}
            	\end{figure}

                The weighting function used depends solely on the true energy of the muons and takes the form of:
                \begin{equation}
                    w = 
                    \begin{cases}
                        1-Sigmoid\left(\frac{E-5000}{300}\right)\quad&E\leq\SI{5000}{\gev},\\
                        1-Sigmoid\left(\frac{E-5000}{600}\right)\quad&E>\SI{5000}{\gev}.\\
                    \end{cases}
                \end{equation}
                This provides both a quick drop-off above the intended range, and a slow tail out to the upper-limit of the training range. Figure~\ref{fig:downweighting} illustrates this weighting function. It should be noted that the above weights correspond to a comparatively smooth modification of the true energy prior; for specific applications where the physics puts hard boundaries on the energy spectrum (such as, {\em e.g.}, a symmetric electron-positron collider, where one may safely assume that muons cannot be produced with energy larger than the beam energy) a sharper prior may be used instead, and significantly improve the resolution at the high end of the spectrum.
                
        %%%%%%%%%%%%%%%%%%%%%%%%%    
        \subsubsection{Optimiser}
\begin{figure*}[ht]
			    \begin{center}
    				\begin{subfigure}[t]{0.6\textwidth}
    					\begin{center}
    					\includegraphics[width=\textwidth]{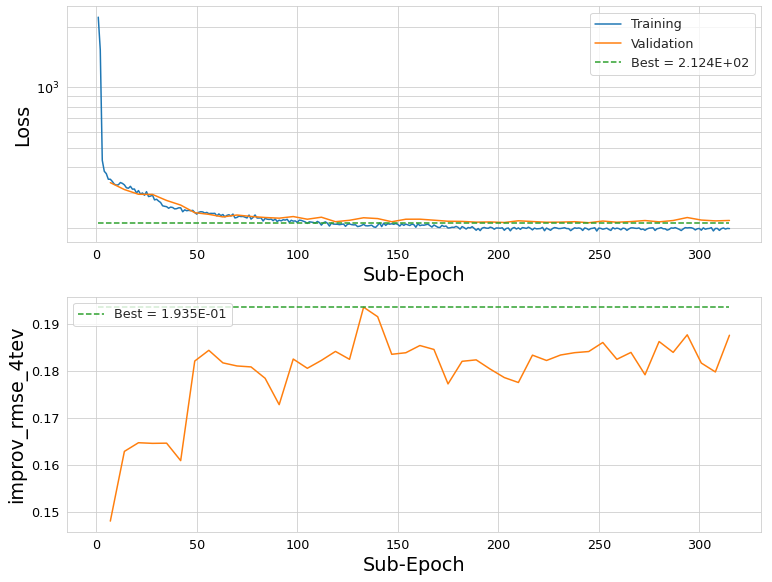}
    						\caption{Typical loss evolution during the training of a single model. The evolution of the Mean Improvement is also shown (described in detail in \autoref{sec:resolution}). Whilst the MI is what we aim to maximise, it fluctuates too much during training to provide a reliable indication of the point of best performance, and instead the validation loss is used to select the best model.}
    						\label{fig:training:history:loss}
    					\end{center}
    				\end{subfigure}
    				\begin{subfigure}[t]{0.6\textwidth}
    					\begin{center}
    						\includegraphics[width=\textwidth]{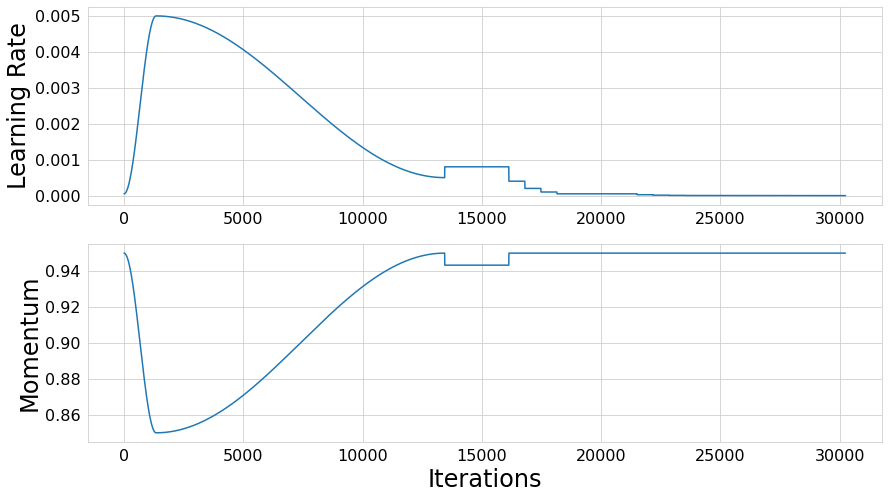}
    						\caption{Learning rate and momentum schedule associated with the training shown in Fig.~\sfCaption{fig:training:history:loss}. Initially, the parameters evolve as per 1cycle with cosine interpolation. Following this fixed period, the best performing models are continually reloaded, and the LR evolves as a step-decay whenever the model fails to improve.}
    					\end{center}
    				\end{subfigure}
    				\caption{Details of a typical training, showing the loss and metric evolution and the associated schedule of the optimiser hyper-parameters.}
    				\label{fig:training:history}
    			\end{center}
    		\end{figure*}

            The \adam optimiser~\cite{kingma2014adam} is used for updating the model weights. The $\epsilon$ and $\beta_2$ parameters are kept constant, at \num{1e-8} and \num{0.999}, respectively. The learning rate (LR) and $\beta_1$ (momentum coefficient) are adjusted during training in two stages. For the first 20 epochs of training, the 1cycle schedule~\cite{Smith_2017,Smith_2018}, with cosine interpolation~\cite{fastai_docs}, is used to train the network quickly at high learning rates; this is followed by up to 30 epochs of a step-decay annealing~\cite{resnet}, which is used to refine the network at small learning rates. For the 1cycle schedule, training begins at an LR of \num{3e-7} and $\beta_1=0.95$. Over the first two epochs of training the LR is increased to \num{3e-5} and $\beta_1$ is decreased to 0.85. Over the next 18 epochs, the LR is decreased to \num{3e-6}, and $\beta_1$ increased back to 0.95. Following this, the best performing model-state, and its associated optimiser state, is loaded and training continues at a fixed LR and $\beta_1$ until two epochs elapse with no improvement in validation loss. At this point, the best performing model-state is again reloaded, $\beta_1$ is set to 0.95, and the LR is halved. This process of training until no improvement, reloading, and halving the LR continues until either all 50 epochs have elapsed, or 10 epochs elapse with no improvement. At this point the best performing model-state is again loaded and saved as the final model. Figure~\ref{fig:training:history} details a typical training with such a schedule.
            
            Explicit, tunable regularisation was not found to be required during training. Instead, overtraining is prevented by continual monitoring of the model performance on a separate validation sample, and saving of the model parameters whenever the validation loss improves.
            
        %%%%%%%%%%%%%%%%%%%%%%%%%%%%%%%%%%%%%%%%%%%%%%%%%%%%%%%%%%%%%%
        \subsubsection{Ensemble training}\label{sec:ensemble_training}
            As mentioned in~\autoref{s:data_description}, the training dataset is split into 36 folds, one of which is retained to provide a comparison between models, and another is used to monitor generalised performance during training. During development and the ablation study (discussed {\em infra}, Appendix~\ref{sec:ablation}), it was useful to obtain an averaged performance of the model architecture from five repeated trainings. Since, however, one training on the full dataset takes about one day, we instead ran these trainings on unique folds of the full dataset, using different folds to monitor generalisation, {\em i.e.} each model is trained on seven folds and monitored on one fold, and no fold is used to train more than one model (but folds can be used to monitor performance for one model and also to train a different model). This allows us to train an ensemble of five models in just one day, and also to get average performance metrics over the unique validation folds, to compare architecture settings. This method of training is referred to as ``unique-fold training''.
            
            For the full, final ensemble, each model is trained on 34 folds and monitored on one fold, which is different for each model. Once trained, the ensemble is formed by weighting the contributions of each model according to the inverse of its validation performance during training. This method of training is referred to as ``all-fold training''.

%\FloatBarrier
%%%%%%%%%%%%%%%%%%%%%%%%%%%%%%%%%%%%%%%
\section {Results \label{s:results}}
    Unless explicitly specified, all results presented in this section refer to the main regression model, in which both raw energy-deposits and the high-level features are used.
    
    %%%%%%%%%%%%%%%%%%%%%%%%%%%%%%%%%%%%%%%%%%%%%%%%%%%%%%
    \subsection{Regressor response and bias correction}
\begin{figure}[ht]
    		\begin{center}
				\includegraphics[width=\sfSmall\textwidth]{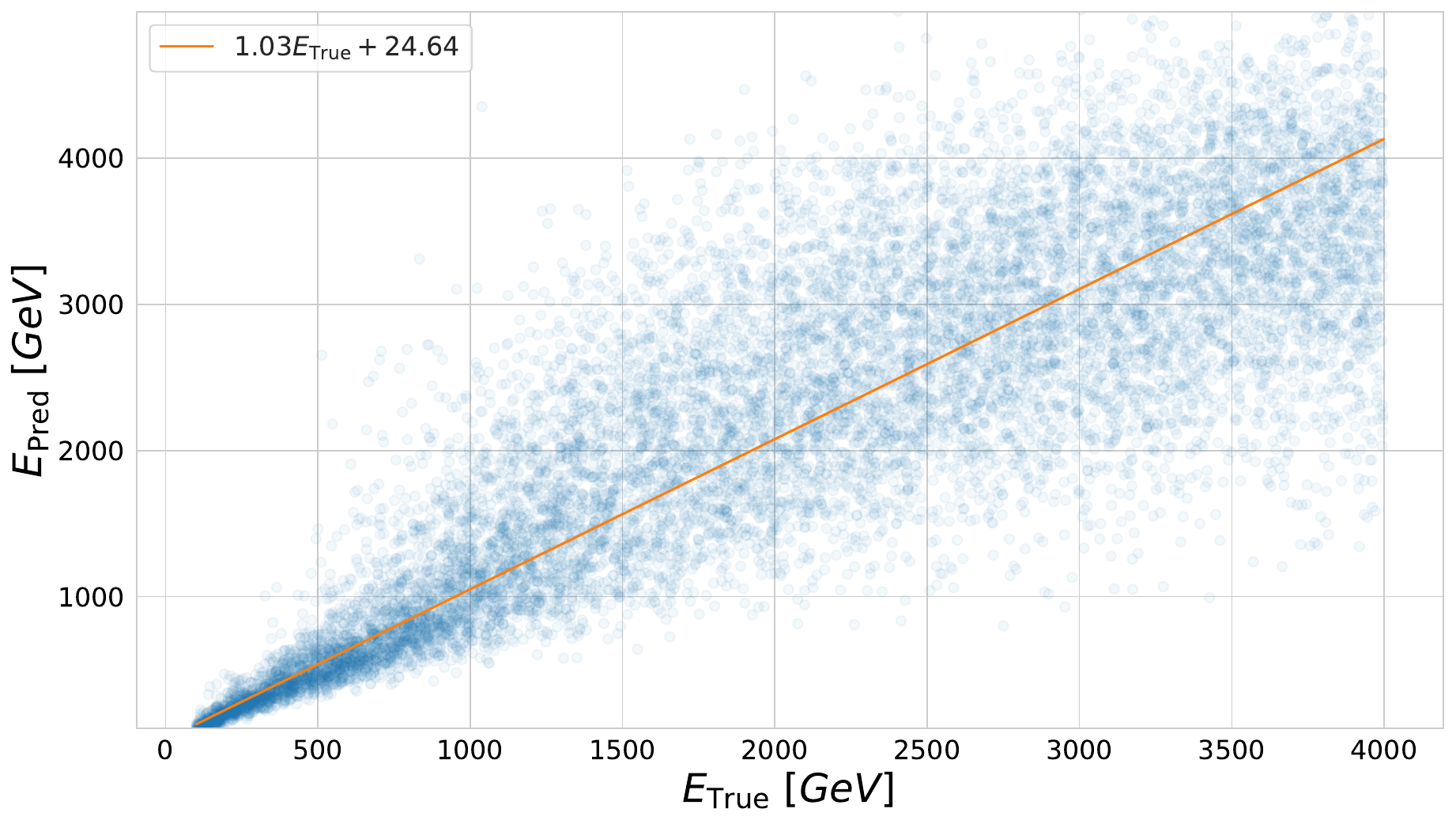}
				\caption{Raw predictions of the regressor ensemble as a function of true energy. The ideal response is for all points to lie on a straight line along $y=x$. The green line shows a linear fit to predictions in bins of true energy.}
				\label{fig:results:raw_response}
			\end{center}
	    \end{figure}

\begin{figure}[ht]
			\begin{center}
				\includegraphics[width=\sfSmall\textwidth]{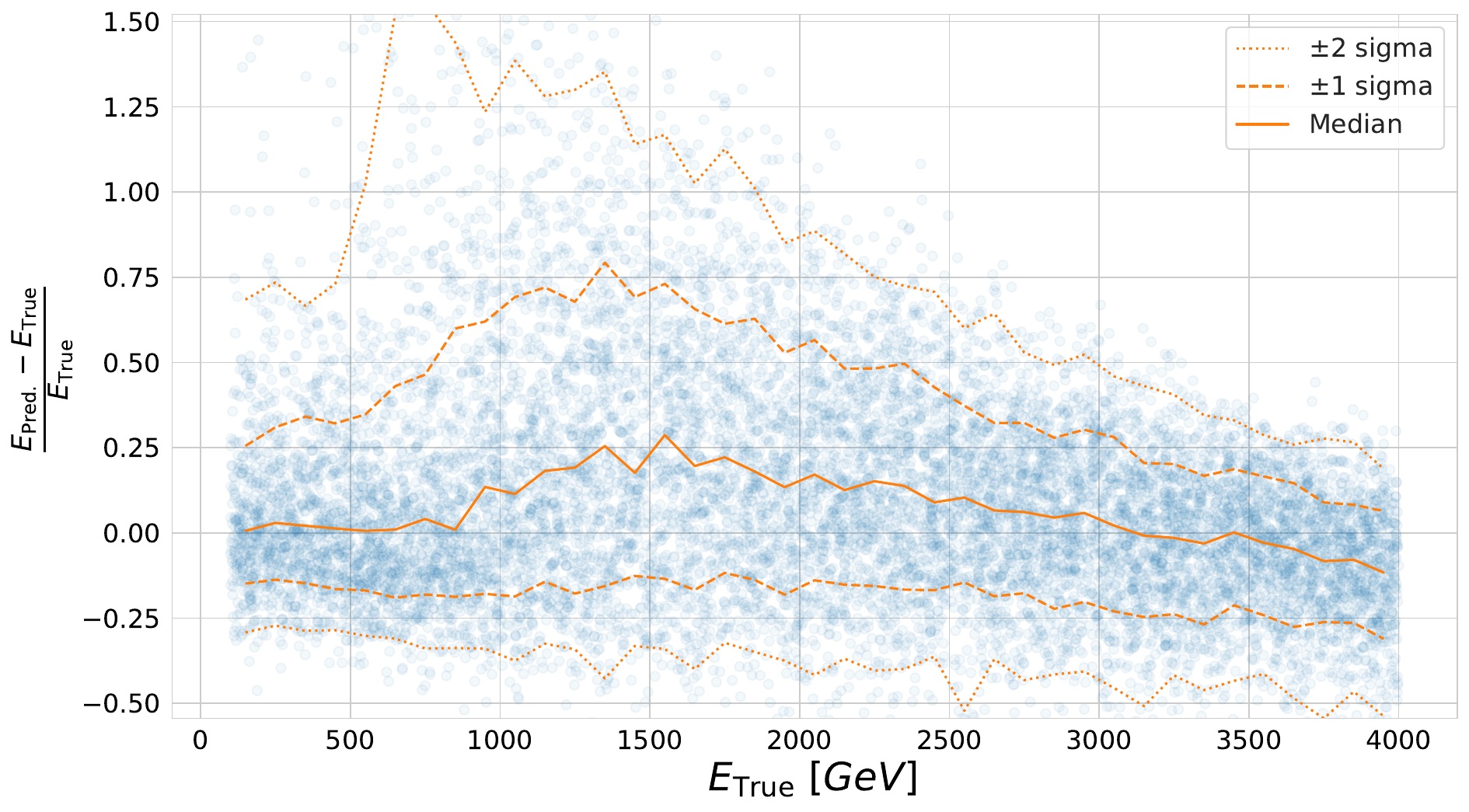}
				\caption{Fractional error of predictions as a function of true energy, along with quantile trends. The ideal response is for all points to lie on a straight line along $y=0$. The one and two sigma lines indicate the $50\pm\SI{34.1}{\%}$ and $50\pm\SI{47.7}{\%}$ percentiles, respectively.}
				\label{fig:results:raw_error}
			\end{center}
	    \end{figure}

        Figure~\ref{fig:results:raw_response} shows the predictions of the regression ensemble as a function of true energy for the holdout-validation dataset. Whilst the general trend is linear, we can see some dispersion, and \autoref{fig:results:raw_error} better details the fractional error as a function of true energy, along with the trends in the quantiles. From this we can see that regressor overestimates medium energies, and underestimates high energies. Low energies are predicted without significant bias. 
            
\begin{figure}[ht]
		    \begin{center}
				\includegraphics[width=\sfSmall\textwidth]{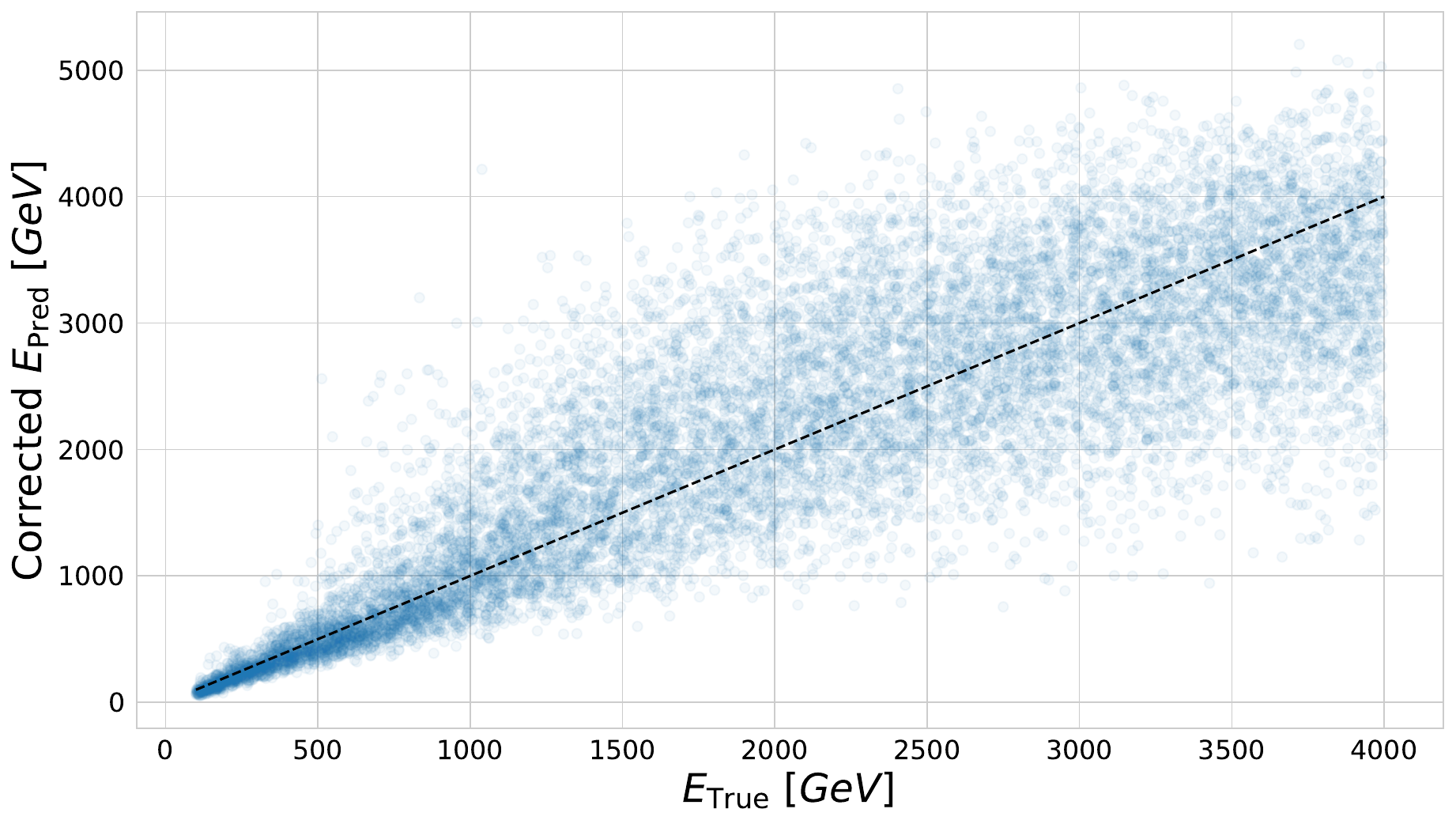}
				\caption{Corrected predictions on validation data resulting from the inversion of the fit as a function of the true energy. The black dashed line indicates the ideal response.}
				\label{fig:results:debias}
			\end{center}
		\end{figure}

	    We can correct for the bias in the prediction, however we must do so in a way that does not assume knowledge of the true energy, such that the correction can also be applied to prediction in actual application. The method used is to fit a function (in this case a linear function - green line in \autoref{fig:results:raw_response}) to the mean of the predictions in bins of true energy, and their uncertainties as estimated from bootstrap resampling. Having fitted the function, the inverse of the function can now be used to look up the true energy of a given prediction, resulting in a \textit{corrected} prediction. Figure~\ref{fig:results:debias} illustrates the corrected predictions on both the continuous validation data. Although the difference is only slight, as we will see later in \autoref{sec:ablation}, the de-biased predictions allow for a better resolution once the residual biases in the predictions are accounted for. To best reproduce actual application, the debiasing correction is fixed using the validation data, and then applied as is to the testing data.
	    
\begin{figure*}[ht]
			\begin{center}
				\includegraphics[width=\sfMid\textwidth]{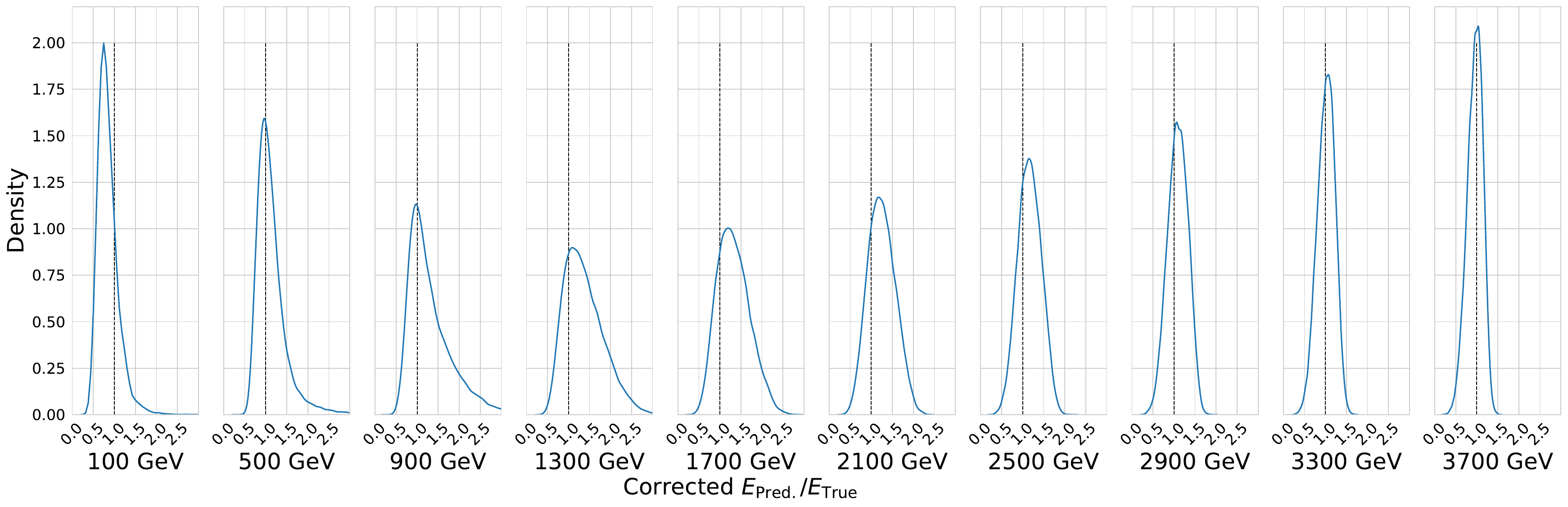}
				\caption{Distributions of the ratios of corrected predictions to true energies in bins of true energy. The ideal response here would be delta distributions centred at one. Distributions not centred at one are indicative of residual bias in the predictions for that energy range.}
				\label{fig:results:pulls}
			\end{center}
	    \end{figure*}

		Figure~\ref{fig:results:pulls} shows the distributions of the ratios of corrected predictions to true energies on the testing data.
		
    %%%%%%%%%%%%%%%%%%%%%%%%%%%%%%%%%%%%%%%%%%%%%%%%%%%%%%%%%%%%%%%%%%%%%%%%%%%%%%%%%%%%%%%%%%%    
    \subsection{Resolution and combination with curvature measurement}\label{sec:resolution}
       From the discussion in \autoref{s:introduction} we can expect that the relative resolution of the energy estimation from the calorimeter should improve as the energy increases, similarly we expect the resolution from magnetic-bending in the tracker will improve as the energy decreases. This difference in energy dependence means that the two measurements are complementary to one another and it would make sense in actual application to use both approaches in a weighted average. 
            
\begin{figure*}[ht]
    			\begin{center}
    				\includegraphics[width=\sfMid\textwidth]{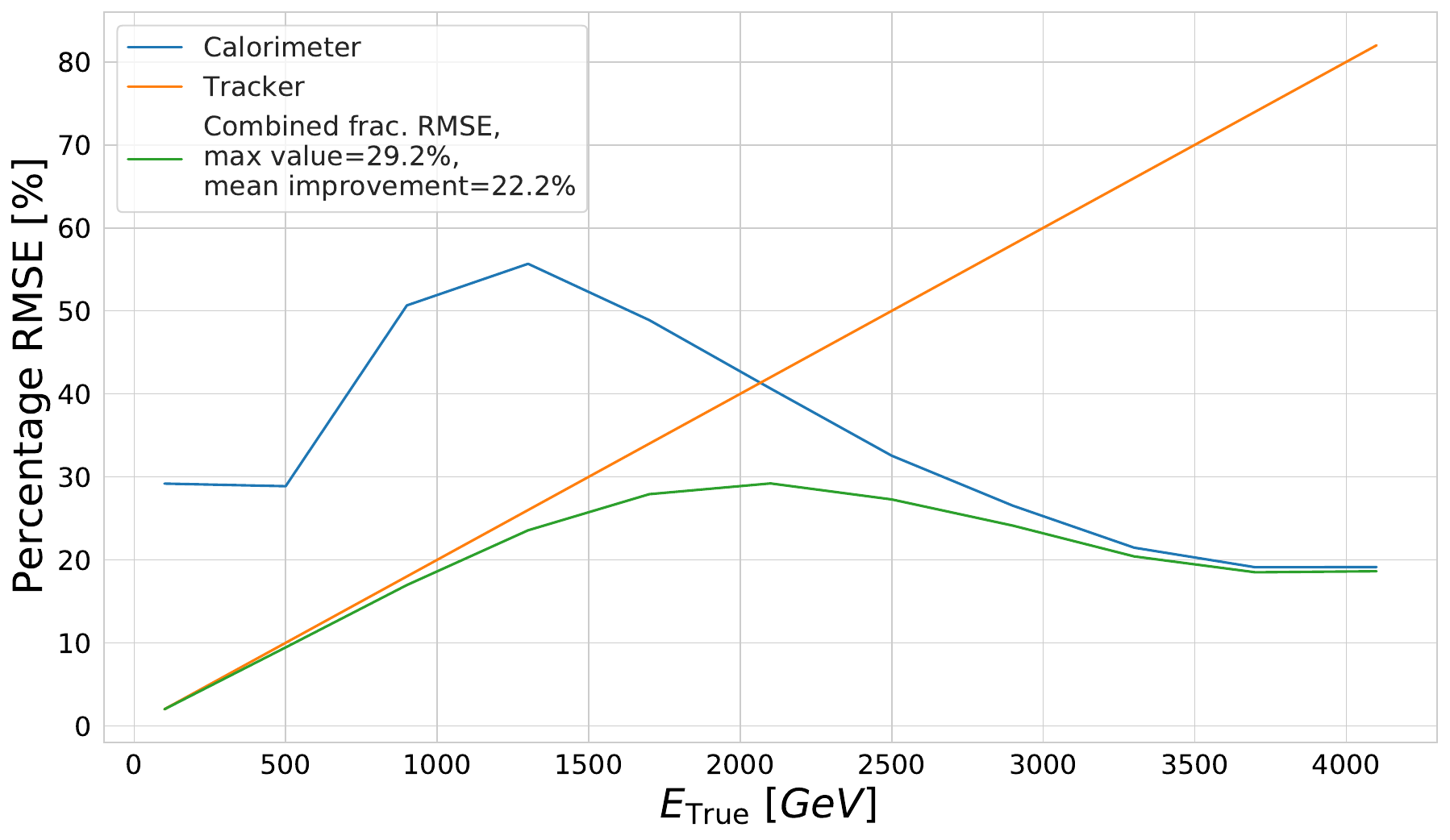}
    				\caption{Resolutions of the energy regression (Calorimeter), the simulated tracker, and their weighted average in a combined measurement. Resolution is computed on testing data at fixed points of true energy. The tracker is assumed to provide a linearly scaling resolution with a relative value of \SI{20}{\%} at \SI{1}{\tev}.}
    				\label{fig:results:rmse}
    			\end{center}
		    \end{figure*}

        Since our setup only includes a calorimeter, we assume that the resolution of a tracking measurement, performed independently by an upstream or downstream detector, scales linearly with energy, and equals \SI{20}{\%} resolution at \SI{1}{\tev}. Figure~\ref{fig:results:rmse} shows the resolution of both the regressor measurement and the simulated tracker measurement, along with the resolution of their weighted average. Resolution here is the fractional root median squared-error computed in bins of true energy according to:
            \begin{equation}
                \text{Resolution} = \frac{\sqrt{\left(\tilde{E_p}-\tilde{E_t}\right)^2+\Delta_{68}\left[E_p\right]^2}}{\tilde{E_t}},
            \end{equation}
            where $\tilde{E_p}$ and $\tilde{E_t}$ are the median predicted and true energies in a given bin of true energy (their difference being the residual bias after the correction via the linear fit), and $\Delta_{68}\left[E_p\right]$ is the difference between the \nth{16} and \nth{84} percentiles of the predicted energy in that bin (the central \nth{68} percentile width). When computing the resolution on the testing data (which are generated at fixed points of true energy), $\tilde{E_t}$ is instead the true energy at a given point.
            
		    It is interesting to note that the regression resolution initially gets worse with energy, rather than starting poor and gradually improving. Good resolution at low energy was not observed during development studies prior to the introduction of the magnetic field, therefore we assume that the CNNs are able to make use of the magnetic bending in the calorimeter to recover performance when there is reduced radiation. As expected, the regressor quickly improves in resolution once the energy reaches a certain threshold at around \SI{1.5}{\tev}.
		    
		    Having established that both the calorimeter and tracker measurements are useful and complementary, for later studies it makes sense to compare models in terms of the performance of the combined measurement. One such metric is the poorest resolution achieved by the combined measurement for the studied energy range (in this case \SI{29.5}{\%} - lower=better). This however relies only on a single point of the response. A more general metric is to compute the improvement of the combined measurement over the tracker-only measurement in bins of true energy, and take the average or sum; this then characterises the improvement due to the regression across the whole spectrum. We will refer to this metric as the Mean Improvement (MI). Considering the 11 points in the range \SIrange[]{100}{4100}{\gev}, our mean improvement is \SI{22.1}{\%} (higher=better). Computation of the MI on the validation data instead uses 20 bins in the \SIrange[]{100}{4000}{\gev}.
		    
	%\FloatBarrier
        
    %%%%%%%%%%%%%%%%%%%%%%%%%%%%%%%%%%%%%%%%%%%%%%%%%%%%%%%%%%%%%%%%%%%%%%%%%%%%%%%%%%%%%%%%%%%
    \paragraph{Input comparison: high-level features and raw inputs}\label{sec:input_study}
        As discussed in \autoref{s:features}, alongside the recorded calorimeter deposits, a range of high-level (HL) features are also fed into the neural network. To better understand what, if anything, the CNN learns extra from the raw information, we can study what happens when the inputs are changed. In \autoref{tab:results:inputs} we show the MI metric values for a range of different inputs. In cases when the raw inputs are not used, the neural network (NN) consists only of the fully connected layers. For this comparison, we use the MI computed during training on the monitoring-validation dataset and average over the five models trained per configuration.
            
            \begin{table}[ht]
            	\begin{center}
            		\begin{tabular}{lll}
            			\hline
            			Inputs & MI & Change in MI [\%]\\
            			\hline
            			Raw inputs + HL feats. & $20.30\pm0.08$ & N/A \\
            			Raw inputs only & $19.53\pm0.06$ & $-3.8\pm0.4$ \\
            			HL-feats. only & $17.60\pm0.08$ & $-13.3\pm0.6$ \\
            			Energy-sum only & $14.98\pm0.05$ & $-26.2\pm0.5$ \\
            			\hline
            		\end{tabular}
                \end{center}
                \caption{Mean Improvements for a range of different input configurations. The MI is computed on the monitoring-validation data and averaged over the training of five models per configuration. The change in MI is computed as the difference between configuration and the nominal model (``Raw inputs + HL feats.") as a fraction of the MI of the nominal model. Energy-sum features are the three features corresponding to the sums of energy in different threshold regions (V[0], V[26], and V[27]).}
            	\label{tab:results:inputs}
            \end{table}
            
\begin{figure*}[ht]
    			\begin{center}
    				\includegraphics[width=\sfMid\textwidth]{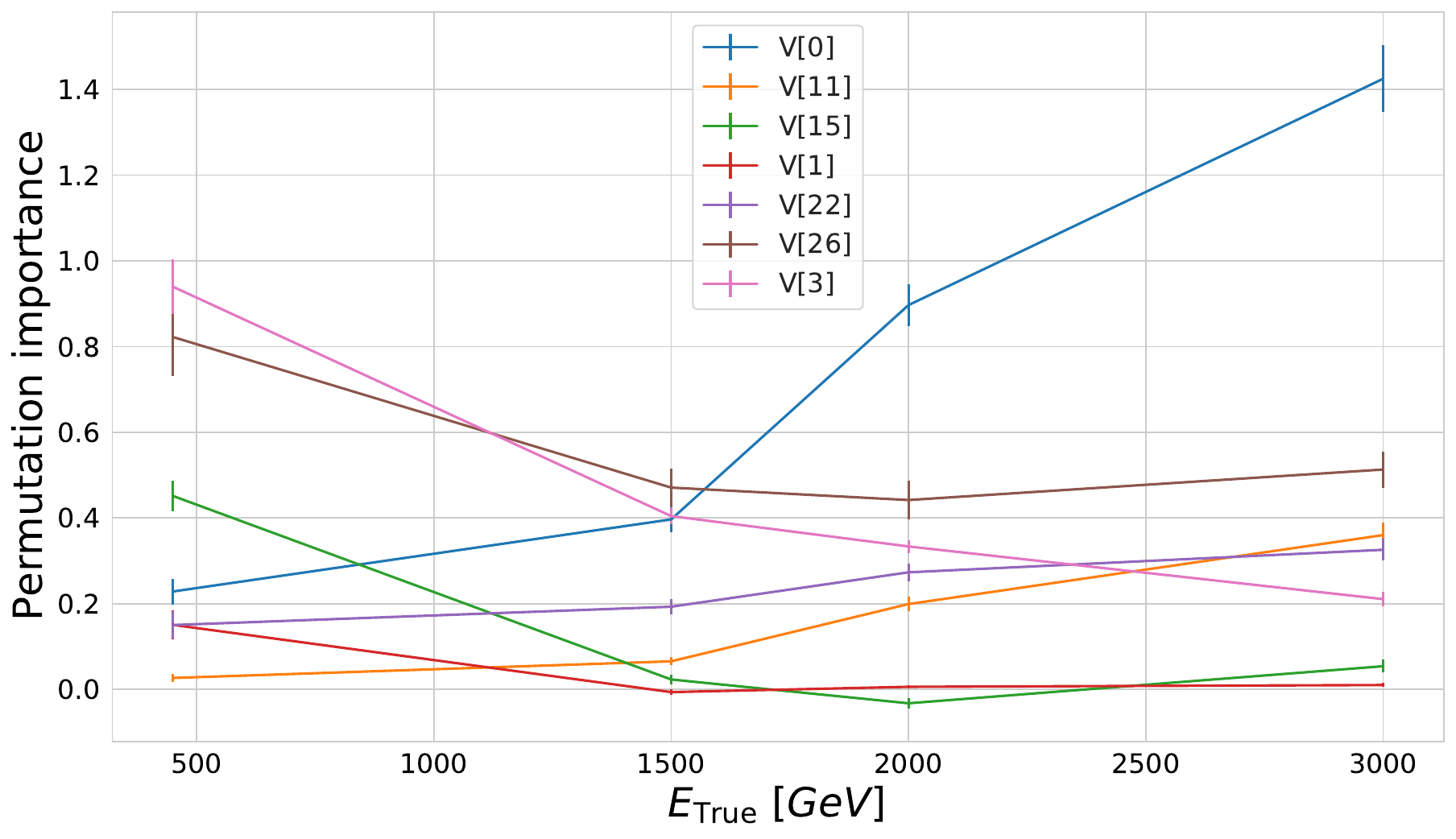}
    				\caption{Permutation importance of the most important features as evaluated using the ``HL-feats. only" model in bins of true energy. The features, described further in \autoref{s:features} are: V[0] - E-sum in cells above \SI{0.1}{\gev}, V[1] - fractional MET, V[3] - overall \nth{2} moment of transverse E distribution, V[11] - maximum total E in clustered deposits, V[15] - maximum energy of cells excluded from clustered deposits, V[22] - relative \nth{1} moment of E distribution along x-axis, V[26] -  E-sum in cells below \SI{0.01}{\gev}.}
    				\label{fig:results:feature_importance}
    			\end{center}
		    \end{figure*}

            From these results we can see the CNN is able to extract more useful information from the raw data that our domain expertise provides, however we are still able to help the model perform better when we also leverage our knowledge. Moreover, we can see that the additionally computed HL-features provide a significant benefit to the energy-sum features. The importance of the top features as a function of true energy is illustrated in \autoref{fig:results:feature_importance}. From this, it is interesting to note the shift in importance between V[0] and V[26] (the sum of energy in cells above \SI{0.1}{\gev} and below \SI{0.01}{\gev}, respectively; see Appendix~\ref{s:features}, {\em infra}). This is due to the increased chance of high-energy deposits as the energy of the muon increases. The fact that the HL-features give access to a finer-grained summation of energy than the energy-pass-through connections in the CNN architecture (which sum all the energy in cells within the kernel-size, regardless of energy) is potentially why the HL-features are still useful; a further extension to the model, then, could be to also perform the binned summation during the energy pass-through.
            
\begin{figure*}[ht]
    			\begin{center}
    				\includegraphics[width=\sfMid\textwidth]{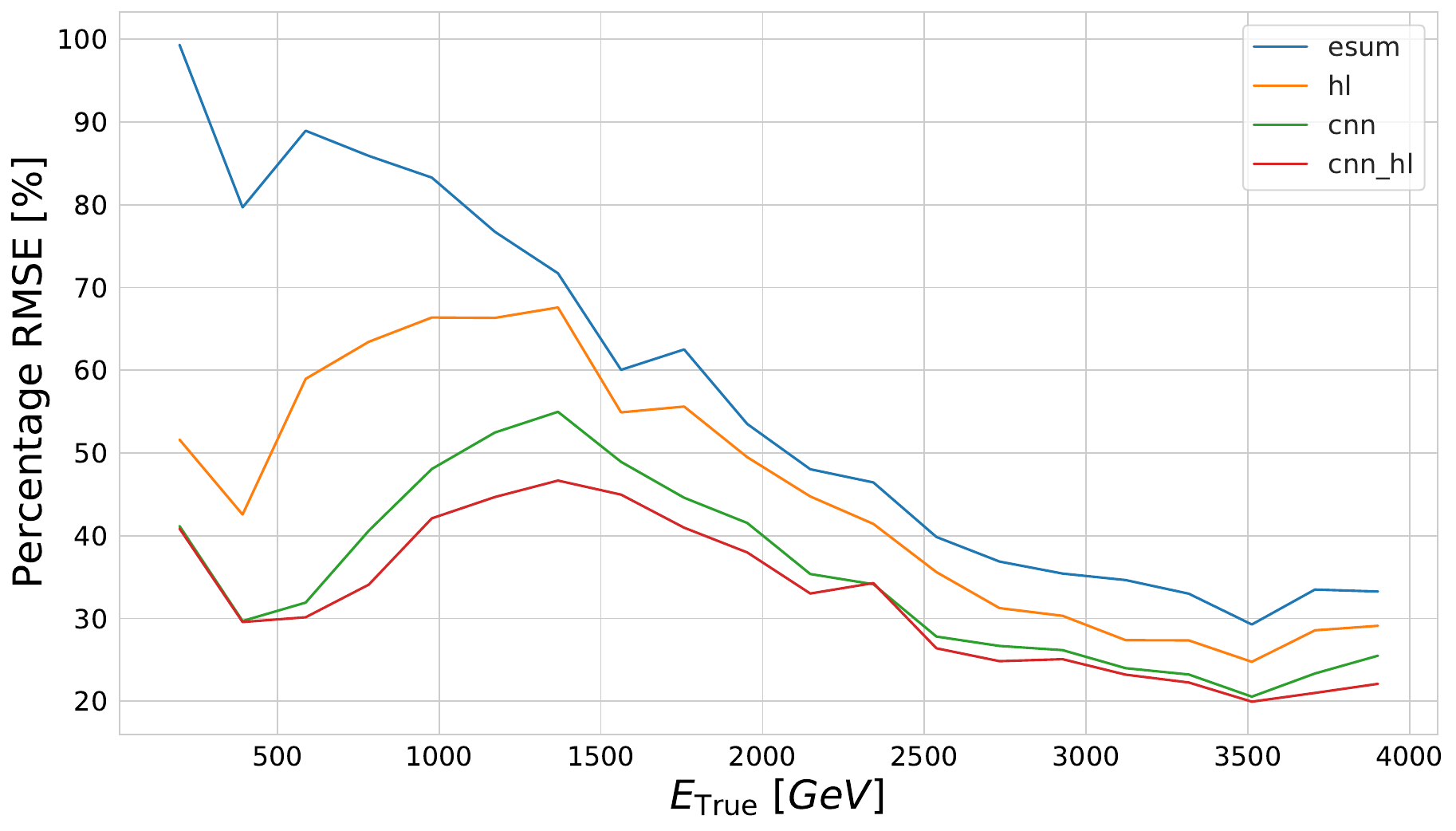}
    				\caption{Resolutions of the models with varying input menus. Resolution is computed on the holdout-validation data in bins of true energy.}
    				\label{fig:results:input_comparison}
    			\end{center}
    	    \end{figure*}

            Figure~\ref{fig:results:input_comparison} shows the resolutions of the four different models on the holdout-validation data. From this we can clearly see the benefits of providing access to the raw-hit data. The benefits of the high-level features are most prominent in the low to medium energy range, where features V[0] and V[26] have very similar importance. 
            
%\FloatBarrier
%\clearpage 

%%%%%%%%%%%%%%%%%%%%%%%%%%%%%%%%%%%%%%%%%
\section{Conclusions}\label{s:conclusion}

As we move towards the investigation of the potential of new accelerators envisioned by the recently published ``2020 Update of the European Strategy for Particle Physics"~\cite{eusupp}, we need to ask ourselves how we plan to determine the energy of multi-\si{\tev} muons in the future detectors which those machines will be endowed with and beyond. As mentioned {\em supra} (\autoref{s:introduction}), the CMS detector is able to achieve relative resolutions in the range of \SIrange{6}{17}{\%} at \SI{1}{\tev}, thanks to its very strong 4-Tesla solenoid. It is important to note that the choice of such a strong magnet for CMS imposed a compact design to the whole central detector; the result proved successful at the LHC, but might be sub-optimal in other experimental situations. Given the linear scaling with momentum of relative momentum resolution as determined with curvature fits, it is clear that complementary estimates of the energy of high-energy muons would be highly beneficial in future experiments.

%breaking rules for calorimeter were necessary here
In this work we investigated, using an idealised {ca\-lo\-ri\-me\-ter} layout, how spatial and energy information on emitted electromagnetic radiation may be exploited to obtain an estimate of muon energy. Given the regularity of the detector configuration, processing of the raw data was possible using 3D convolutional neural networks. These allowed us to exploit the granular information of the deposited energy pattern to learn high-level representations of the detector readout, which we could also combine with high-level information produced by physics-inspired statistical summaries. We found the use of deep learning and domain-driven feature engineering to both be beneficial. In \autoref{sec:ablation} we further explore the CNN architecture and training loss, finding there too, that using knowledge of the physical task can help inspire more performant solutions.

Our studies show that the fine-grained information on the radiation patterns allows for a significant improvement of the precision of muon energy estimates. {\em E.g.} for muons in the \SIrange{1}{3}{\tev} range, which are the ones of higher interest for future applications, the relative resolution improves approximately by a factor of two with respect to what can be achieved by only using the total energy release (see \autoref{fig:results:input_comparison}). A combination of such information with that offered by a curvature measurement, such as a resolution term of the form $\delta P = 0.2 P$ (with P in \si{\tev}) which can typically be enabled by tracking in a $B=\SI{2}{\tesla}$ magnetic field, may keep the overall relative resolution of multi-TeV muons below \SI{30}{\%} across the spectrum, and achieve values below \SI{20}{\%} at \SI{4}{\tev} (see \autoref{fig:results:rmse}).

\section*{Acknowledgements}
    A significant fraction of computational and storage resources used in this investigation were provided by CloudVeneto; we thank the them and their support team not only for the compute offered, but the high up-time.

\section*{Data Availability Statement}
This manuscript has associated data in a
data repository.  A public version of the research code is available from Ref.~\cite{muon_reg_repo}. The pre-processed datasets are available from Ref.~\cite{muon_reg_dataset}, and are designed to be used directly with the code-base.

\section*{Open Access}
This article is licensed under a Creative Commons Attribution 4.0 International License, which permits use, sharing, adaptation,
distribution and reproduction in any medium or format, as long as you
give appropriate credit to the original author(s) and the source, provide a link to the Creative Commons licence, and indicate if changes
were made. The images or other third party material in this article
are included in the article’s Creative Commons licence, unless indicated otherwise in a credit line to the material. If material is not
included in the article’s Creative Commons licence and your intended
use is not permitted by statutory regulation or exceeds the permitted use, you will need to obtain permission directly from the copyright holder. To view a copy of this licence, visit \\ http://creativecommons.org/licenses/by/4.0/.
\\ Funded by SCOAP3.

%\FloatBarrier
%\clearpage

\section*{Appendix}
    %%%%%%%%%%%%%%%%%%%%%%%%%%%%%%%%%%%%%%%%%%%%%%%%%%%%%%%%%%%%%%%%%
    \section{High-level muon features \label{s:features}}
        The regression task we set up in \autoref{s:regressor} uses 28 global features extracted by combining spatial and energy information collected in the calorimeter cells. In this section we describe how those features are calculated.
        
        Some of the features describe general properties of the energy deposition ({\em e.g.}, the sum of the signal in all cells recording energy above or below a $E_{thr}=\SI{0.1}{\gev}$ threshold), while others are fully reliant on fine-grained information (moments of the energy distribution, in five regions of detector depth: $z<\SI{400}{\mm}$, $400<z<\SI{800}{\mm}$, $800<z<\SI{1200}{\mm}$, $1200<z<\SI{1600}{\mm}$, and $z>\SI{1600}{\mm}$; and imbalance of the deposited energy in the transverse plane). A few more variables describe the result of a clustering of the energy deposits, which is briefly described in \autoref{s:cluster} {\em infra}. A final set of features described in \autoref{s:bendingvar} are specifically constructed to leverage the magnetic field and estimate the curvature of muons by detecting the spread of the radiation pattern along the x coordinate caused by the small bending along $x$ that muons of sub-TeV energy follow as they penetrate in the calorimeter. Below we discuss in detail how the features are computed.
        
        %%%%%%%%%%%%%%%%%%%%%%%%%%%%%%%%%%%%%%%%%%%%%%%%%%%%%%%%%%%%%%
        \subsection{Clustering of calorimeter cells \label{s:cluster}}
        
        The small size of calorimeter cells (which span 0.24 radiation lengths in $x$ and $y$, and 4.5 radiation lengths in $z$) implies that photons of energy large enough to produce showers by pair production will produce a signal in multiple cells, especially if they are emitted with non-null angles with respect to the $z$ direction. Given that all the information on radiation emission by the muon is possessed by primary photons, it seems reasonable to try and decipher the pattern of emitted radiation by aggregating the granular cell-based information into
        clusters, whose properties may constitute useful statistical summaries to complement the full resolution of the calorimeter.
        
        We set a minimum threshold $E_{thr}=\SI{0.1}{\gev}$ for the energy recorded in cells elected as seeds for the clustering procedure. The search for clusters starts with seed cells belonging to the column of same transverse coordinates $x$ and $y$ of the incident muon\footnote{The impact position of muons is well determined as that of the $xy$ position whose $z$-integrated recorded energy is the highest, but we assume here that we know it from a tracking detector located upstream, with no loss of generality.}, and performs the following calculations:
        
        \begin{enumerate}
            \item The highest-energy cell is selected as a seed if it has $E>E_{\text{thr}}$;
            \item The six calorimeter cells adjacent in either $x$, $y$, or $z$ to the seed cell are added to the cluster if they recorded a non-null energy deposition;
            \item Cells with non-null energy deposition are progressively added to the cluster if they are adjacent to already included cells; 
            \item The final cluster is formed when there are no more cells passing the above criteria; at that point, features such as the number of included cells and the total cluster energy are computed (see below). 
            \item All cells belonging to the cluster are removed from the list of unassigned cells;
            \item The algorithm returns to step 1 to form other clusters.
        \end{enumerate}
        
        \noindent
        Once clusters seeded by the column of cells along the muon trajectory are formed by the above procedure, a second set of clusters is constructed using cells yet unassigned to any cluster:\par
        
        \begin{enumerate}
            \item The highest-energy cell above $E_{\text{thr}}$ is considered, irrespective of its $x$, $y$ coordinates;
            \item The six calorimeter towers adjacent in either $x$, $y$, or $z$ to the seed cell are added to the cluster if they recorded a non-null energy deposition;
            \item Cells with non-null energy deposition are progressively added to the cluster if they are adjacent to cells already included;
            \item The final cluster is formed when there are no more cells passing the above criteria; features are then computed for the identified cluster;
            \item All cells belonging to the cluster are removed from the list of unassigned cells;
            \item The algorithm returns to step 1) to search for additional clusters.
        \end{enumerate}
        
        \noindent
        Using the results of the above two-step clustering procedure, we define the following high-level features:\par
        
        \begin{itemize}
            \item V[9]: The number of muon trajectory-seeded clusters (type-1 clusters);
            \item V[10]: The maximum number of cells among type-1 clusters;
            \item V[11]: The maximum total energy among type-1 clusters;
            \item V[12]: The maximum extension along x of type-1 clusters;
            \item V[13]: The maximum extension along y of type-1 clusters;
            \item V[14]: The maximum extension along z of type-1 clusters;
            %\item V[15]: Ratio between maximum energy and maximum number of cells of type-1 clusters;
            \item V[16]: Average number of cells included in type-1 clusters.
            \item V[17]: The number of clusters seeded by a cell not belonging to the muon trajectory (type-2 clusters);
            \item V[18]: The maximum number of cells among type-2 clusters;
            \item V[19]: The maximum total energy among type-2 clusters;
            \item V[20]: Ratio between maximum energy and maximum number of cells of type-2 clusters;
            \item V[21]: Average number of cells included in type-2 clusters.
        \end{itemize}
        
\begin{figure*}[h!]
            \begin{center}
                \begin{subfigure}[t]{0.40\textwidth}
                    \begin{center}
                        \includegraphics[width=\textwidth]{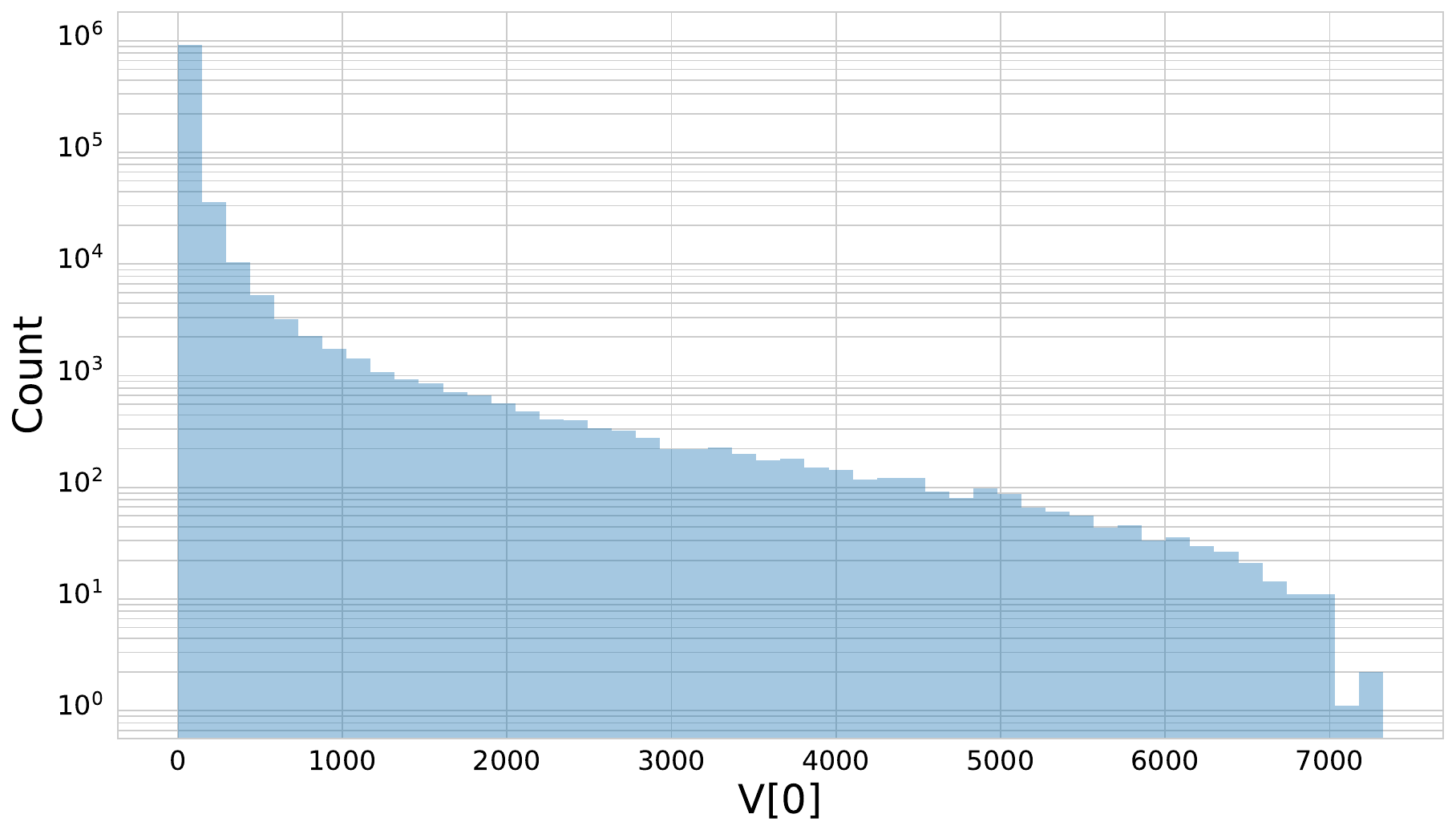}
                    \end{center}
                \end{subfigure}
                \begin{subfigure}[t]{0.40\textwidth}
                    \begin{center}
                        \includegraphics[width=\textwidth]{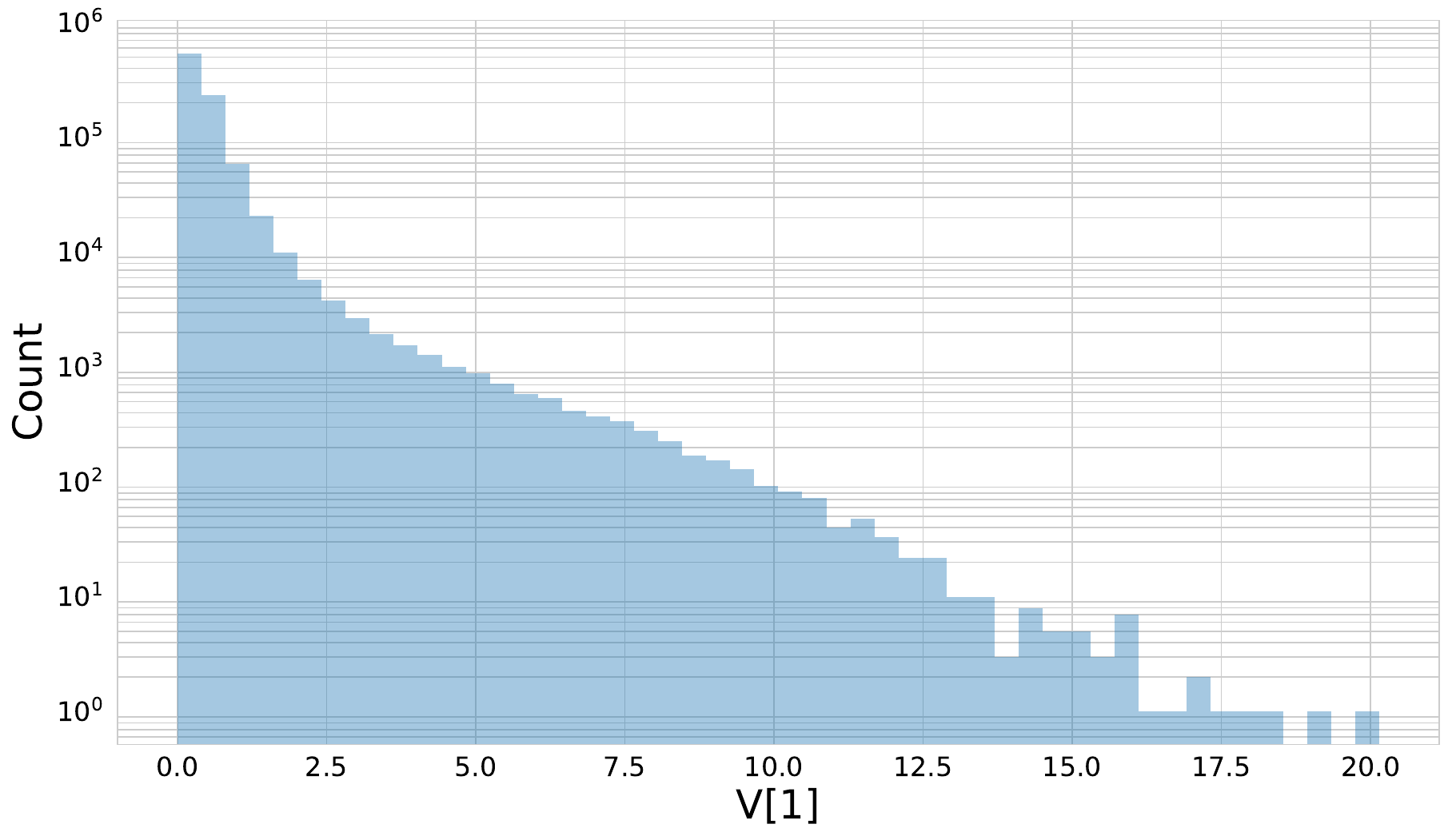}
                    \end{center}
                \end{subfigure}
                \begin{subfigure}[t]{0.40\textwidth}
                    \begin{center}
                        \includegraphics[width=\textwidth]{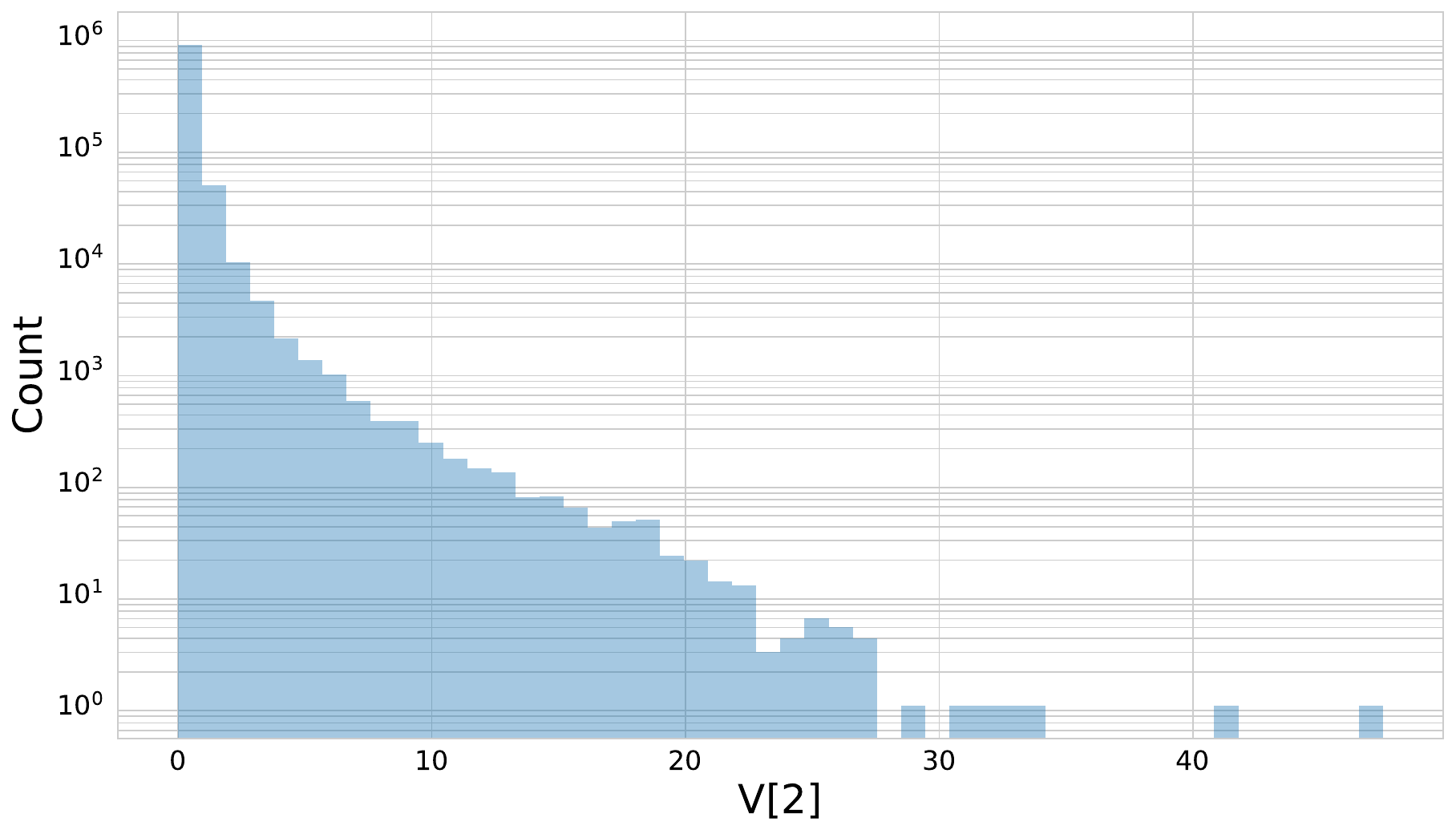}
                    \end{center}
                \end{subfigure}
                \begin{subfigure}[t]{0.40\textwidth}
                    \begin{center}
                        \includegraphics[width=\textwidth]{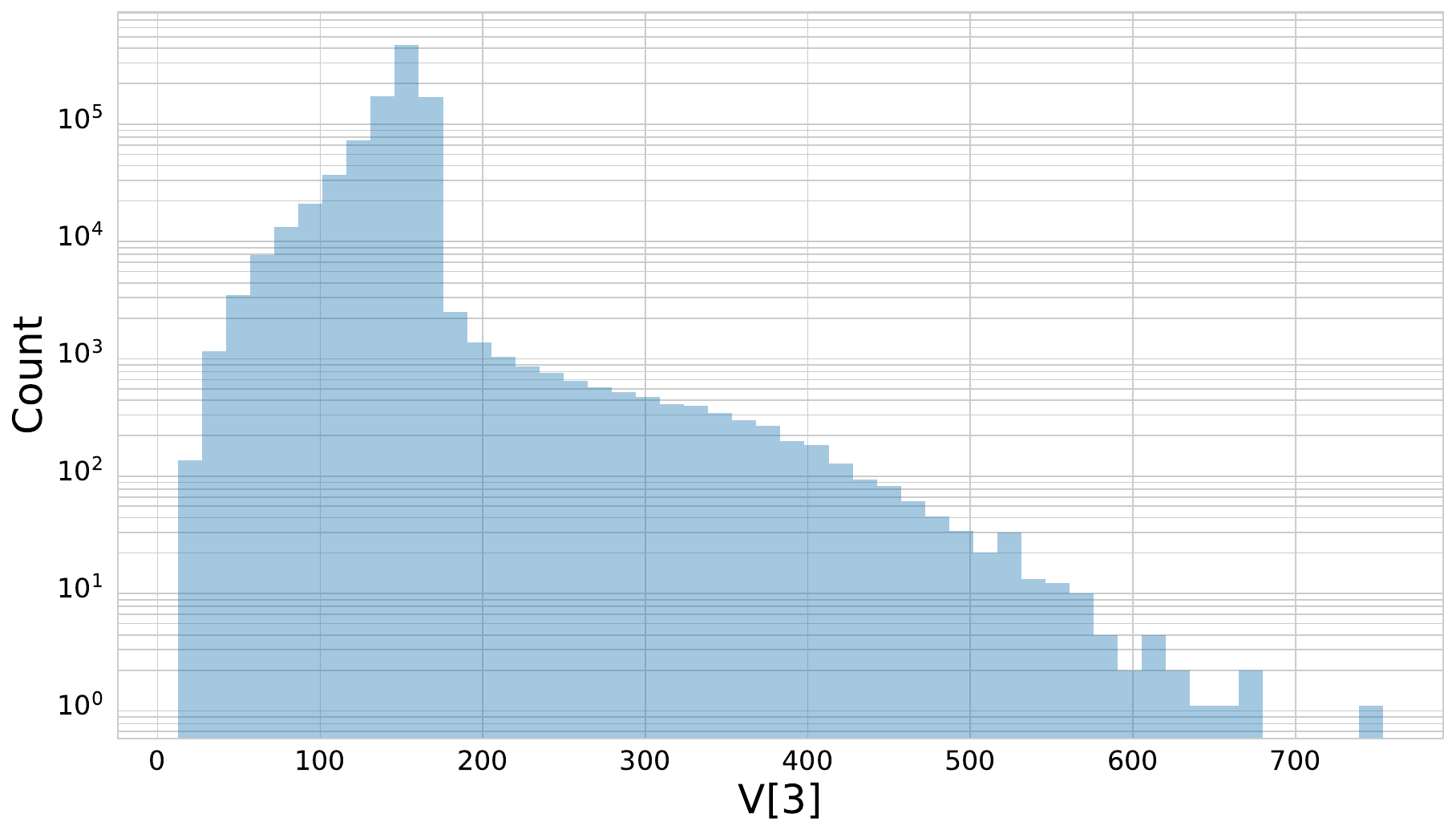}
                    \end{center}
                \end{subfigure}
                \begin{subfigure}[t]{0.40\textwidth}
                    \begin{center}
                        \includegraphics[width=\textwidth]{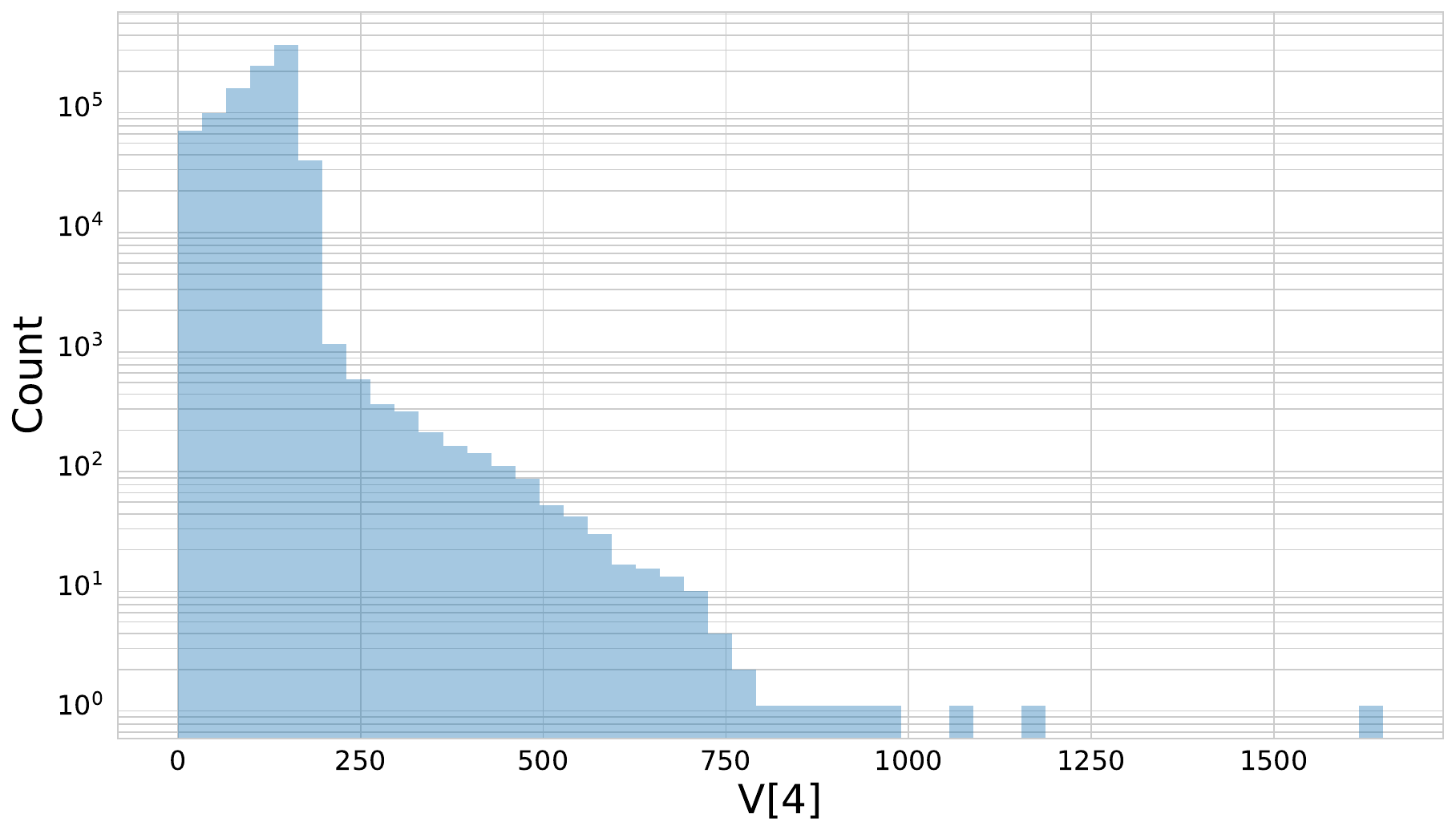}
                    \end{center}
                \end{subfigure}
                \begin{subfigure}[t]{0.40\textwidth}
                    \begin{center}
                        \includegraphics[width=\textwidth]{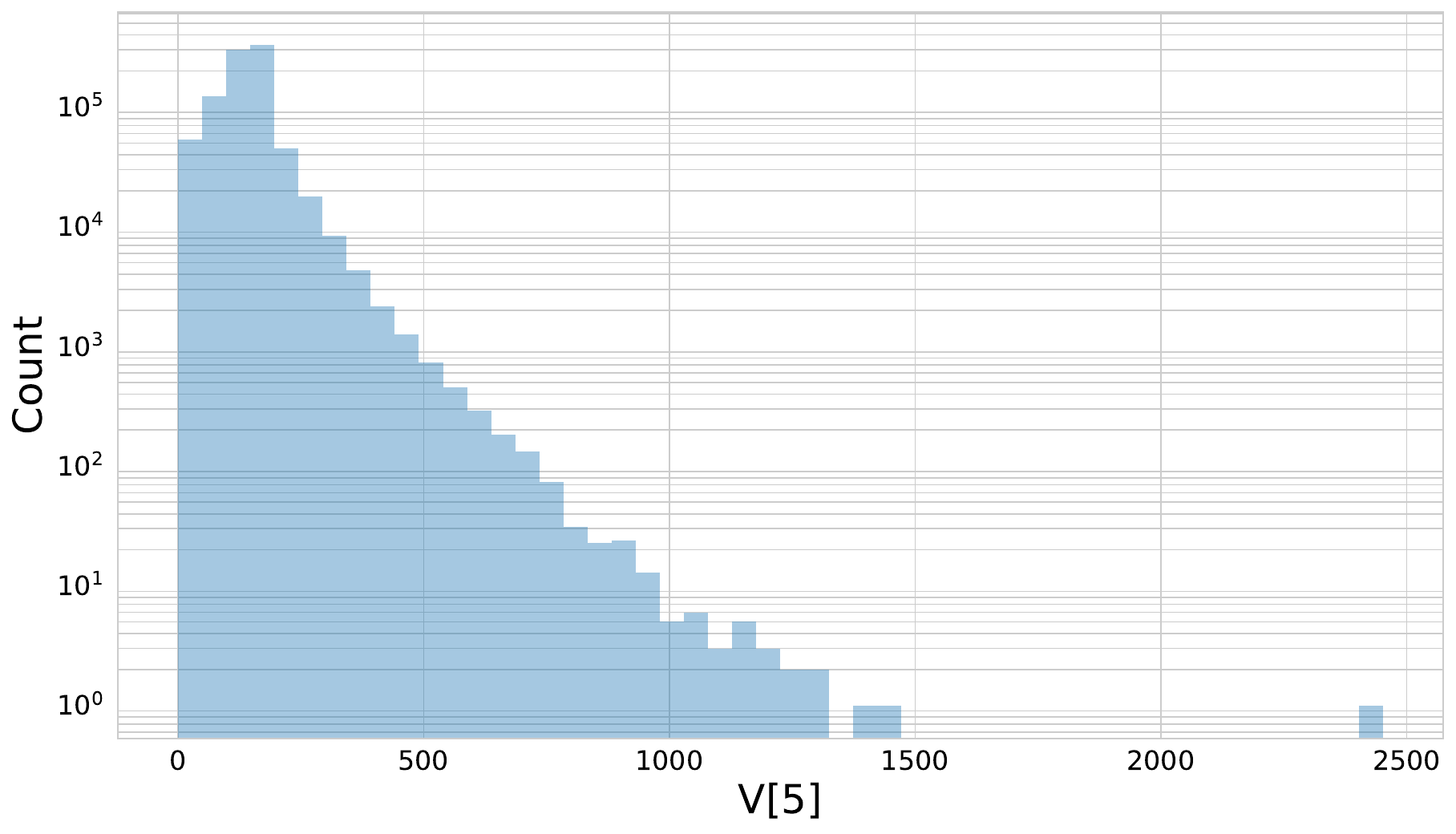}
                    \end{center}
                \end{subfigure}
                \begin{subfigure}[t]{0.40\textwidth}
                    \begin{center}
                        \includegraphics[width=\textwidth]{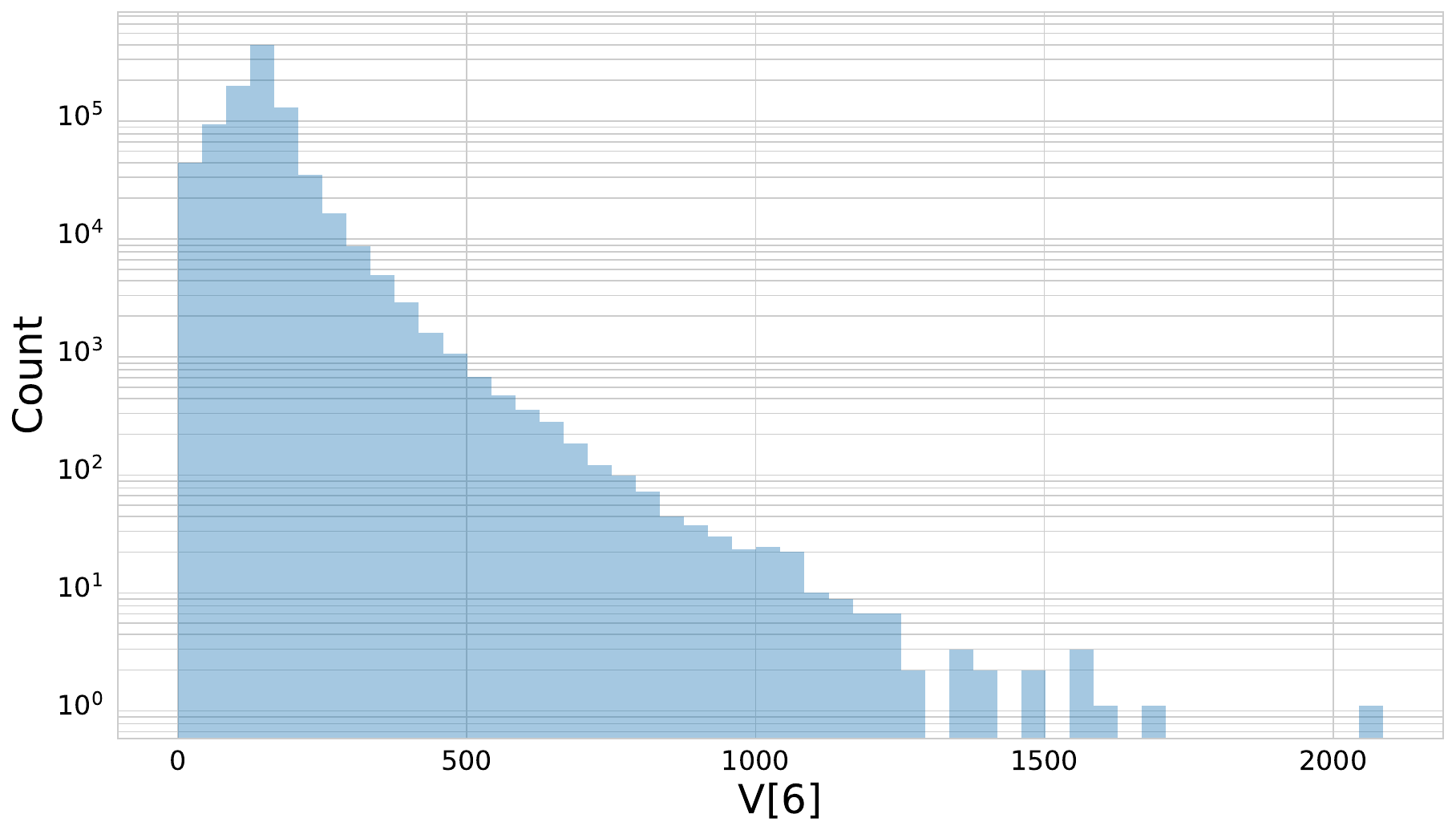}
                    \end{center}
                \end{subfigure}
                \begin{subfigure}[t]{0.40\textwidth}
                    \begin{center}
                        \includegraphics[width=\textwidth]{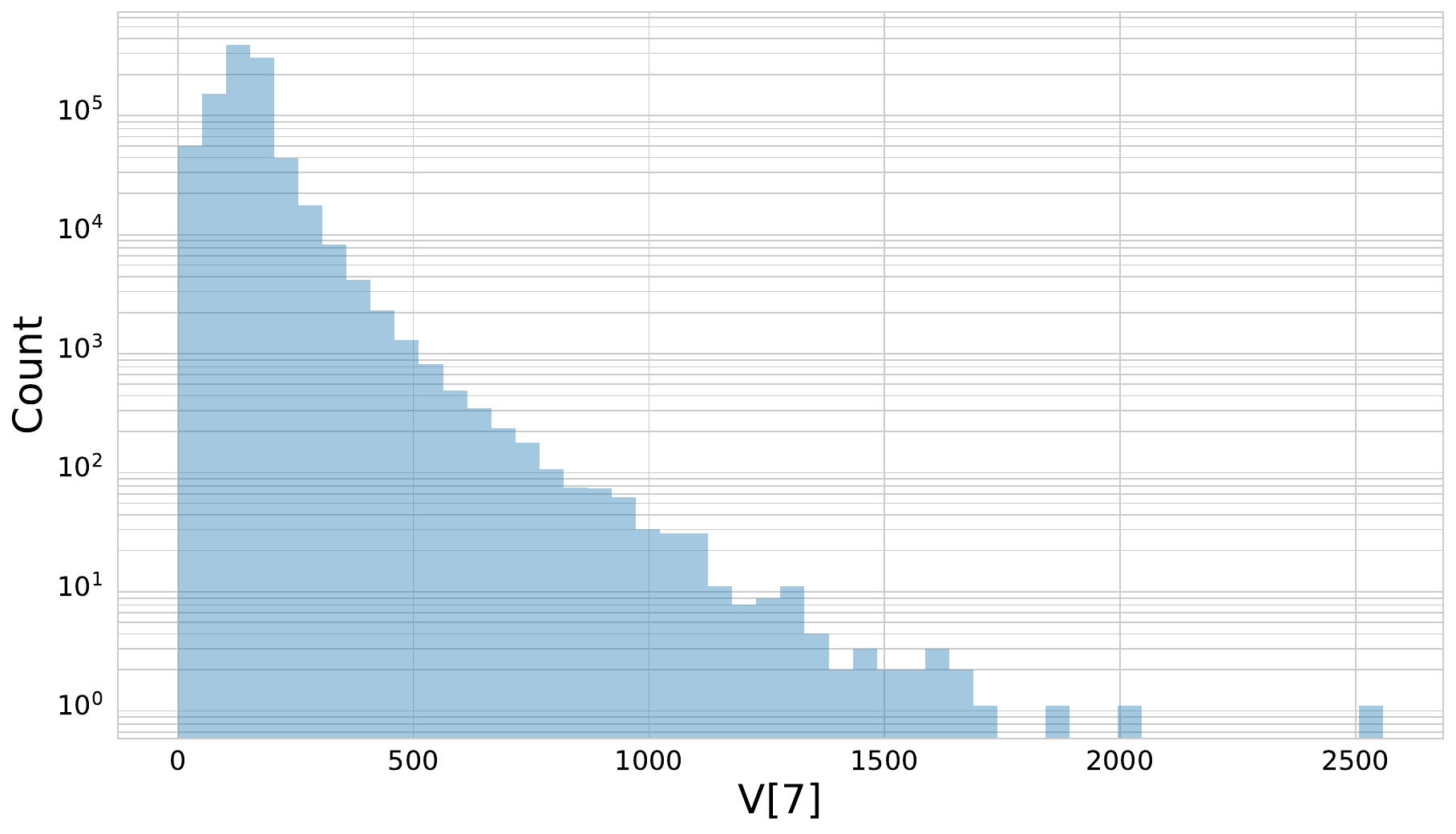}
                    \end{center}
                \end{subfigure}
                \begin{subfigure}[t]{0.40\textwidth}
                    \begin{center}
                        \includegraphics[width=\textwidth]{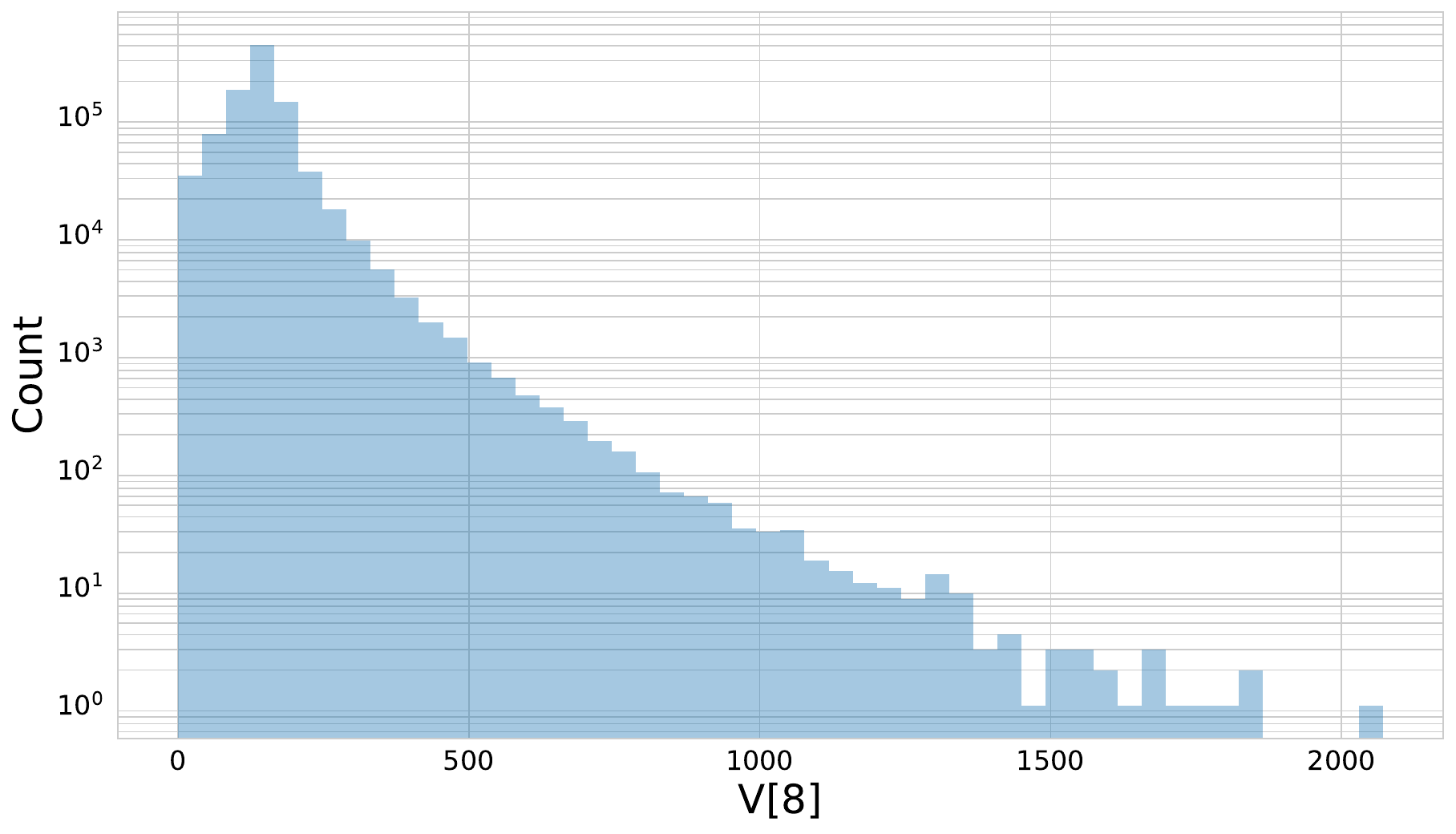}
                    \end{center}
                \end{subfigure}
                \begin{subfigure}[t]{0.40\textwidth}
                    \begin{center}
                        \includegraphics[width=\textwidth]{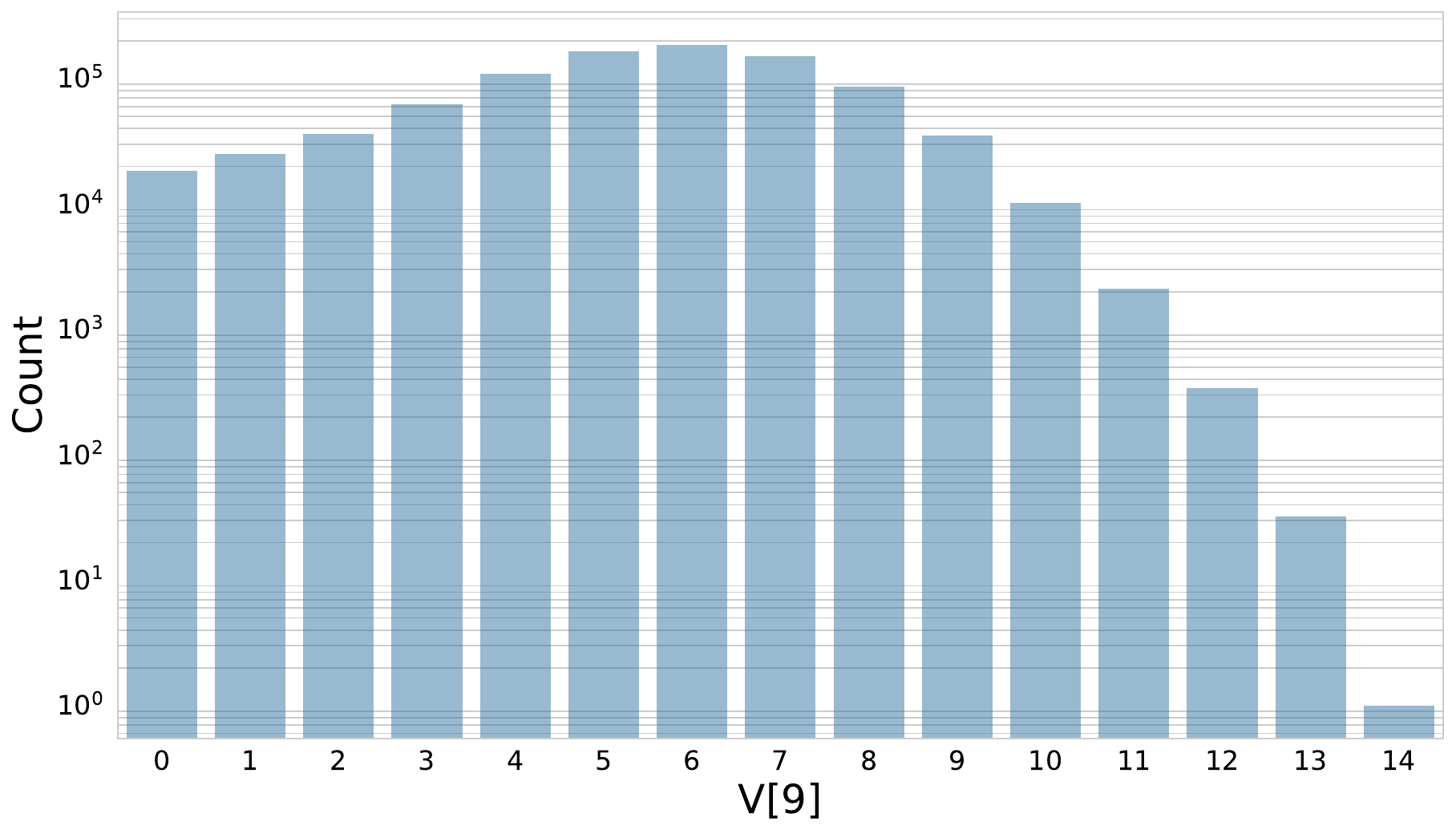}
                    \end{center}
                \end{subfigure}
                \caption{1D density-distributions of features V[0] to V[9]. Features are defined in Section~\ref{s:features_description}.}
                \label{f:1d_feats:0-9}
            \end{center}
        \end{figure*}

\begin{figure*}[h!]
            \begin{center}
                \begin{subfigure}[t]{0.40\textwidth}
                    \begin{center}
                        \includegraphics[width=\textwidth]{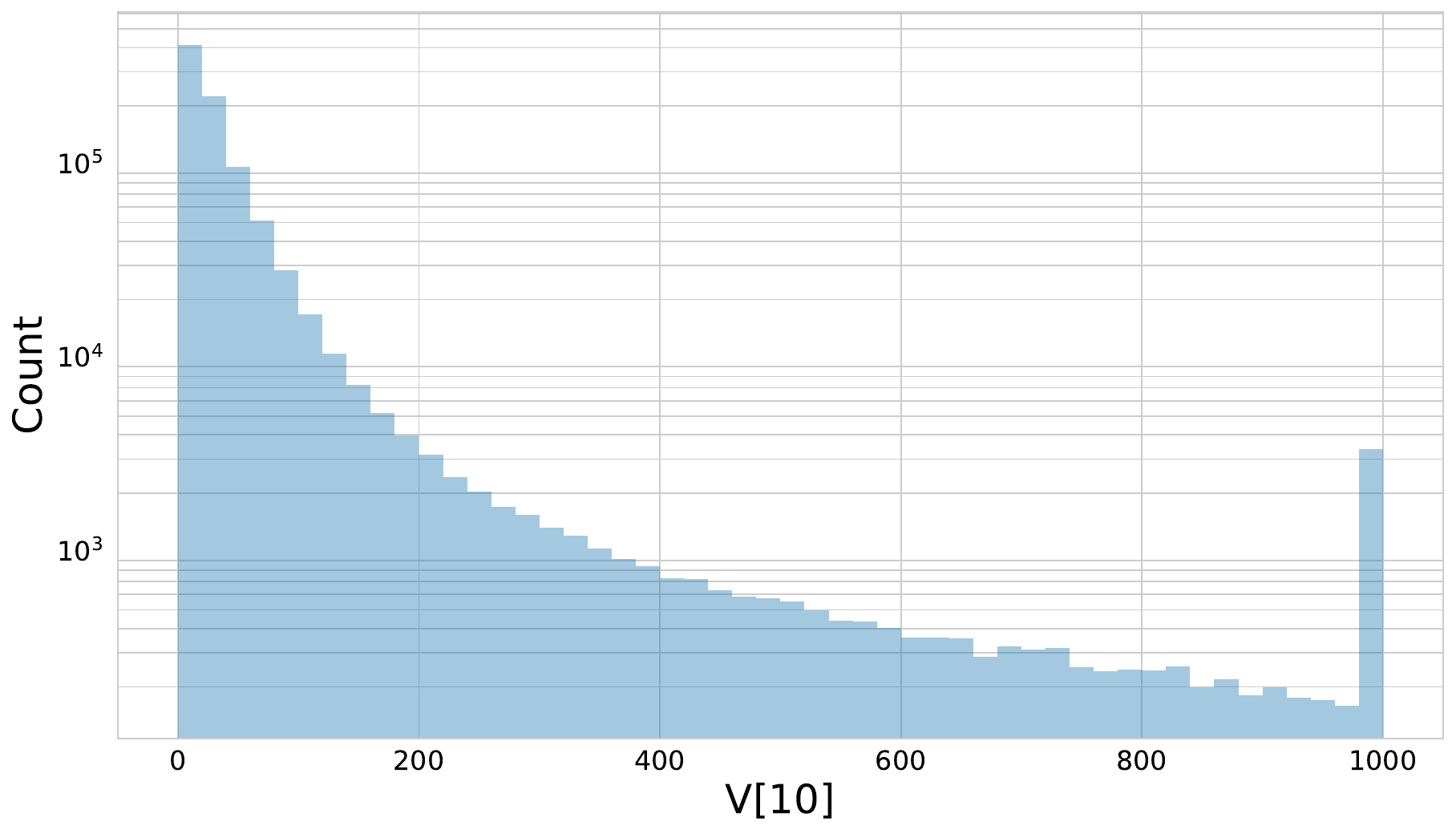}
                    \end{center}
                \end{subfigure}
                \begin{subfigure}[t]{0.40\textwidth}
                    \begin{center}
                        \includegraphics[width=\textwidth]{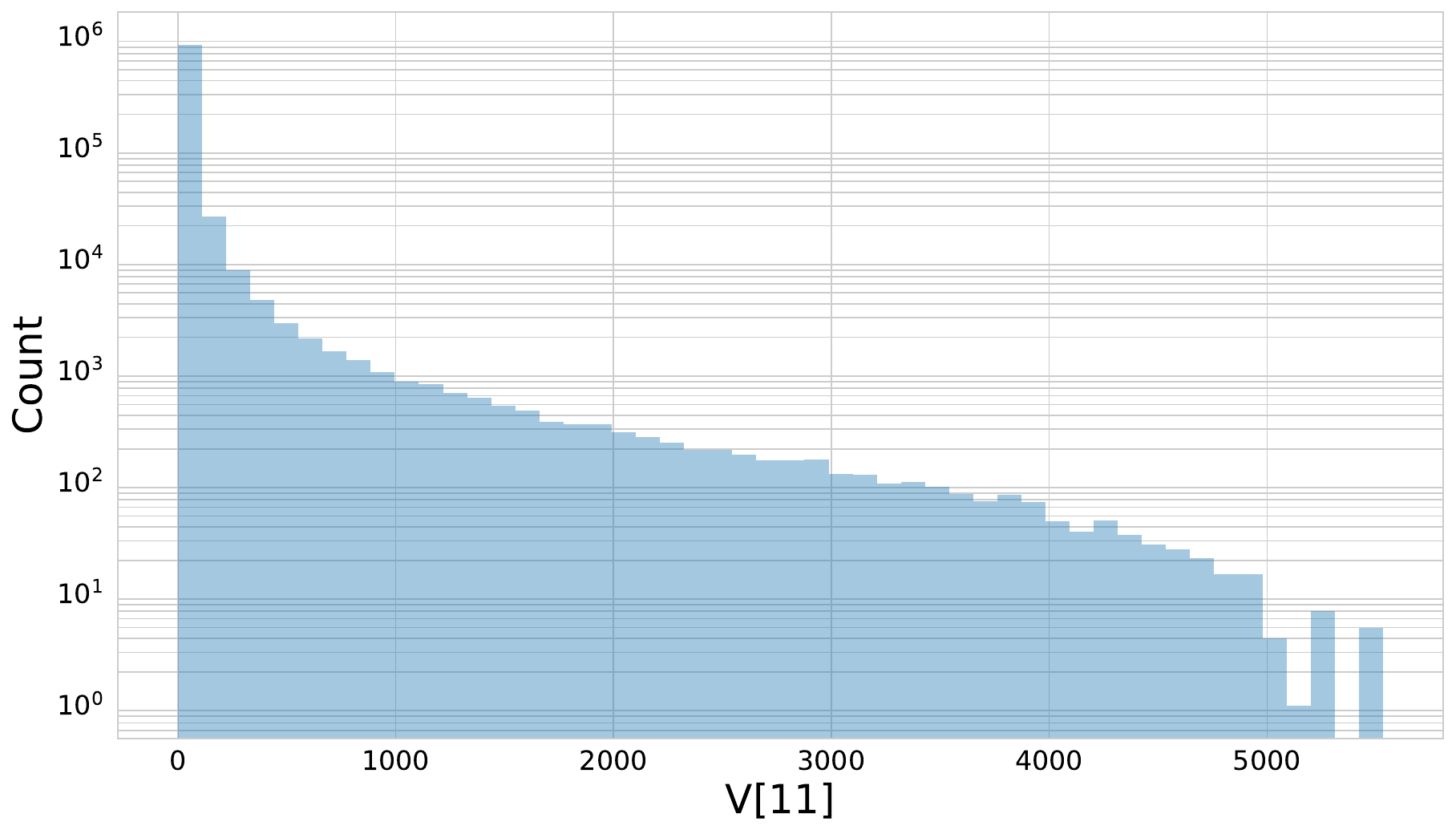}
                    \end{center}
                \end{subfigure}
                \begin{subfigure}[t]{0.40\textwidth}
                    \begin{center}
                        \includegraphics[width=\textwidth]{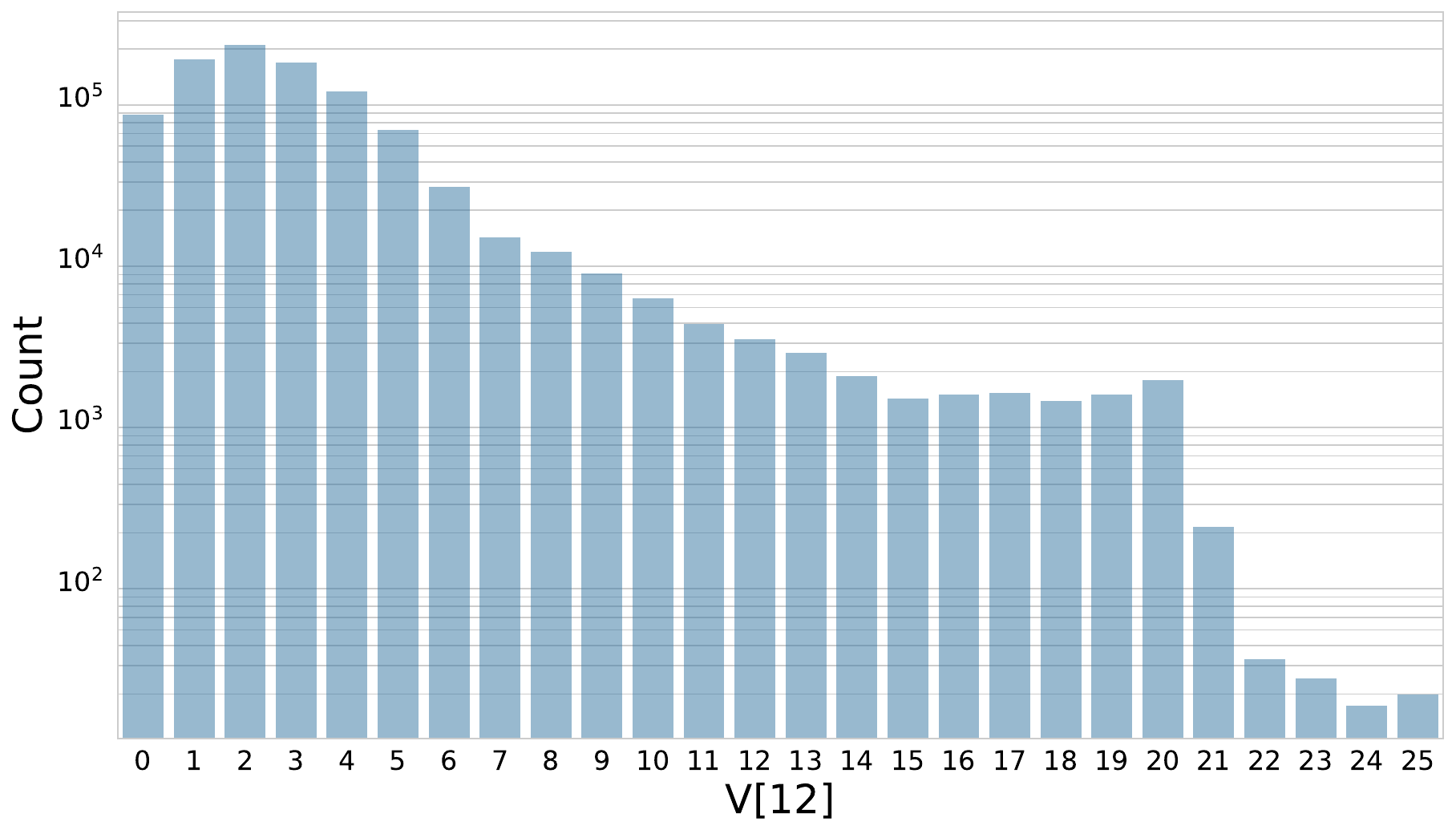}
                    \end{center}
                \end{subfigure}
                \begin{subfigure}[t]{0.40\textwidth}
                    \begin{center}
                        \includegraphics[width=\textwidth]{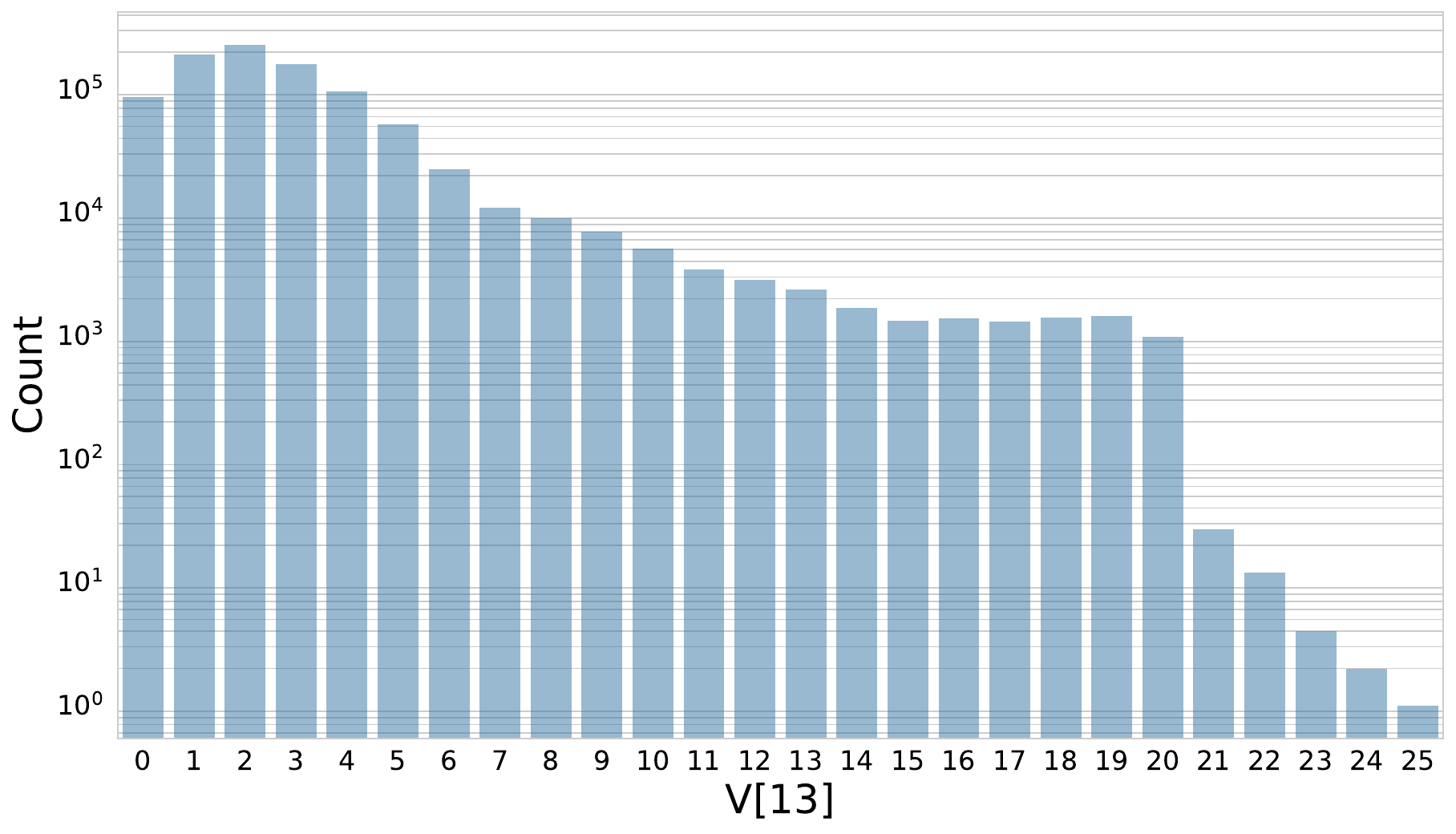}
                    \end{center}
                \end{subfigure}
                \begin{subfigure}[t]{0.40\textwidth}
                    \begin{center}
                        \includegraphics[width=\textwidth]{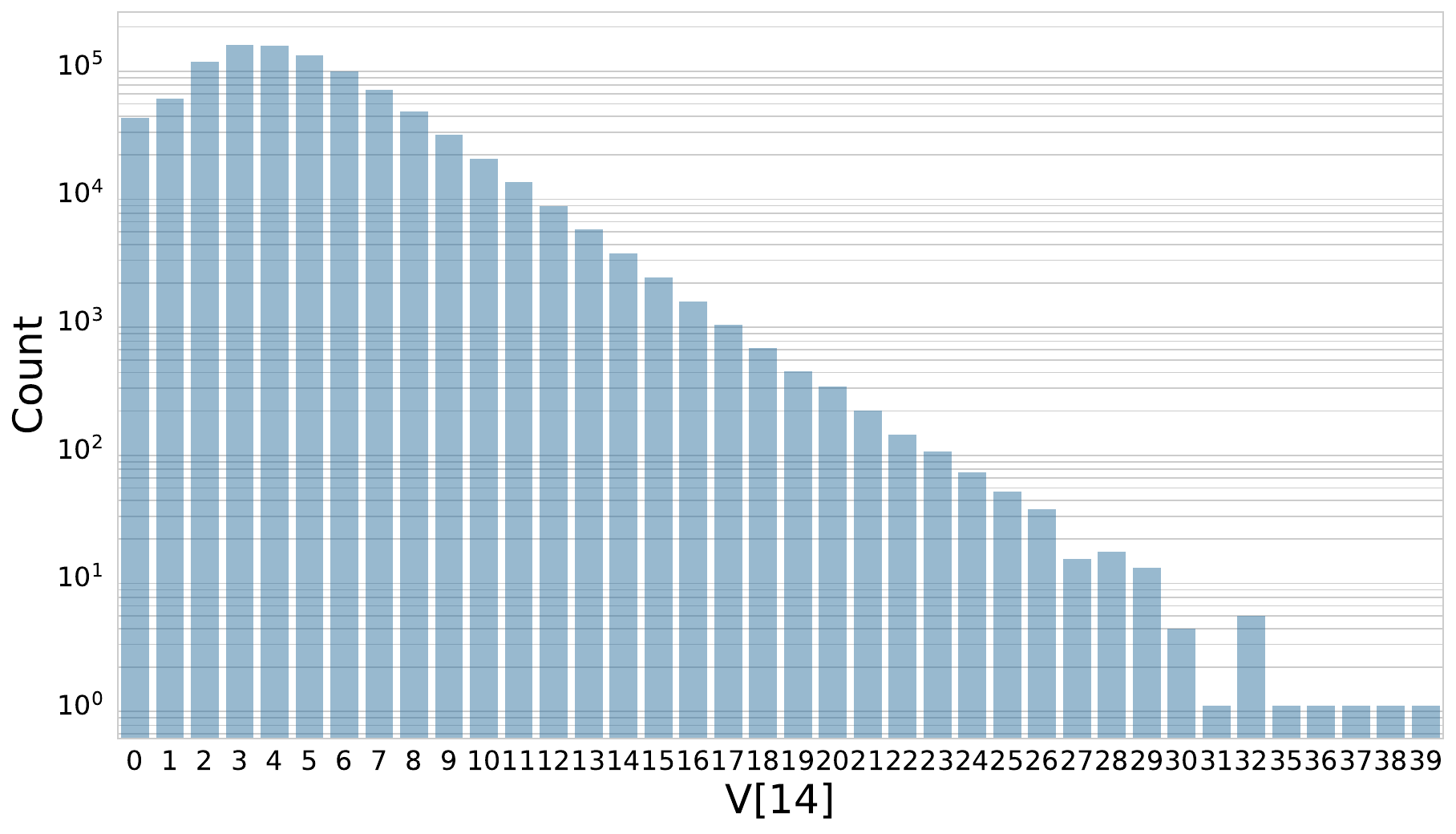}
                    \end{center}
                \end{subfigure}
                \begin{subfigure}[t]{0.40\textwidth}
                    \begin{center}
                        \includegraphics[width=\textwidth]{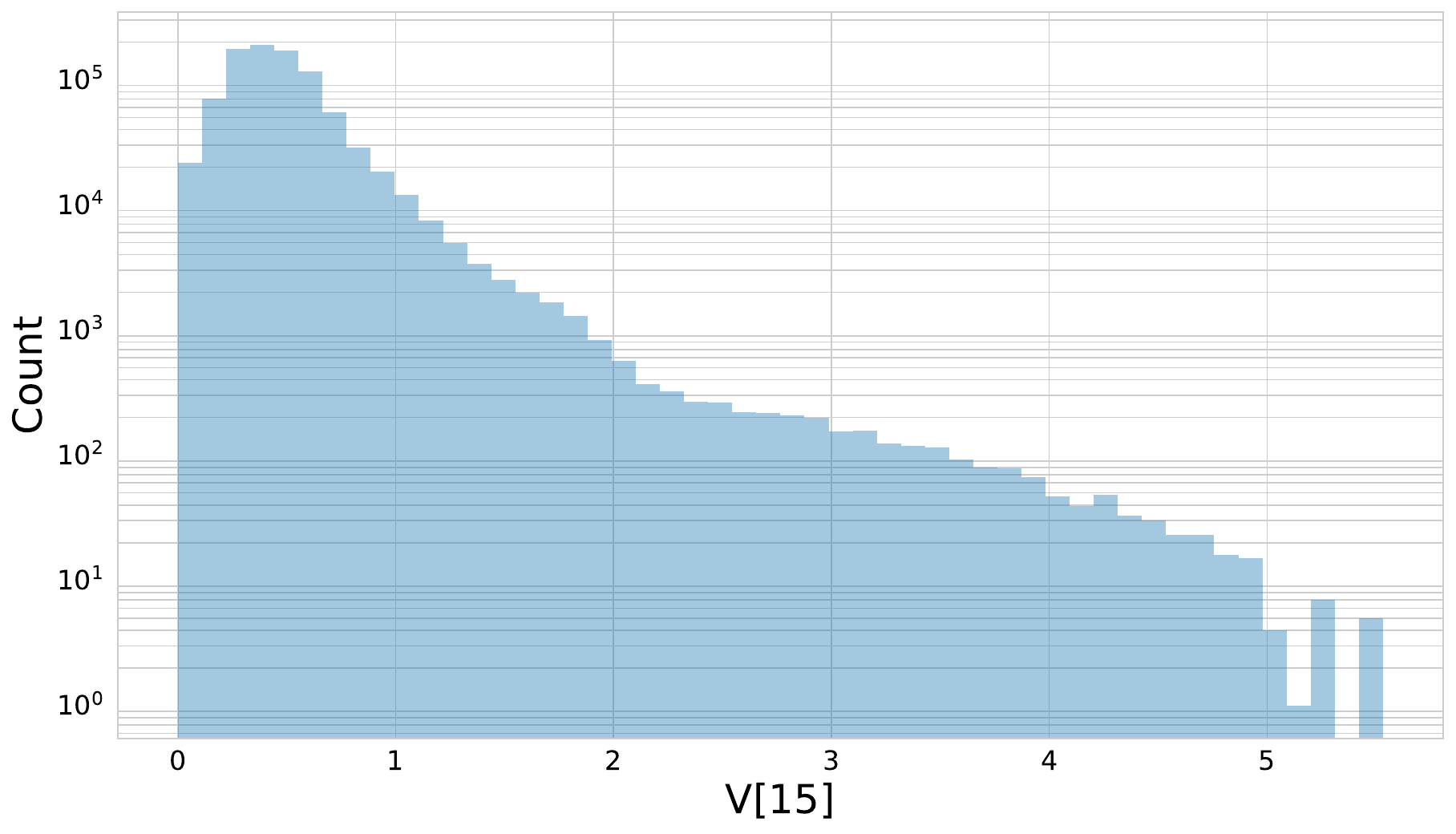}
                    \end{center}
                \end{subfigure}
                \begin{subfigure}[t]{0.40\textwidth}
                    \begin{center}
                        \includegraphics[width=\textwidth]{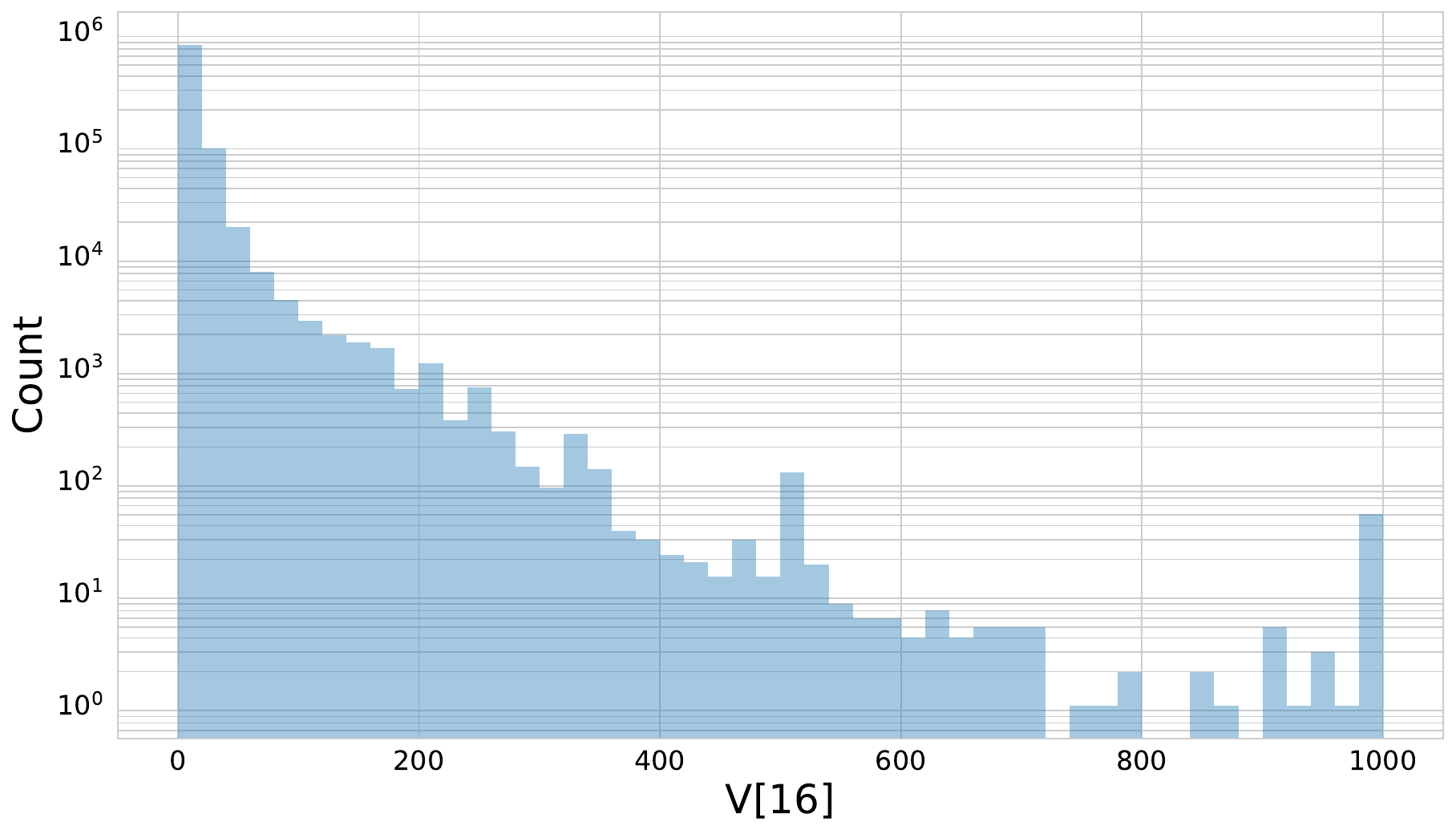}
                    \end{center}
                \end{subfigure}
                \begin{subfigure}[t]{0.40\textwidth}
                    \begin{center}
                        \includegraphics[width=\textwidth]{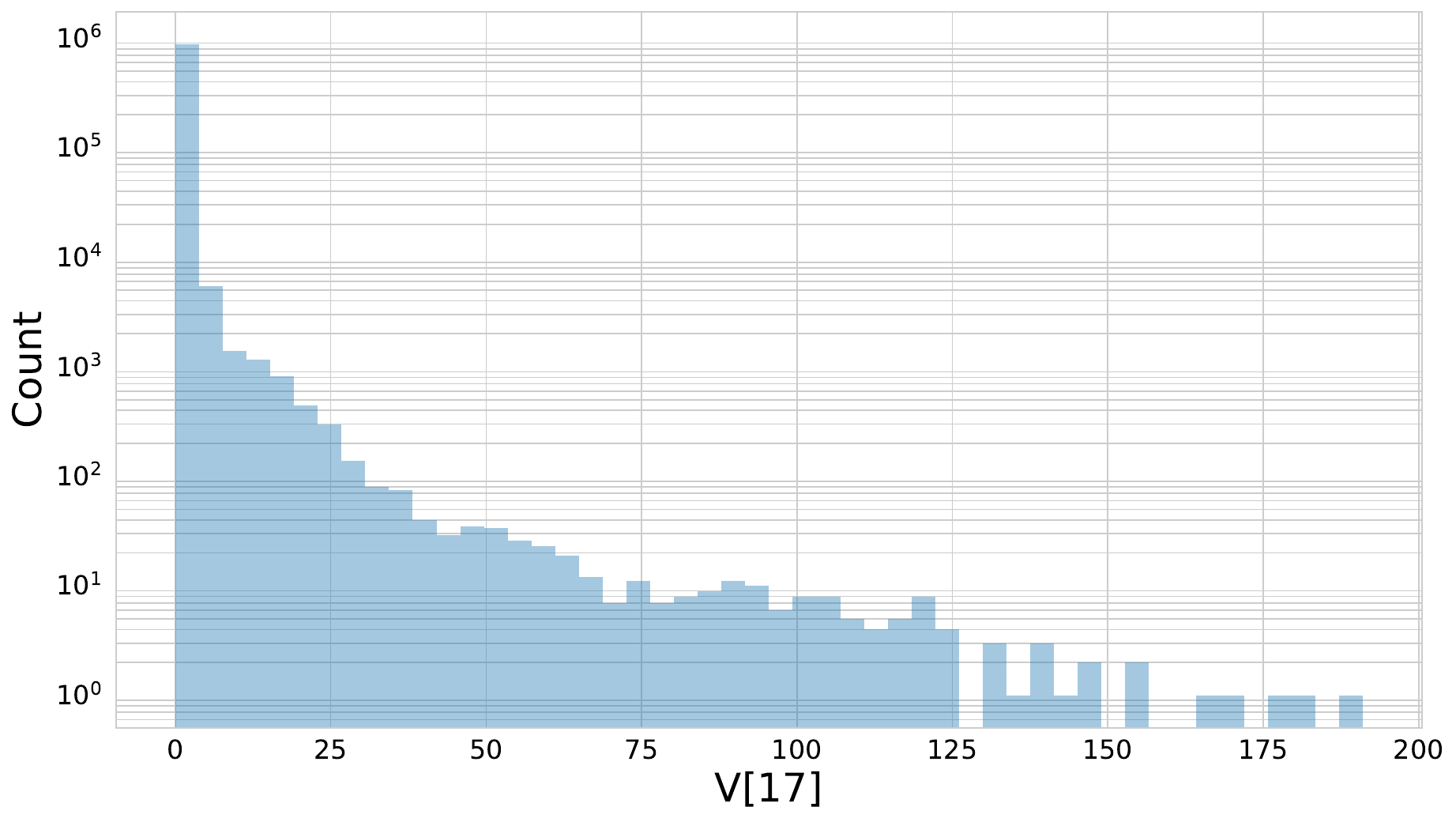}
                    \end{center}
                \end{subfigure}
                \begin{subfigure}[t]{0.40\textwidth}
                    \begin{center}
                        \includegraphics[width=\textwidth]{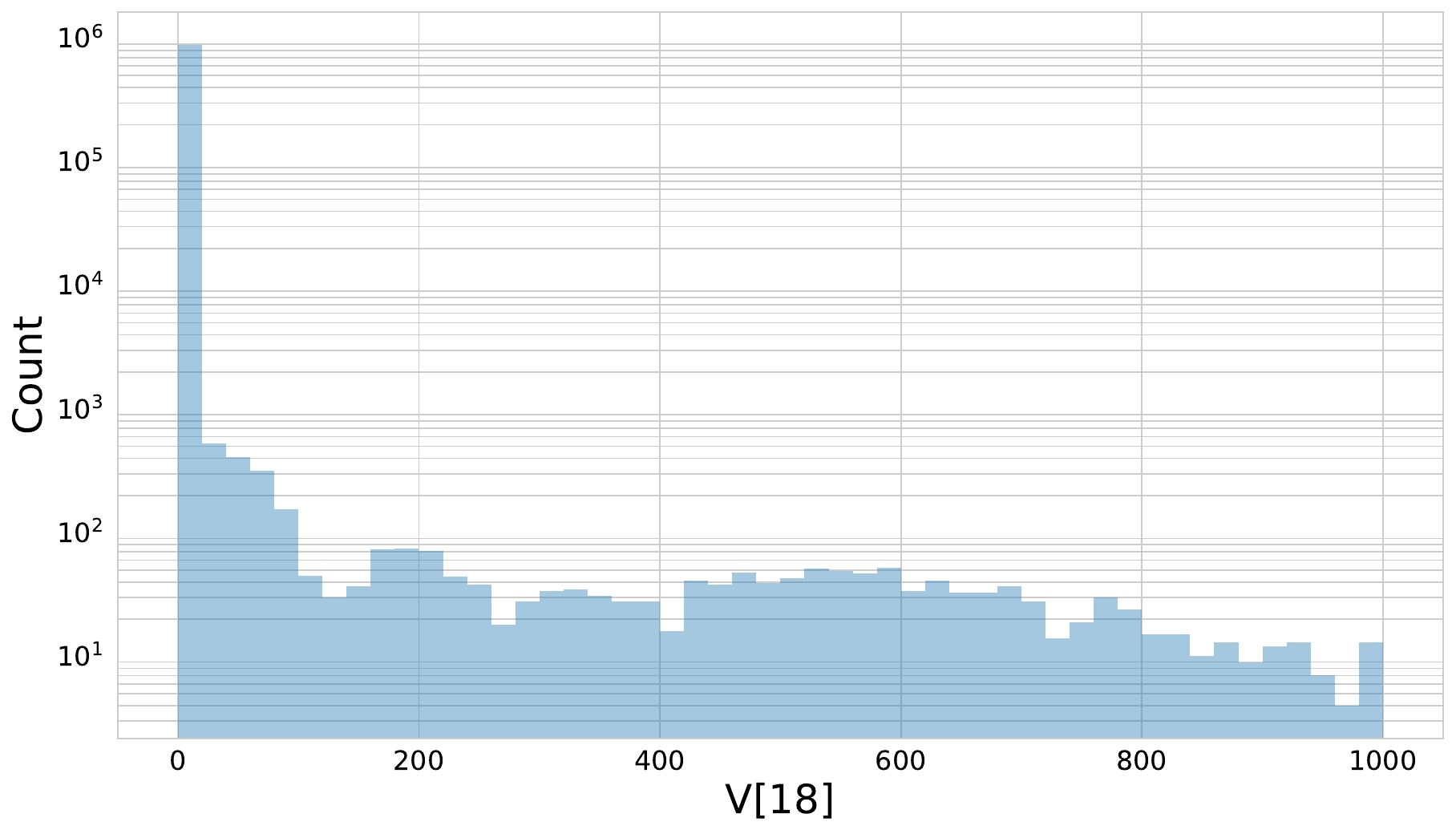}
                    \end{center}
                \end{subfigure}
                \begin{subfigure}[t]{0.40\textwidth}
                    \begin{center}
                        \includegraphics[width=\textwidth]{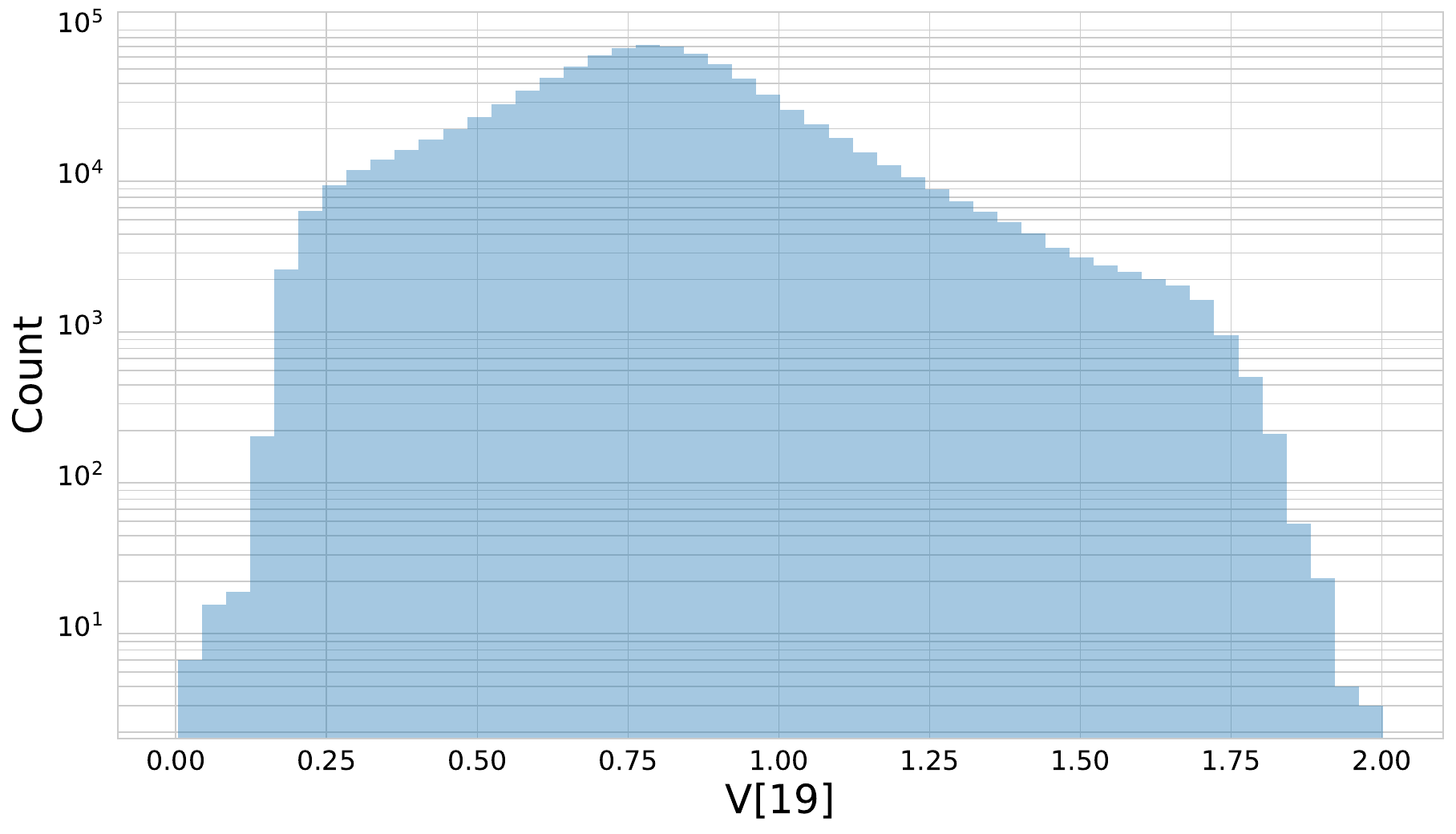}
                    \end{center}
                \end{subfigure}
                \caption{1D density-distributions of features V[10] to V[19]. Features are defined in Section~\ref{s:features_description}.}
                \label{f:1d_feats:10-19}
            \end{center}
        \end{figure*}

\begin{figure*}[h!]
            \begin{center}
                \begin{subfigure}[t]{0.40\textwidth}
                    \begin{center}
                        \includegraphics[width=\textwidth]{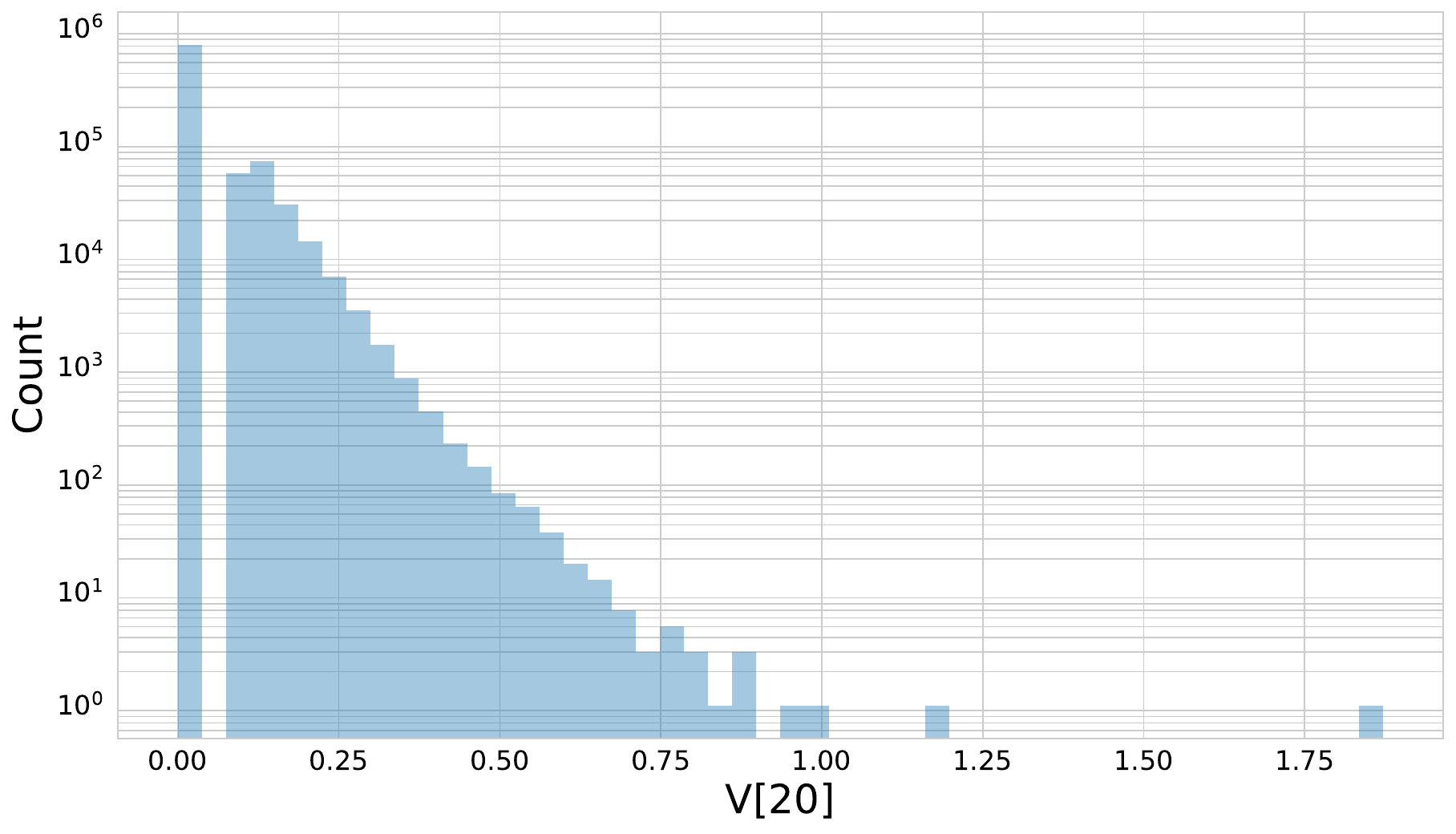}
                    \end{center}
                \end{subfigure}
                \begin{subfigure}[t]{0.40\textwidth}
                    \begin{center}
                        \includegraphics[width=\textwidth]{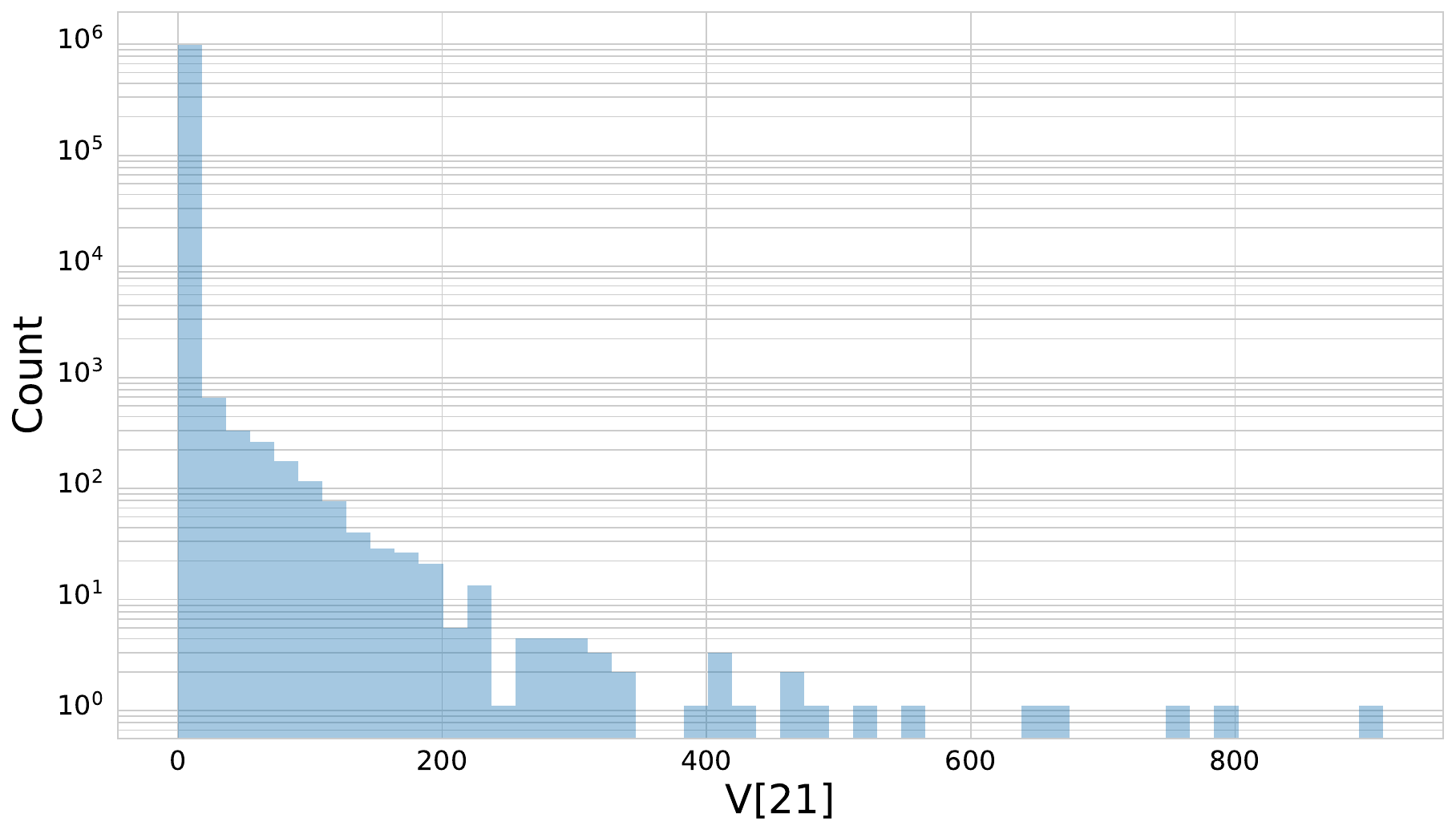}
                    \end{center}
                \end{subfigure}
                \begin{subfigure}[t]{0.40\textwidth}
                    \begin{center}
                        \includegraphics[width=\textwidth]{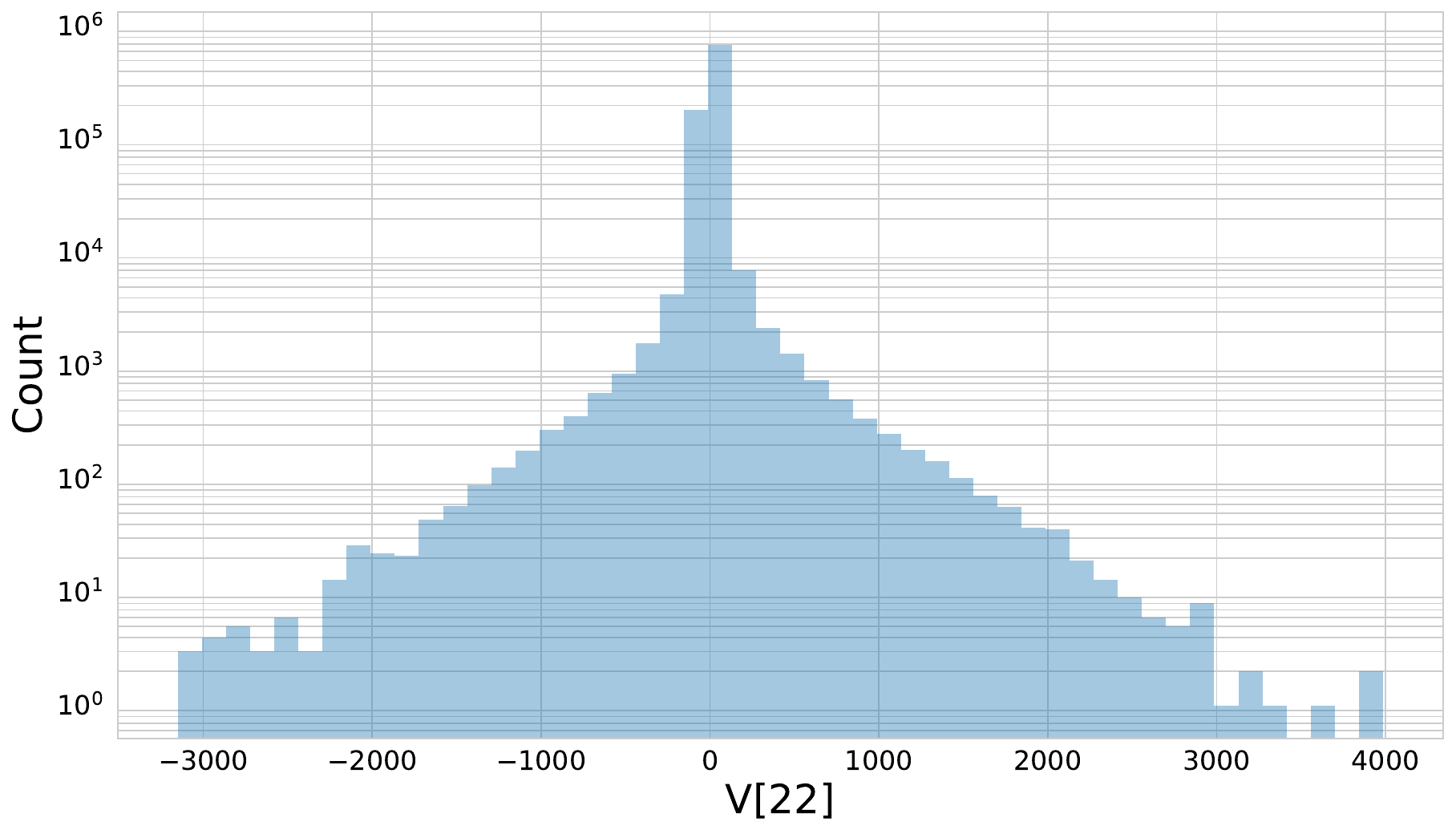}
                    \end{center}
                \end{subfigure}
                \begin{subfigure}[t]{0.40\textwidth}
                    \begin{center}
                        \includegraphics[width=\textwidth]{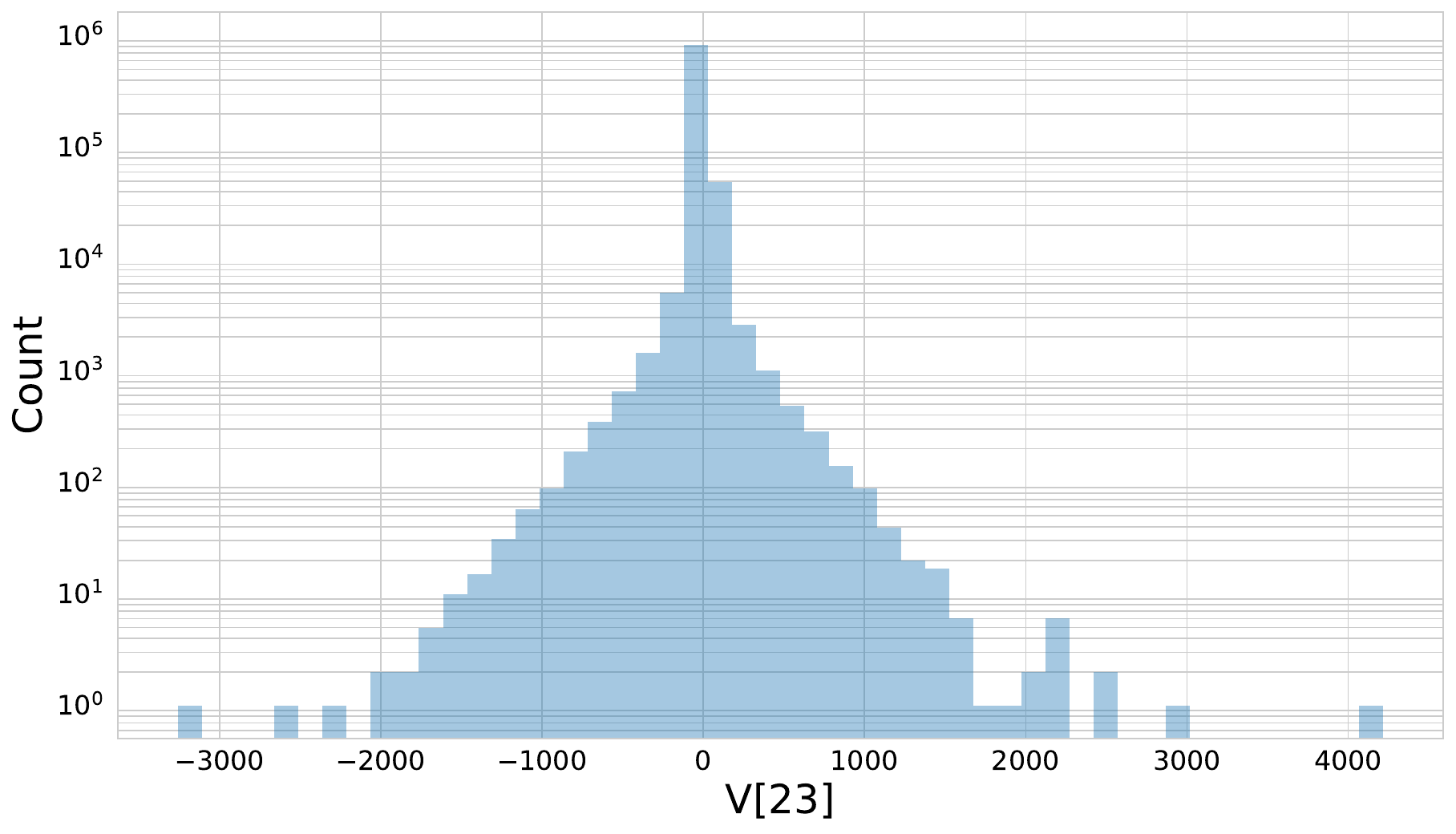}
                    \end{center}
                \end{subfigure}
                \begin{subfigure}[t]{0.40\textwidth}
                    \begin{center}
                        \includegraphics[width=\textwidth]{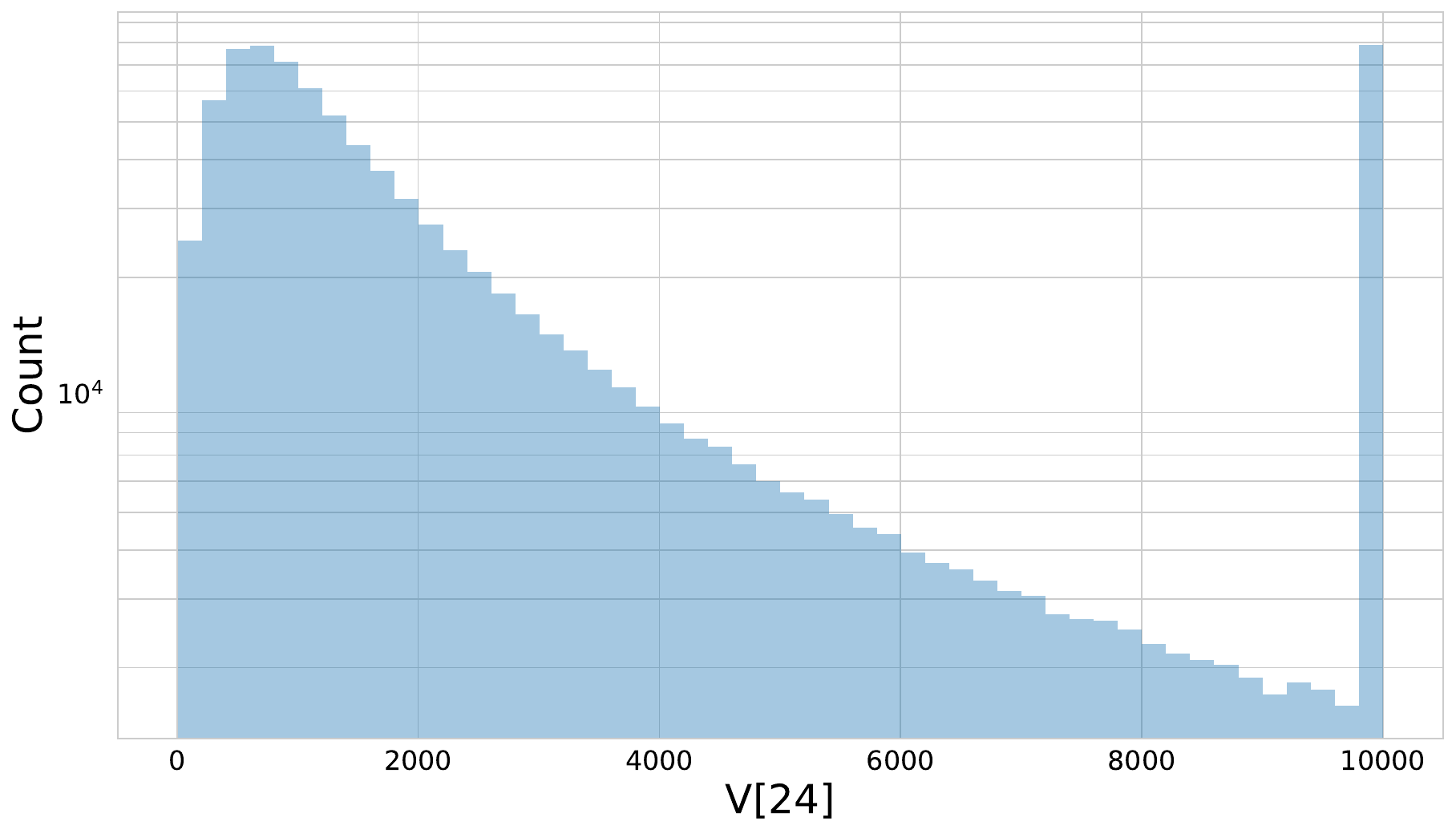}
                    \end{center}
                \end{subfigure}
                \begin{subfigure}[t]{0.40\textwidth}
                    \begin{center}
                        \includegraphics[width=\textwidth]{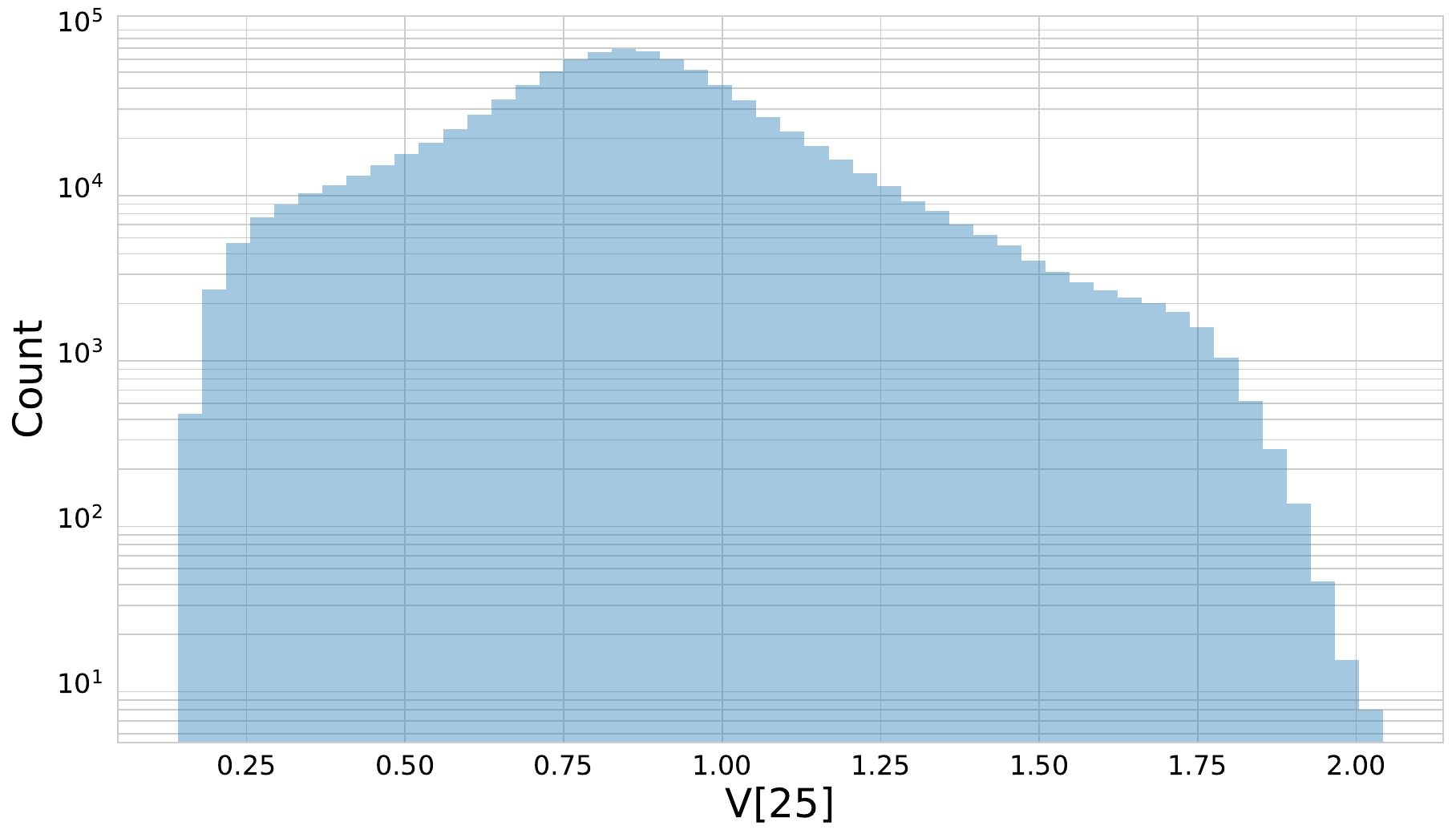}
                    \end{center}
                \end{subfigure}
                \begin{subfigure}[t]{0.40\textwidth}
                    \begin{center}
                        \includegraphics[width=\textwidth]{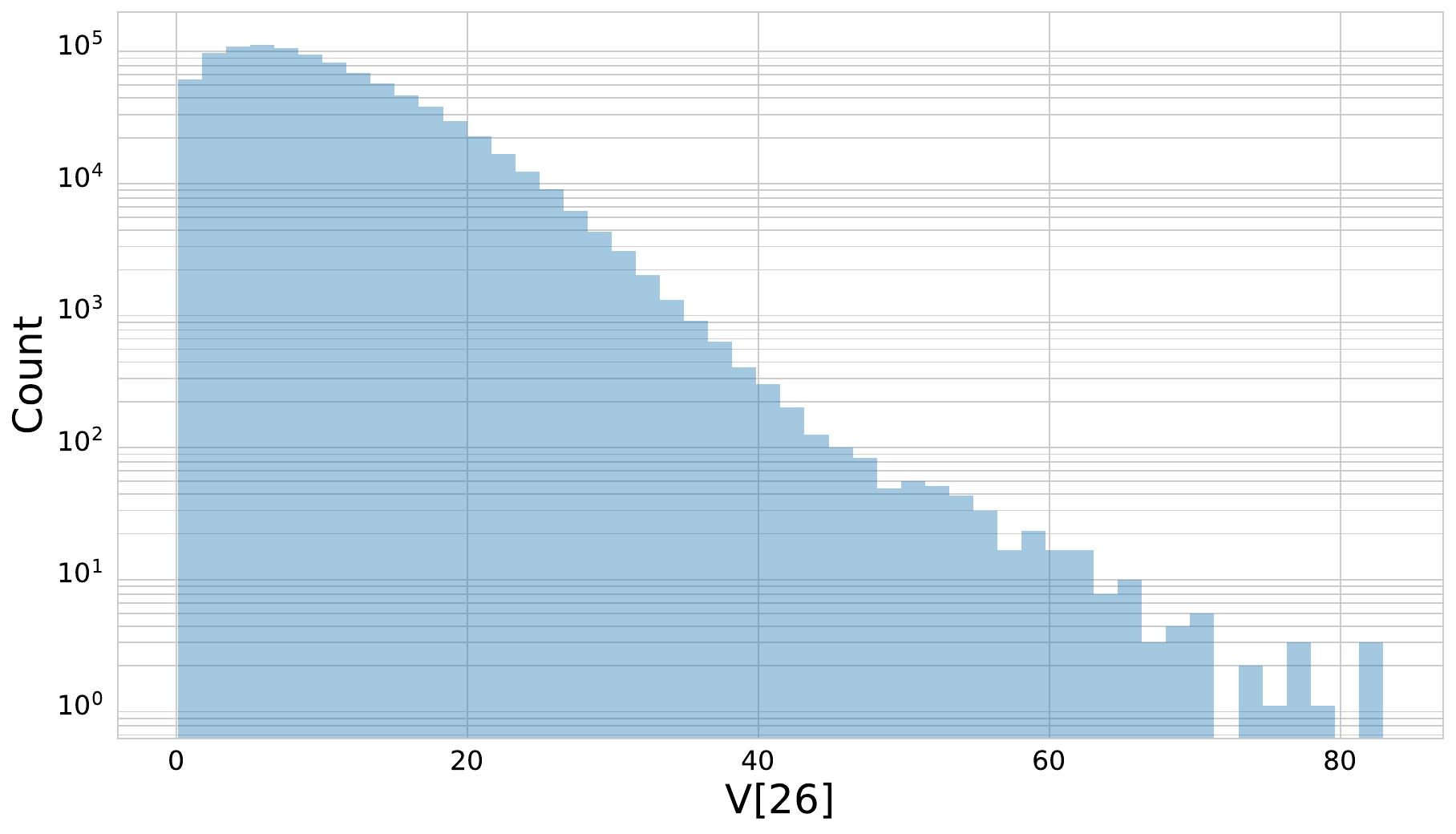}
                    \end{center}
                \end{subfigure}
                \begin{subfigure}[t]{0.40\textwidth}
                    \begin{center}
                        \includegraphics[width=\textwidth]{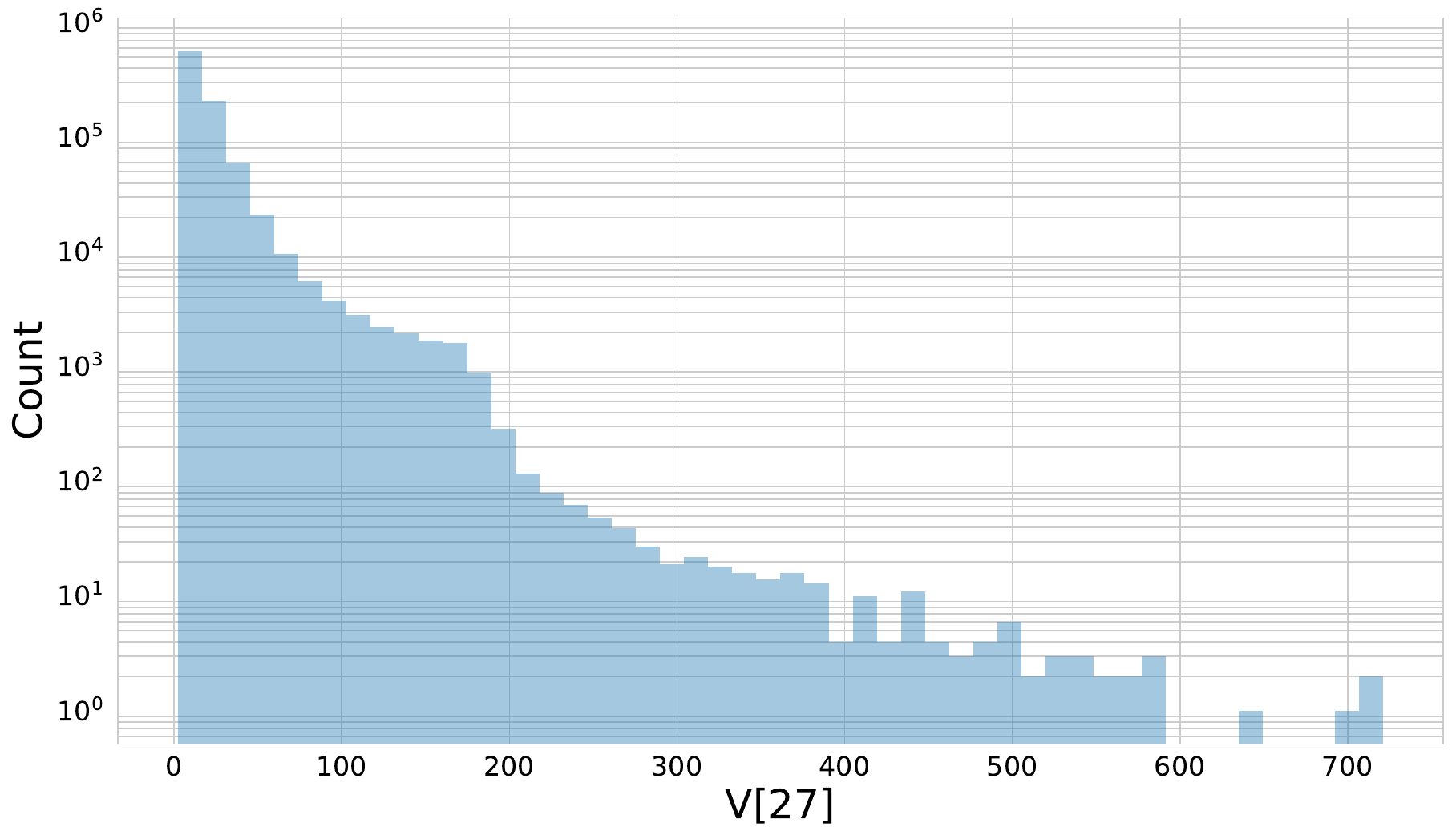}
                    \end{center}
                \end{subfigure}
                \caption{1D density-distributions of features V[20] to V[27]. Features are defined in Section~\ref{s:features_description}.}
                \label{f:1d_feats:20-27}
            \end{center}
        \end{figure*}

\begin{figure*}[h!]
            \begin{center}
                \begin{subfigure}[t]{0.30\textwidth}
                    \begin{center}
                        \includegraphics[width=\textwidth]{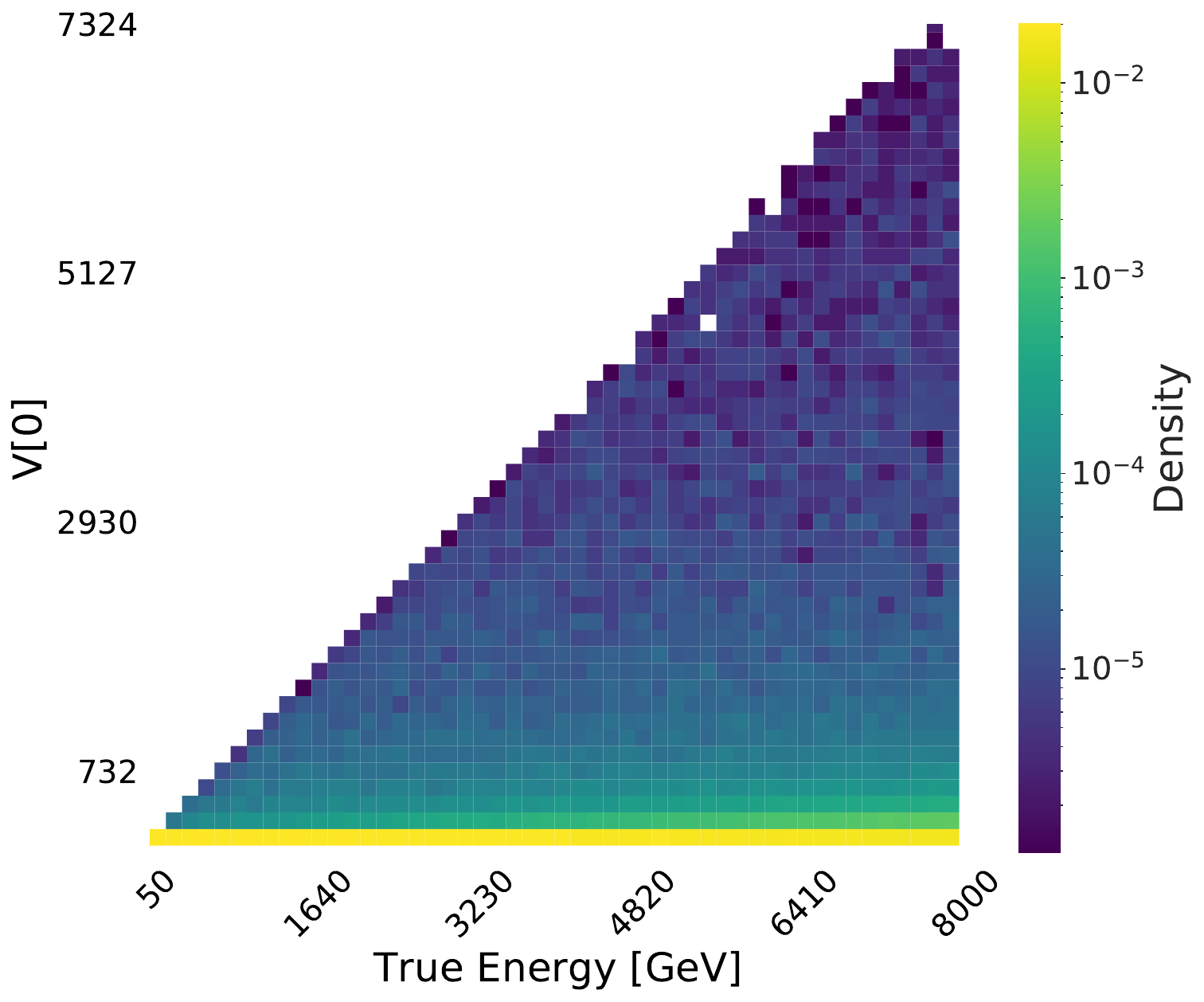}
                    \end{center}
                \end{subfigure}
                \begin{subfigure}[t]{0.30\textwidth}
                    \begin{center}
                        \includegraphics[width=\textwidth]{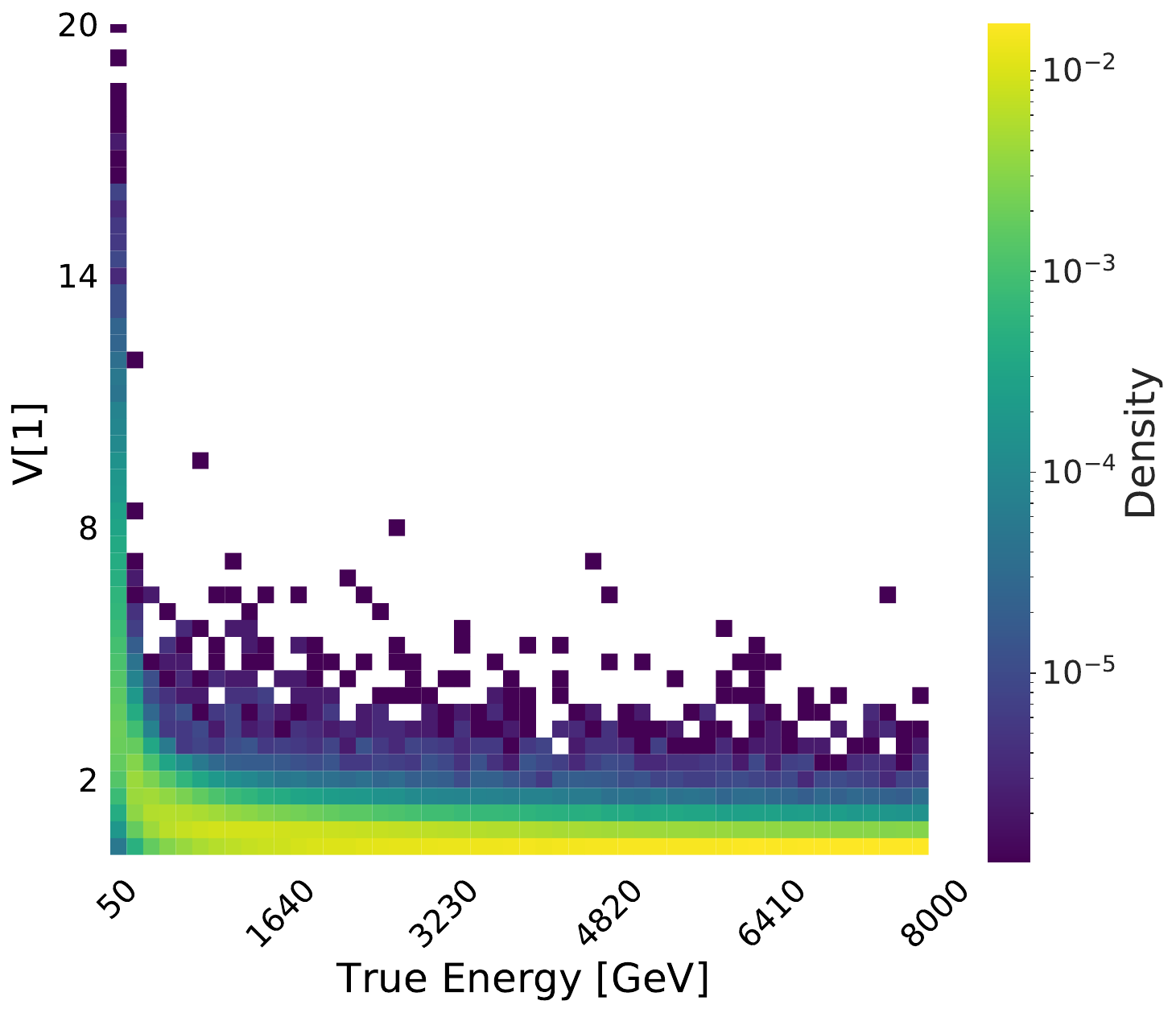}
                    \end{center}
                \end{subfigure}
                \begin{subfigure}[t]{0.30\textwidth}
                    \begin{center}
                        \includegraphics[width=\textwidth]{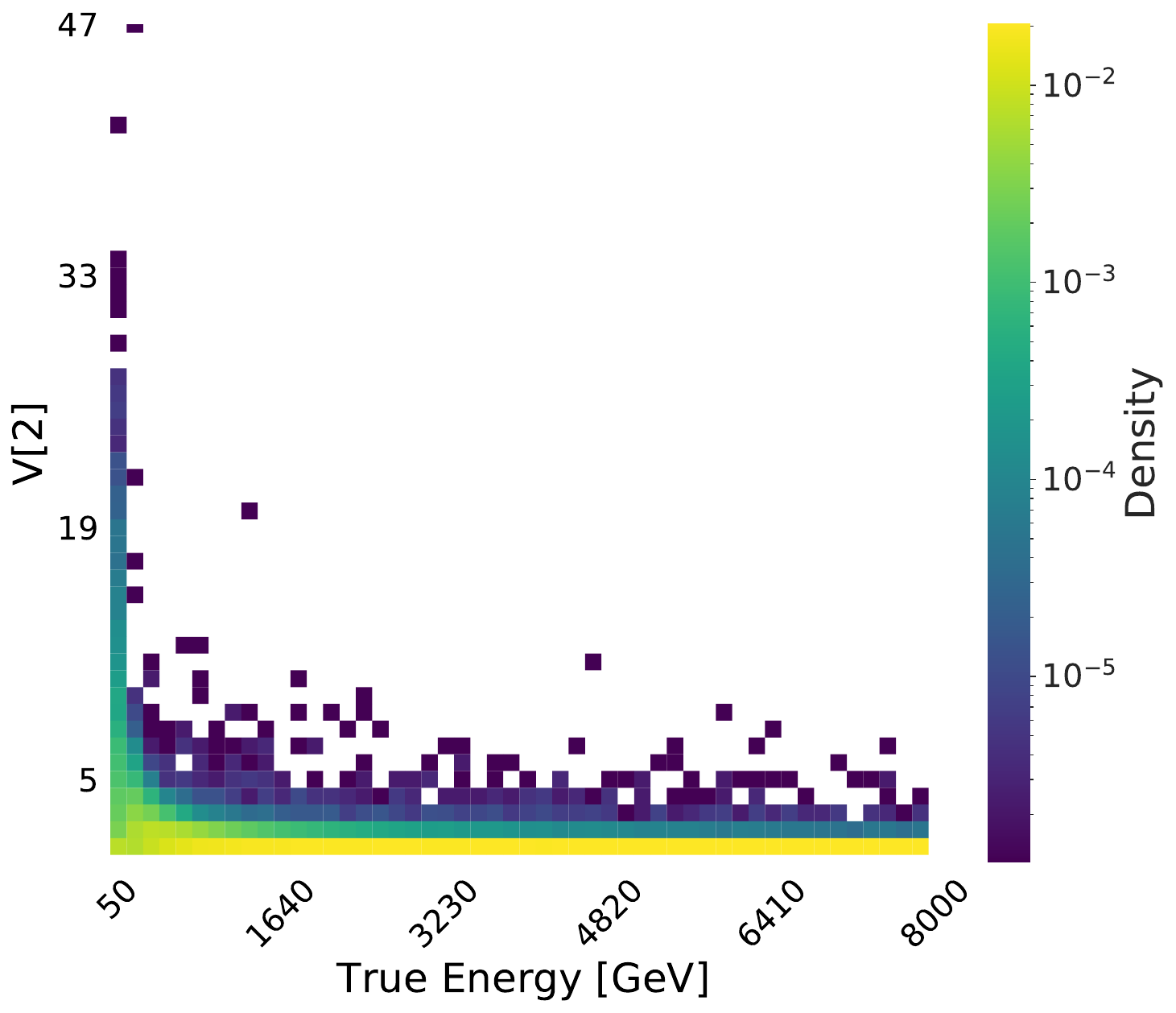}
                    \end{center}
                \end{subfigure}
                \begin{subfigure}[t]{0.30\textwidth}
                    \begin{center}
                        \includegraphics[width=\textwidth]{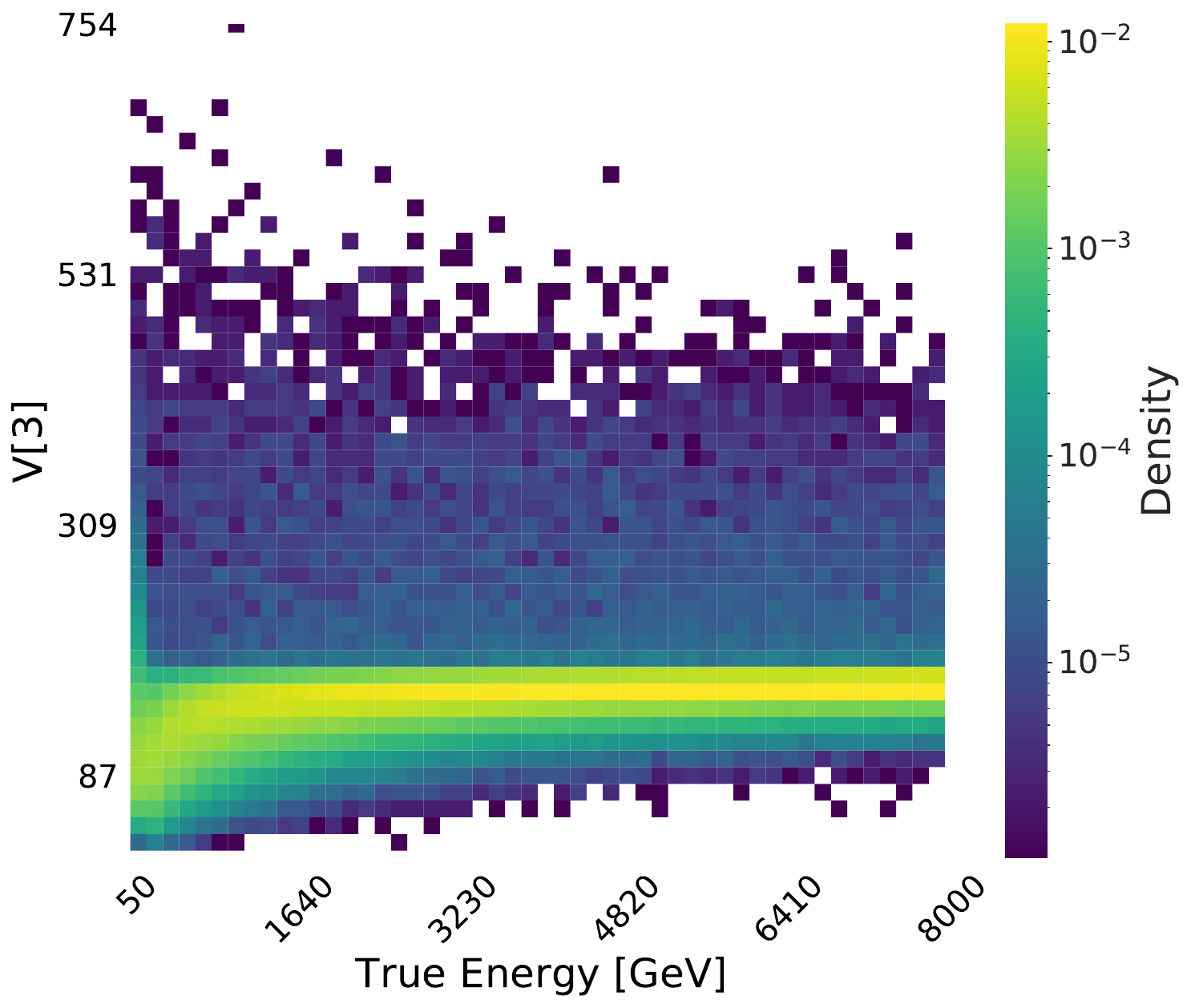}
                    \end{center}
                \end{subfigure}
                \begin{subfigure}[t]{0.30\textwidth}
                    \begin{center}
                        \includegraphics[width=\textwidth]{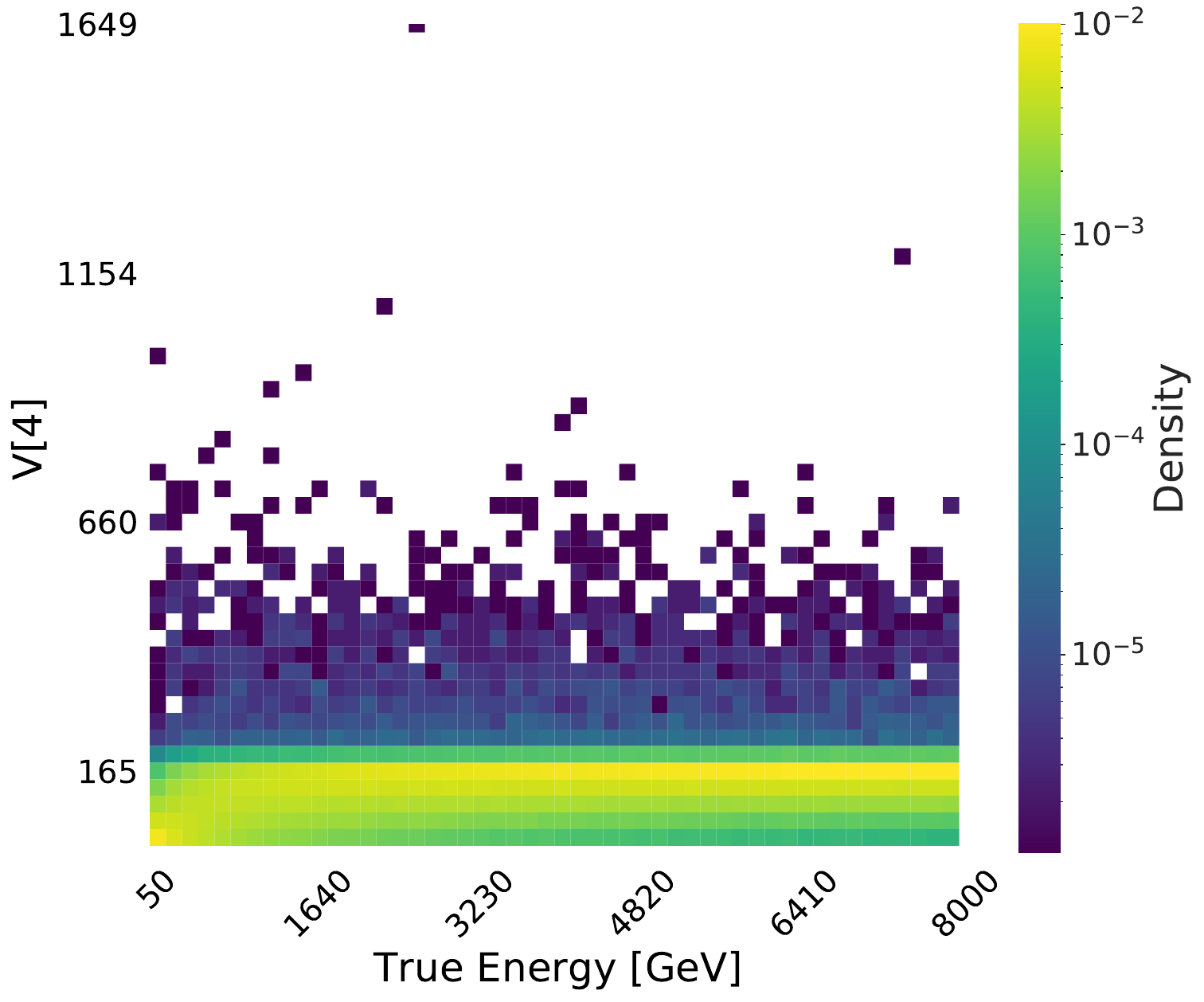}
                    \end{center}
                \end{subfigure}
                \begin{subfigure}[t]{0.30\textwidth}
                    \begin{center}
                        \includegraphics[width=\textwidth]{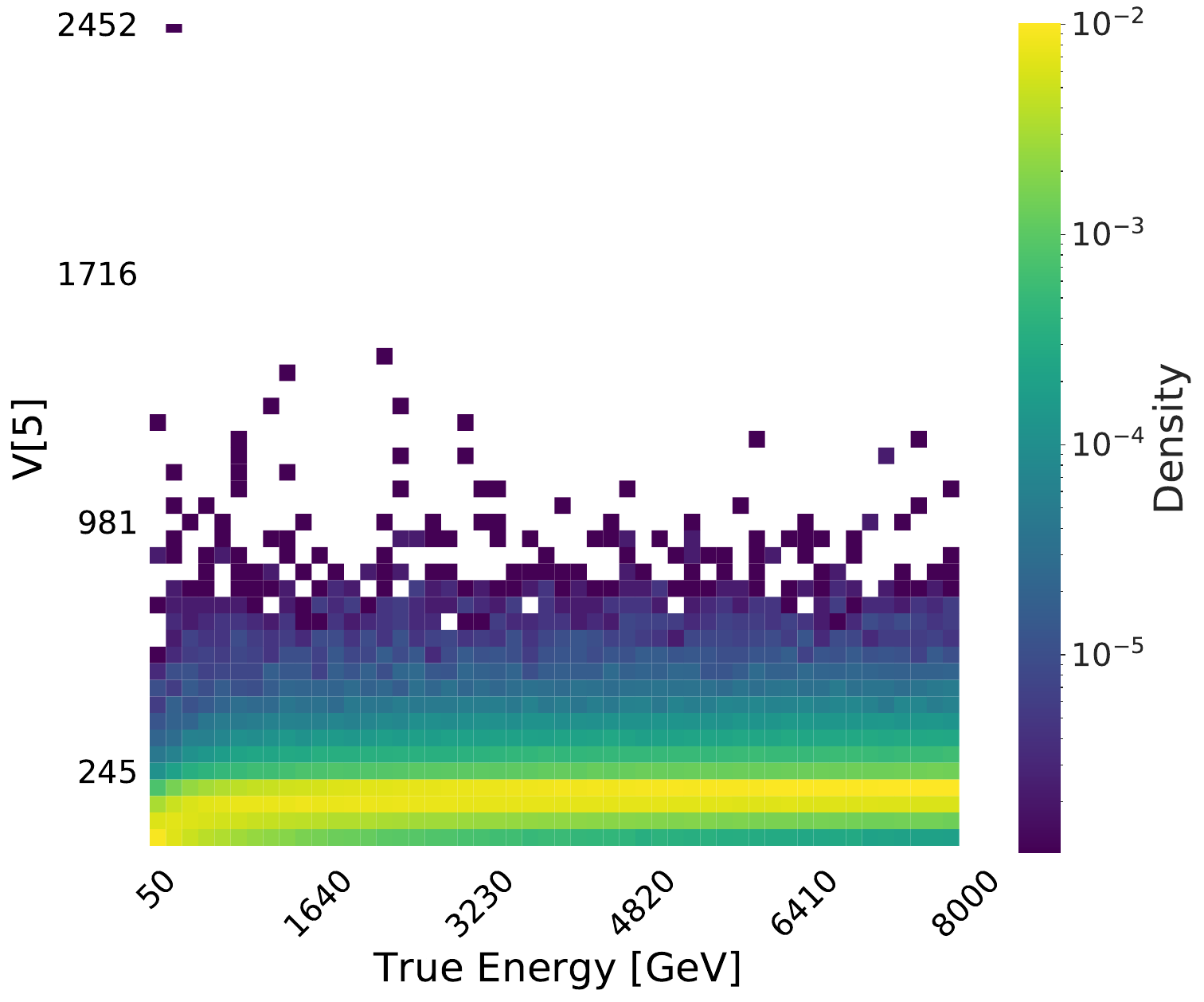}
                    \end{center}
                \end{subfigure}
                \begin{subfigure}[t]{0.30\textwidth}
                    \begin{center}
                        \includegraphics[width=\textwidth]{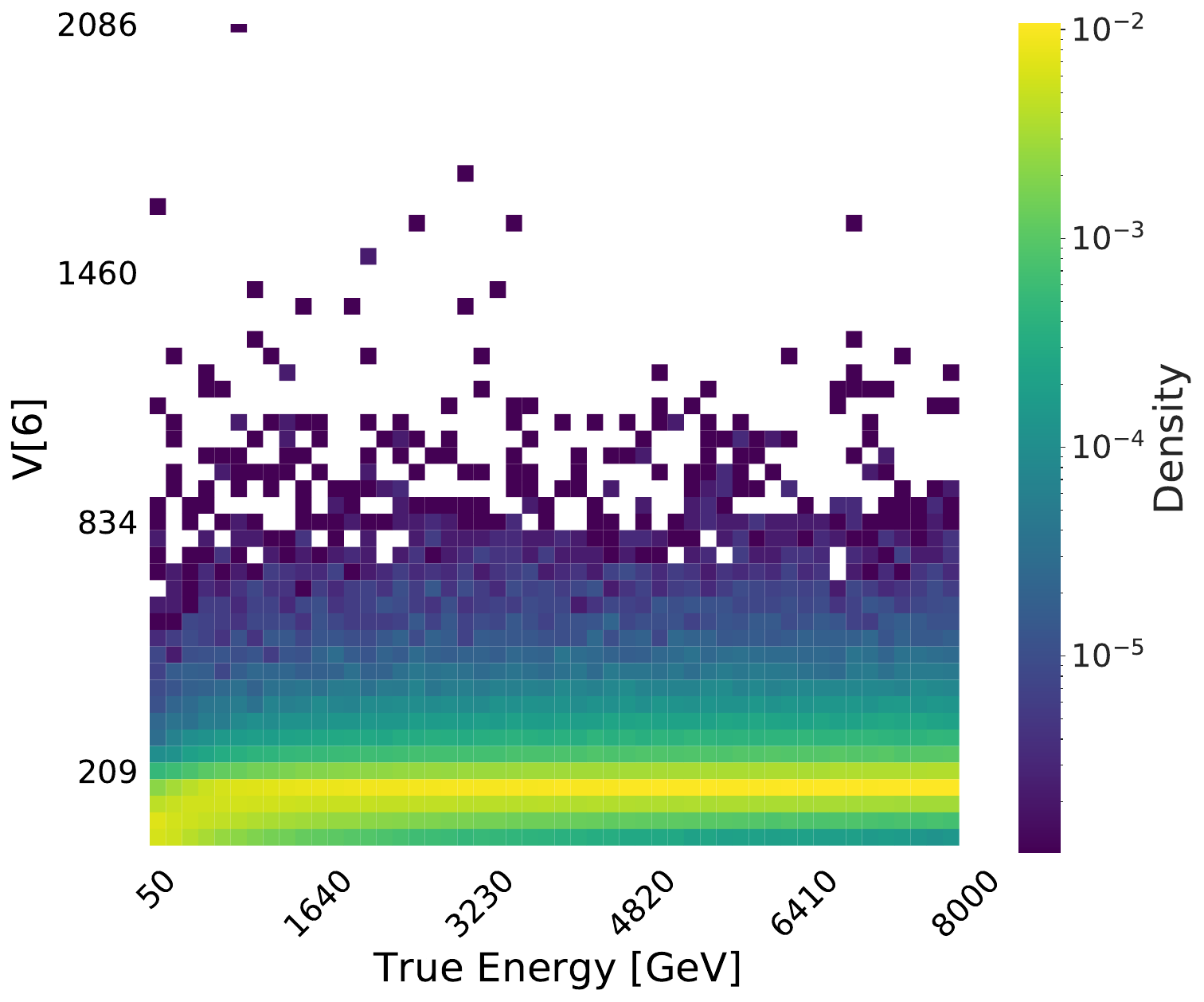}
                    \end{center}
                \end{subfigure}
                \begin{subfigure}[t]{0.30\textwidth}
                    \begin{center}
                        \includegraphics[width=\textwidth]{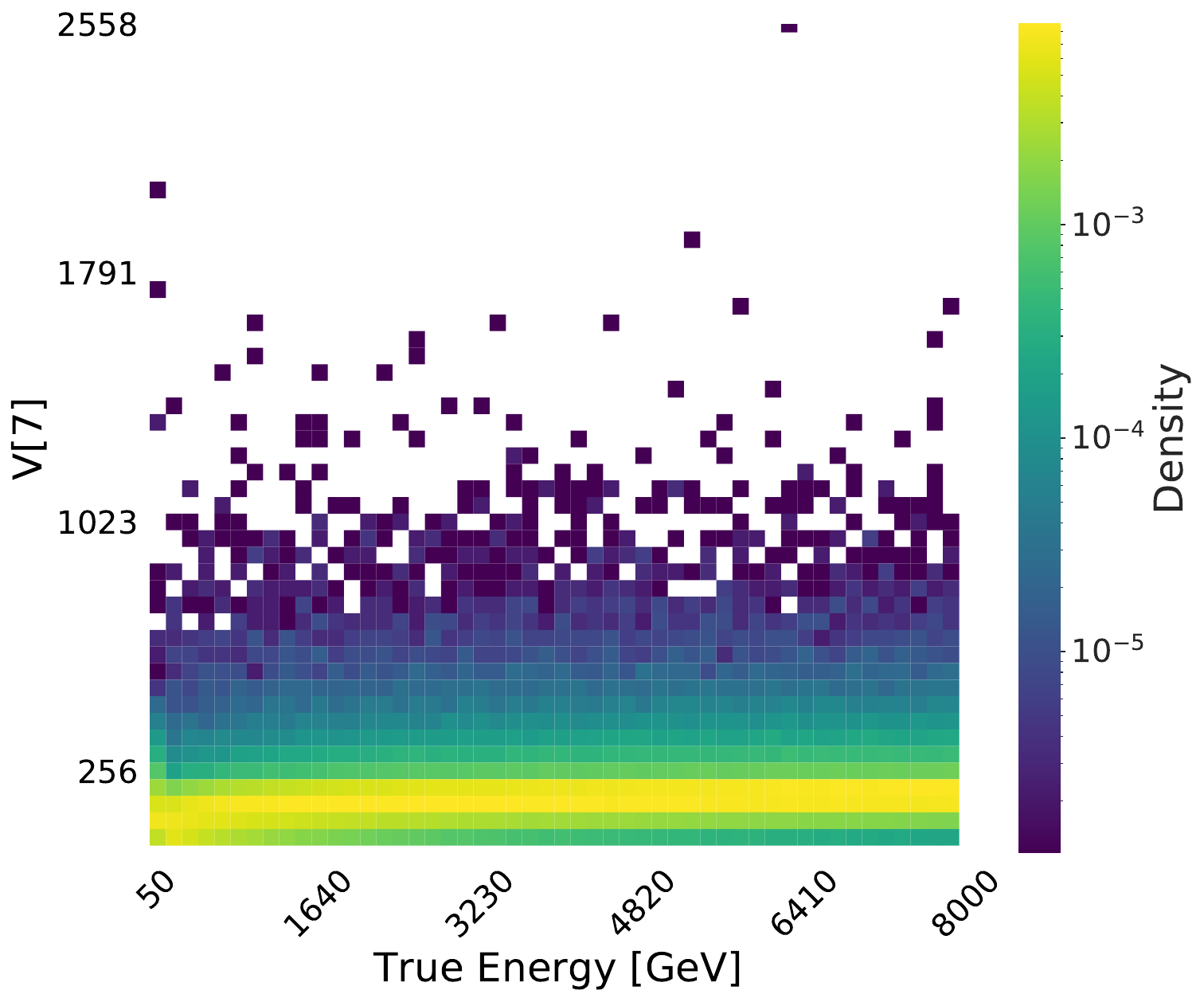}
                    \end{center}
                \end{subfigure}
                \begin{subfigure}[t]{0.30\textwidth}
                    \begin{center}
                        \includegraphics[width=\textwidth]{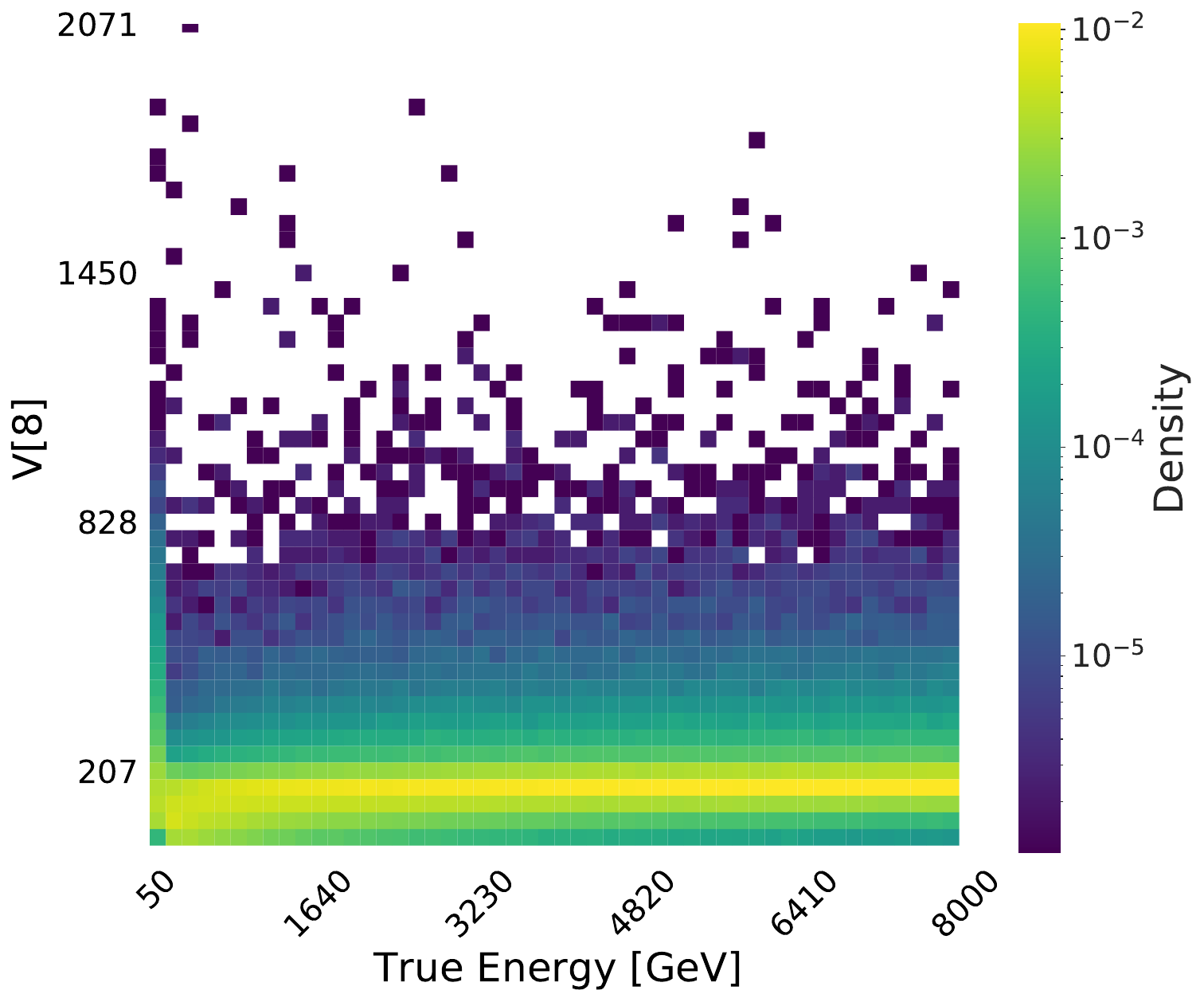}
                    \end{center}
                \end{subfigure}
                \begin{subfigure}[t]{0.30\textwidth}
                    \begin{center}
                        \includegraphics[width=\textwidth]{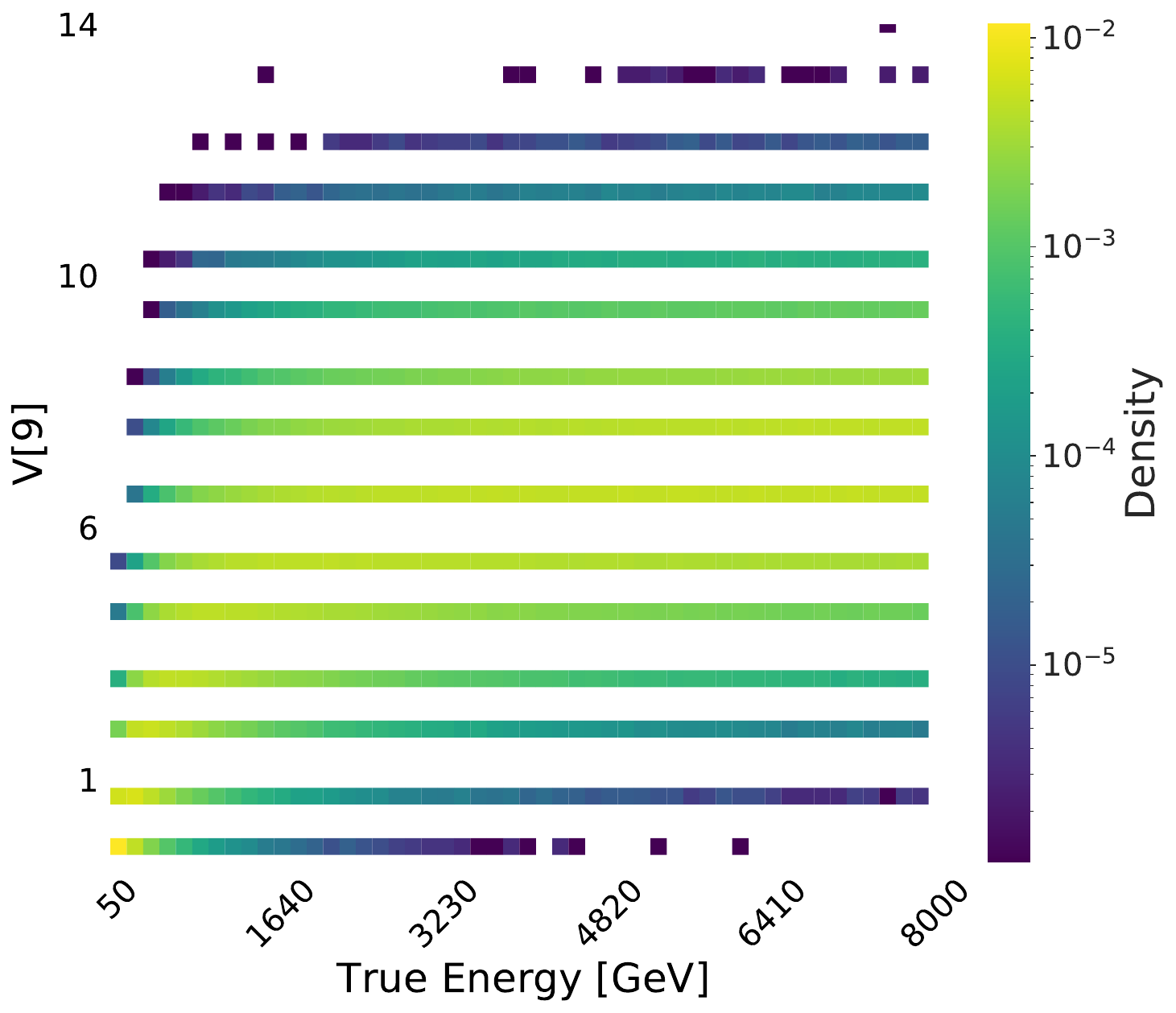}
                    \end{center}
                \end{subfigure}
                \caption{2D histograms showing the dependence of features V[0] to V[9] (on the y axes) on true muon energy (on the x axes). Features are defined in Section~\ref{s:features_description}.}
                \label{f:2d_feats:0-9}
            \end{center}
        \end{figure*}

\begin{figure*}[h!]
            \begin{center}
                \begin{subfigure}[t]{0.30\textwidth}
                    \begin{center}
                        \includegraphics[width=\textwidth]{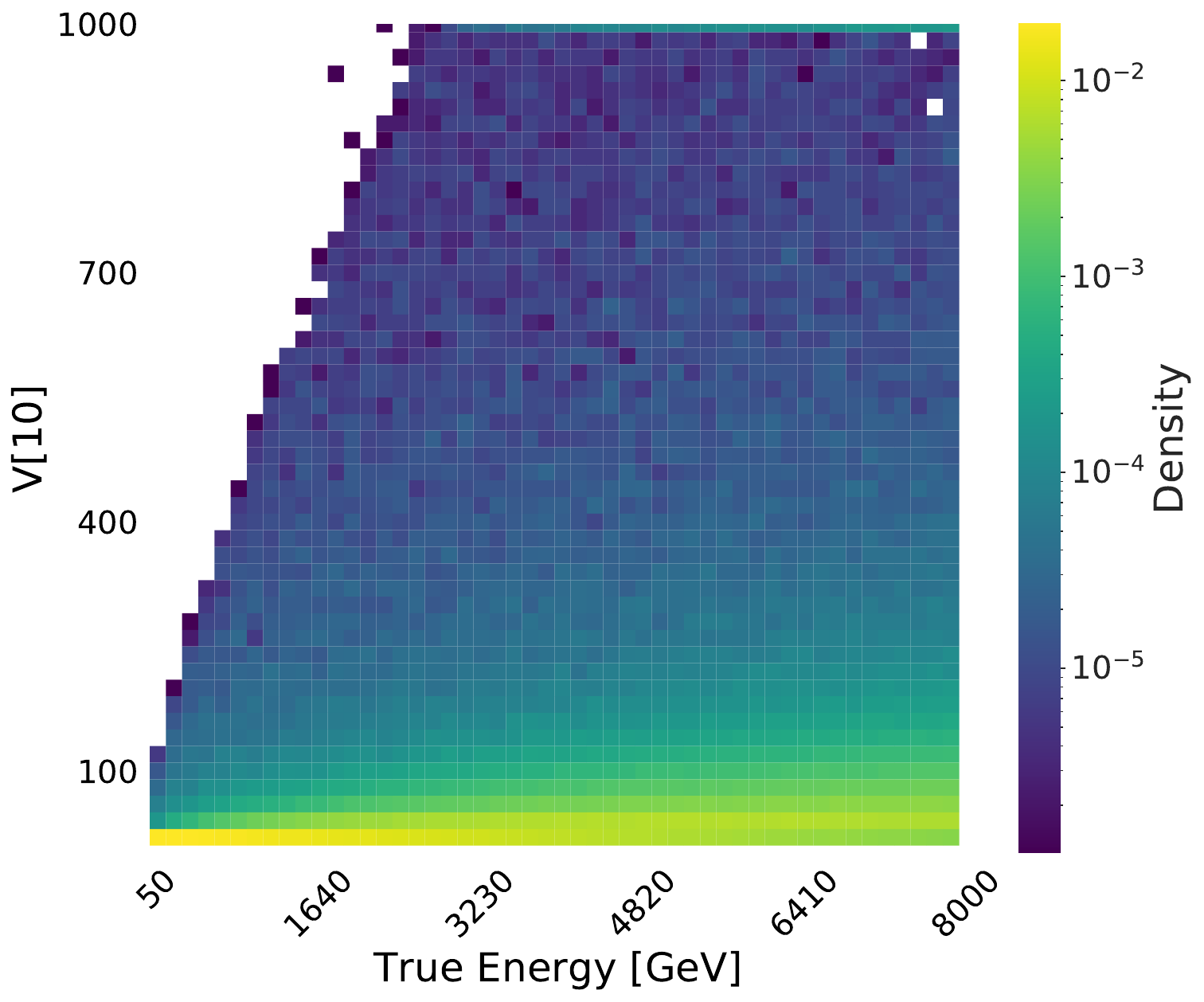}
                    \end{center}
                \end{subfigure}
                \begin{subfigure}[t]{0.30\textwidth}
                    \begin{center}
                        \includegraphics[width=\textwidth]{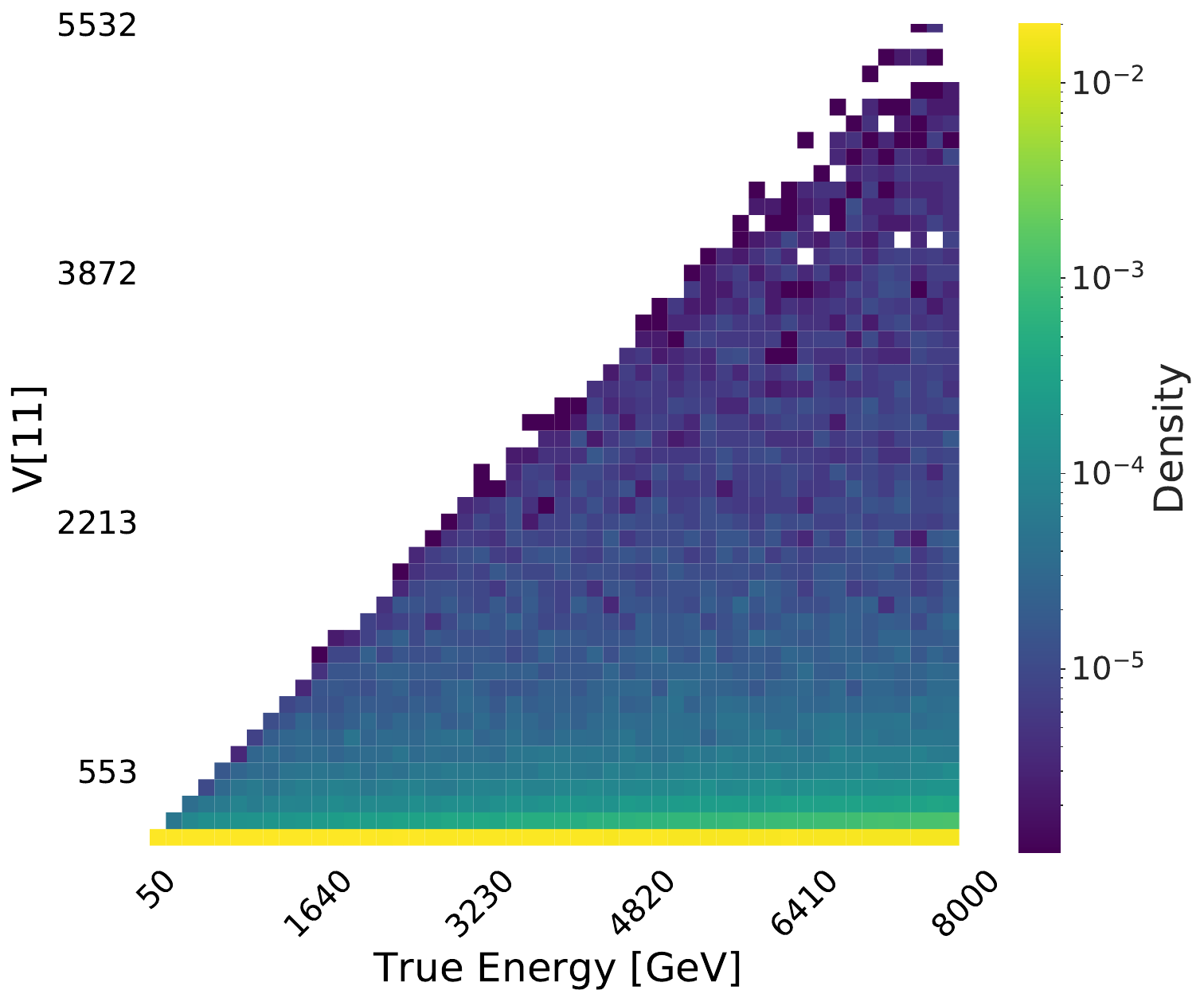}
                    \end{center}
                \end{subfigure}
                \begin{subfigure}[t]{0.30\textwidth}
                    \begin{center}
                        \includegraphics[width=\textwidth]{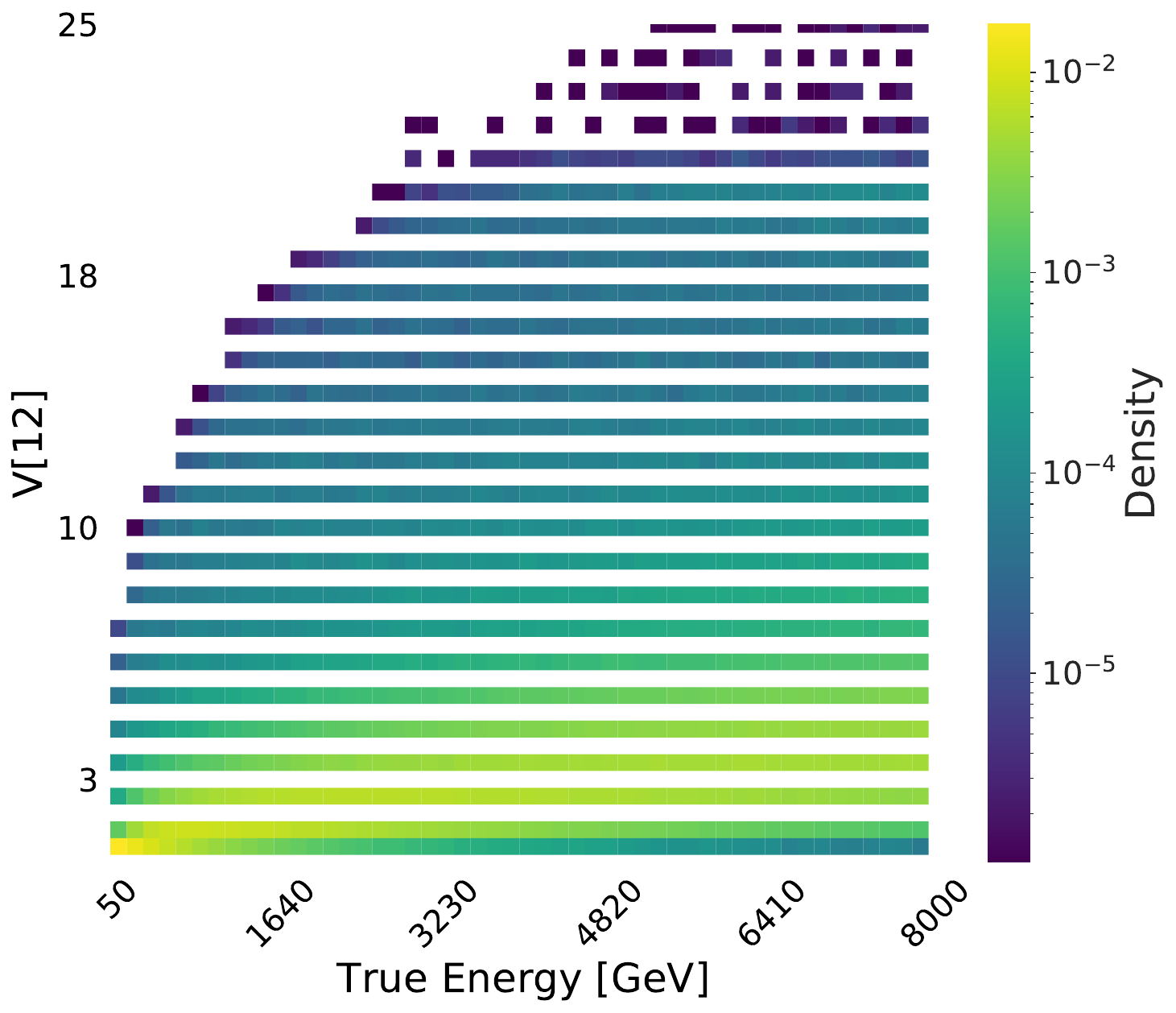}
                    \end{center}
                \end{subfigure}
                \begin{subfigure}[t]{0.30\textwidth}
                    \begin{center}
                        \includegraphics[width=\textwidth]{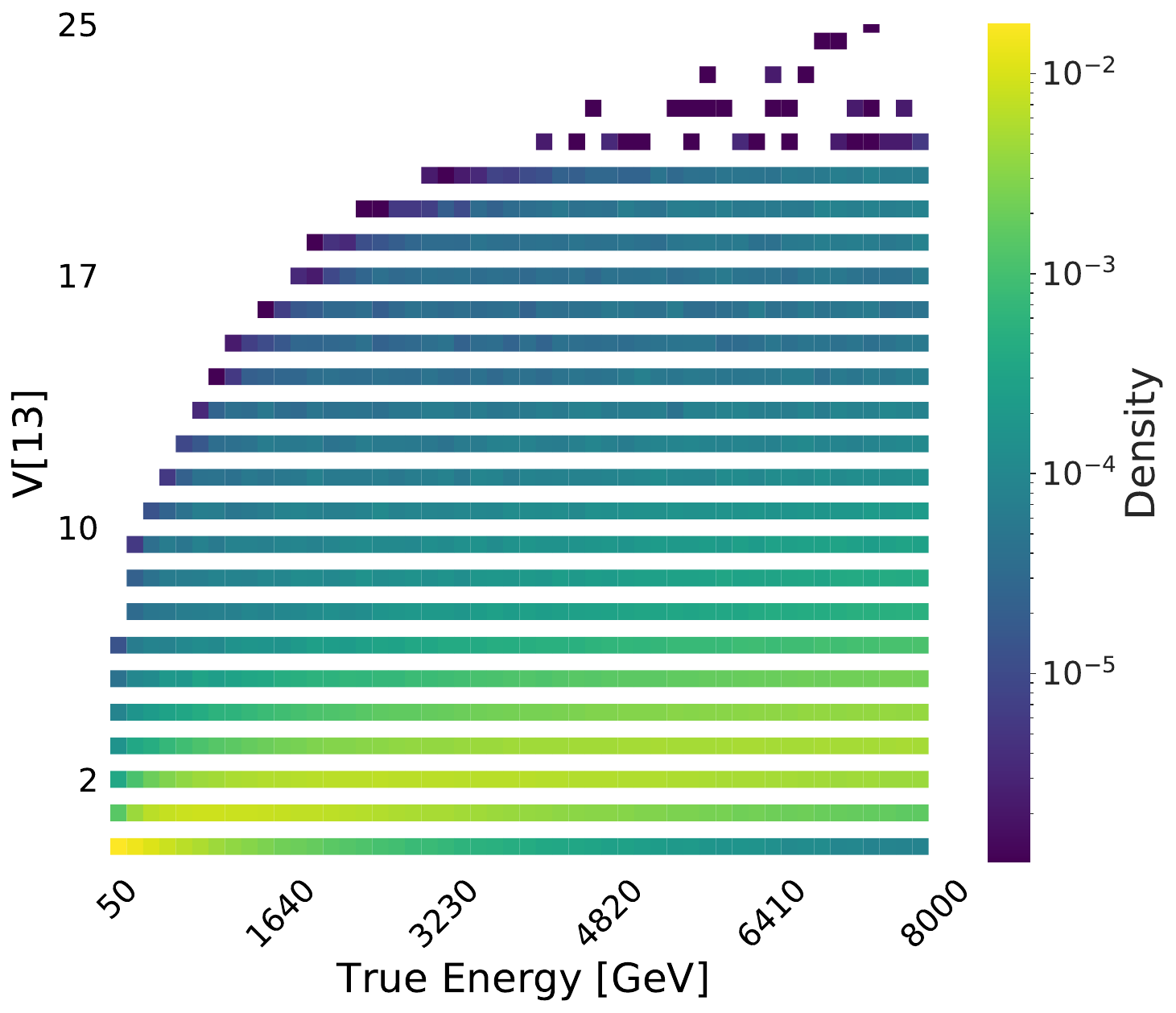}
                    \end{center}
                \end{subfigure}
                \begin{subfigure}[t]{0.30\textwidth}
                    \begin{center}
                        \includegraphics[width=\textwidth]{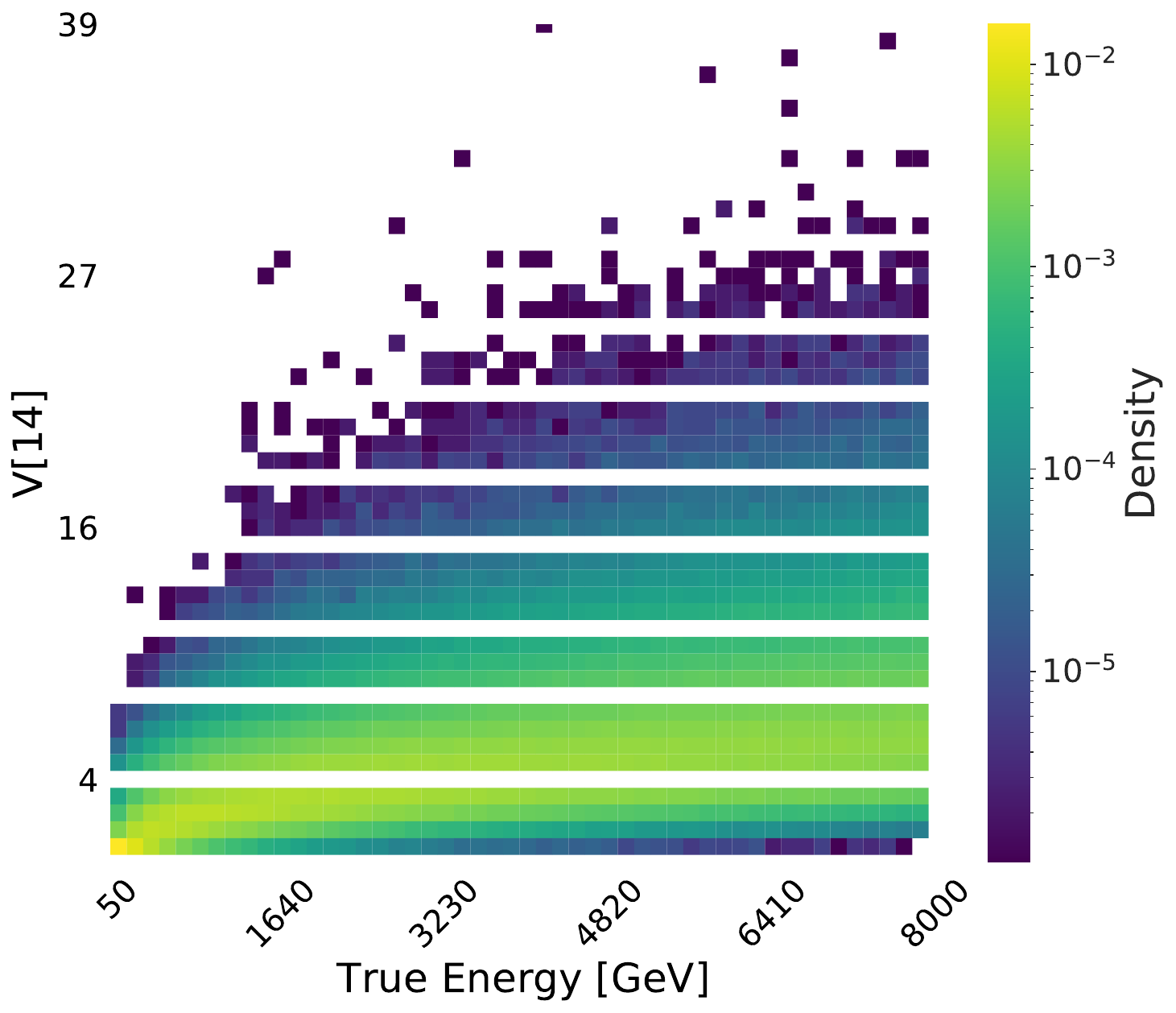}
                    \end{center}
                \end{subfigure}
                \begin{subfigure}[t]{0.30\textwidth}
                    \begin{center}
                        \includegraphics[width=\textwidth]{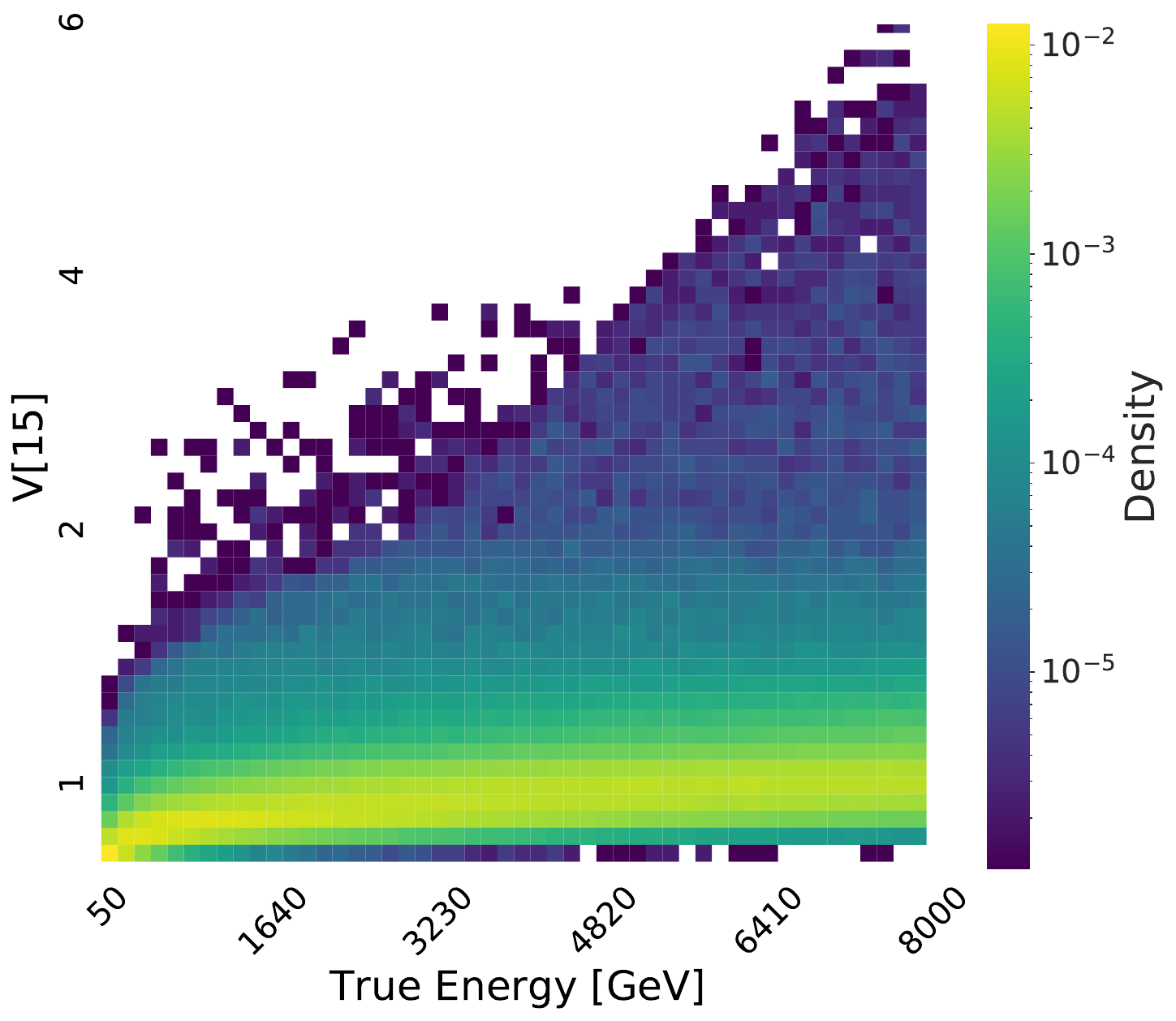}
                    \end{center}
                \end{subfigure}
                \begin{subfigure}[t]{0.30\textwidth}
                    \begin{center}
                        \includegraphics[width=\textwidth]{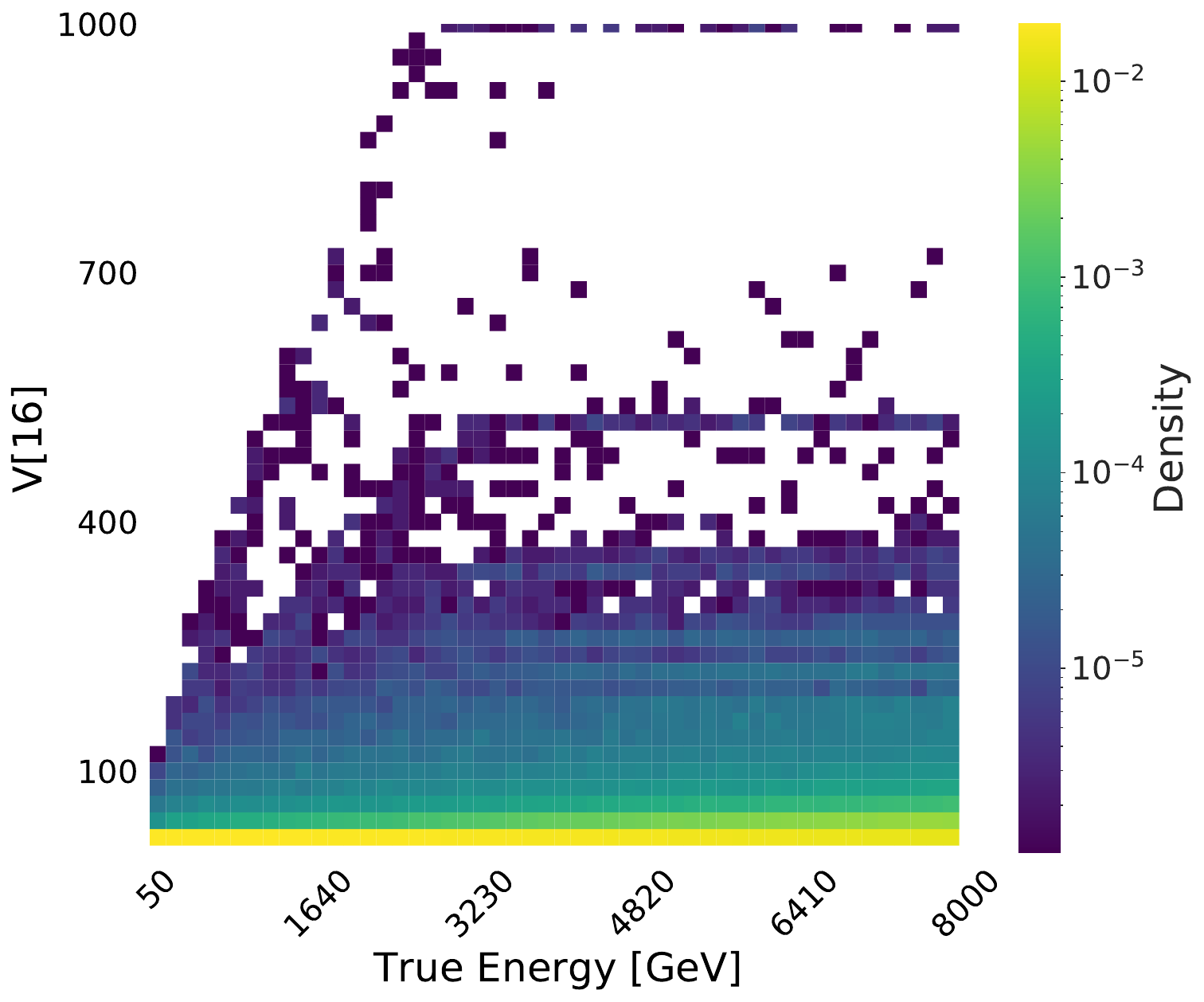}
                    \end{center}
                \end{subfigure}
                \begin{subfigure}[t]{0.30\textwidth}
                    \begin{center}
                        \includegraphics[width=\textwidth]{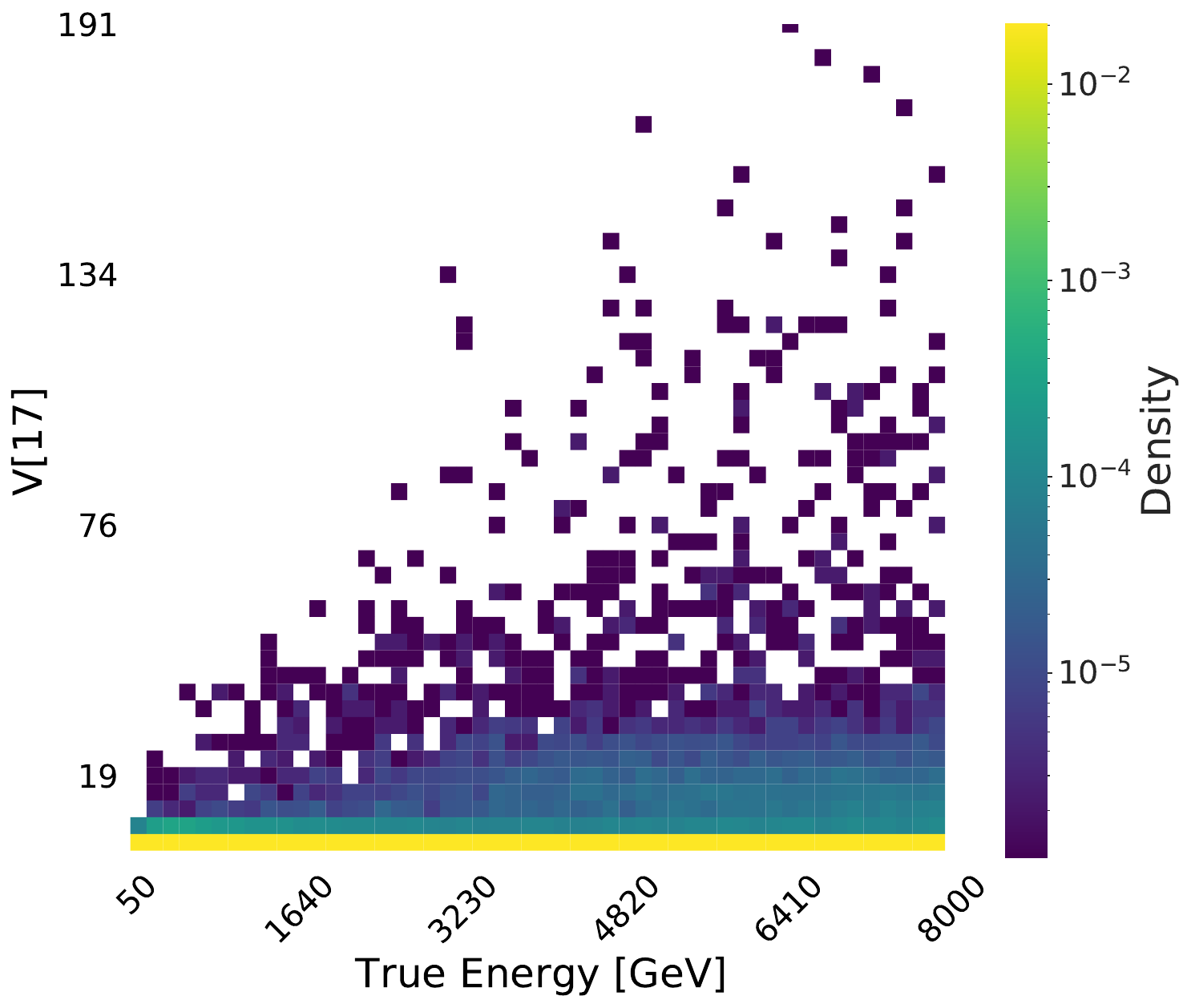}
                    \end{center}
                \end{subfigure}
                \begin{subfigure}[t]{0.30\textwidth}
                    \begin{center}
                        \includegraphics[width=\textwidth]{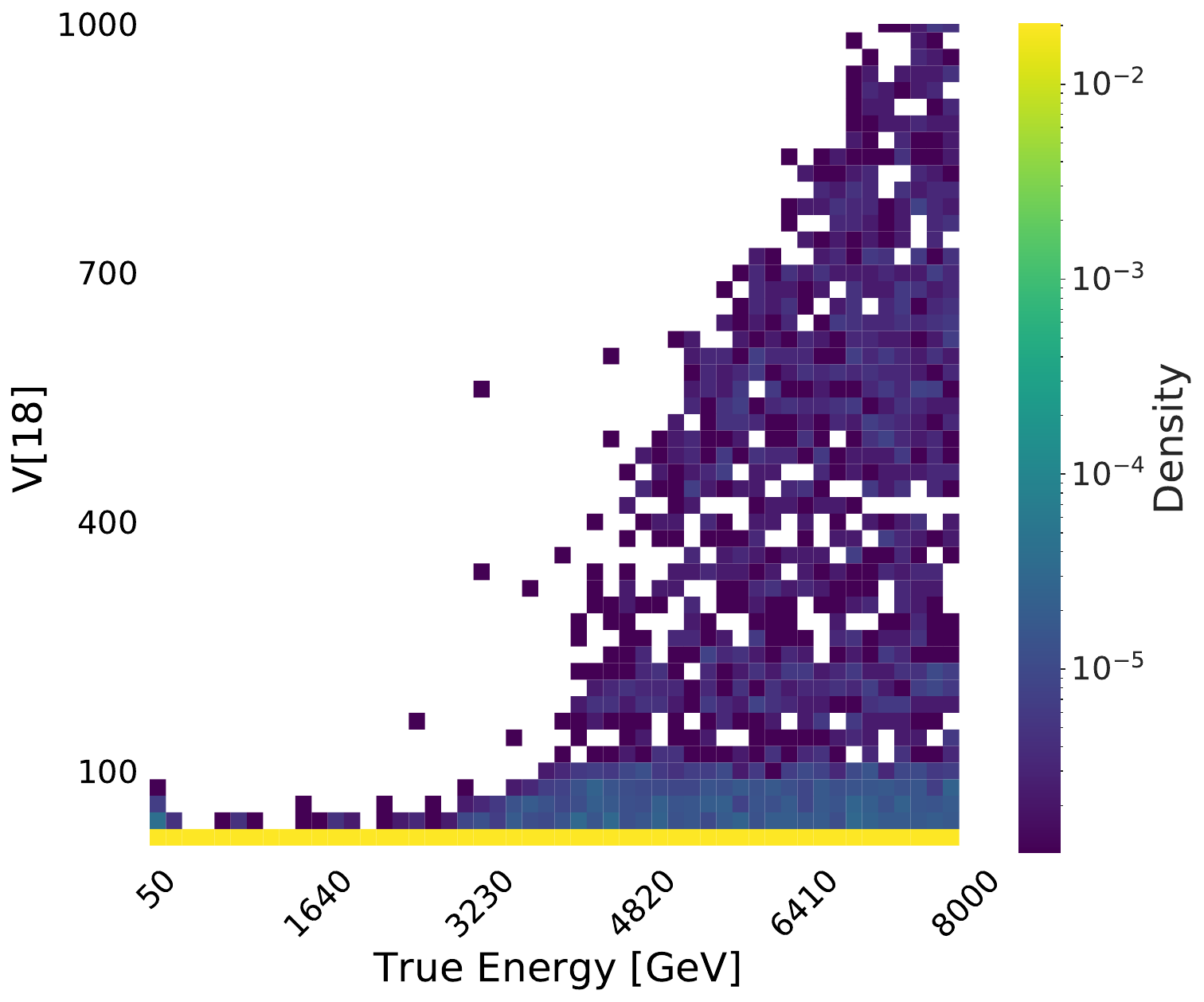}
                    \end{center}
                \end{subfigure}
                \begin{subfigure}[t]{0.30\textwidth}
                    \begin{center}
                        \includegraphics[width=\textwidth]{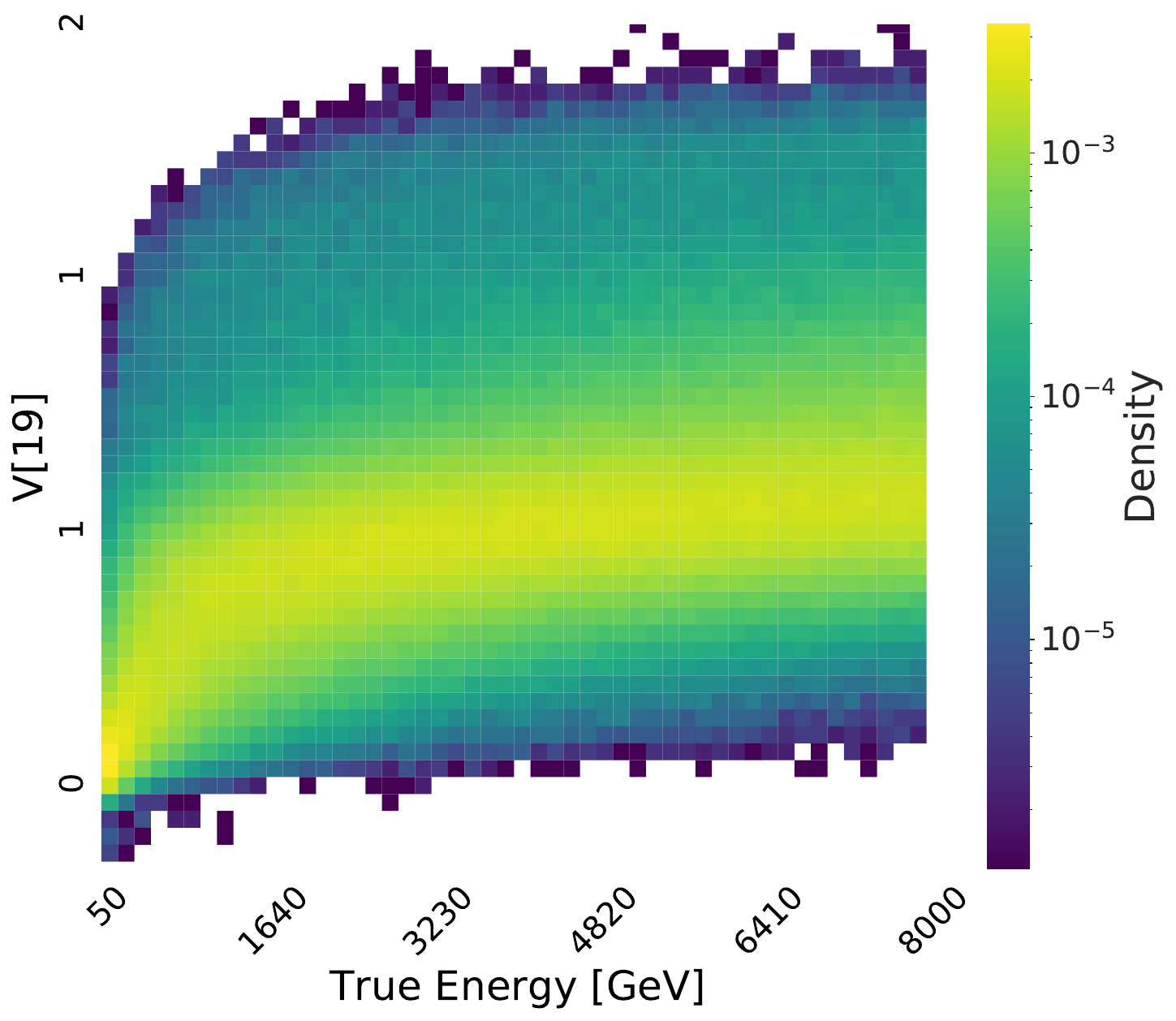}
                    \end{center}
                \end{subfigure}
                \caption{2D histograms showing the dependence of features V[10] to V[19] (on the y axes) on true muon energy (on the x axes). Features are defined in Section~\ref{s:features_description}.}
                \label{f:2d_feats:10-19}
            \end{center}
        \end{figure*}

        \noindent
        Finally, some cells may remain non associated to any type-1 or type-2 clusters. To extract further information from them, we search for the $3\!\times\!3\times\!3$ cube of 27 cells in $x$,$y$,$z$ which captures the highest total energy among cells still not included in clusters (V[25]), and the second-highest total energy (V[15]). These two features are listed {\em infra}. Standardised distribution of these features, along with the others defined in this Appendix, are shown in Figs.~\ref{f:1d_feats:0-9}-\ref{f:1d_feats:20-27}; correlations with muon energy are shown in Figs.~\ref{f:2d_feats:0-9}-\ref{f:2d_feats:20-27}.
        
        %%%%%%%%%%%%%%%%%%%%%%%%%%%%%%%%%%%%%%%%%%%%%%%%%%%%%%%%%%%%%%%%%%%%%%%%%%
        \subsection{Measuring curvature with energy deposits \label{s:bendingvar}}
        
        Muons entering our simulated calorimeter do so with an initial trajectory orthogonal to the calorimeter front face~\footnote{ A very small initial bending is produced for low-energy muons in traversing the first 50mm from the point of origin to the calorimeter front face.}. From that point on, they undergo interactions with the material, as well as a bending Lorentz force. If we ignore all physical effects except the magnetic bending, which we wish to estimate, we may model the muon track as an arc of a circle in the $xz$ plane. At the back face of the calorimeter, the expected deviation of such a circumference from a straight line oriented along $z$  is very small in absolute terms: for a muon of momentum $P$ in GeV in a magnetic field $B$ in Tesla, the curvature of the trajectory is $R=P/(0.3 B)$ meters, hence the estimated deviation is $\Delta_x = R-\sqrt{R^2-\Delta_z^2}$, where $\Delta_z=\SI{2}{\meter}$ is the calorimeter depth along $z$. Assuming, {\em e.g.}, $P=\SI{600}{GeV}$ we find a curvature $R=\SI{1000}{\meter}$ and from it a displacement $\Delta_x=\SI{2}{\mm}$, which is already smaller than the calorimeter granularity. 
        
        In constructing a variable sensitive to curvature, we observe that circular trajectories that start orthogonal at the front face of the calorimeter may in principle be determined by measuring any two points along their path in the lead tungstate material. We further notice that while calorimeter cells traversed by the muon track usually collect a detectable amount of energy from ionization processes, they are not the only ones carrying information on the muon trajectory. In fact, for muons that bend very little in the magnetic field, the process of muon radiation in a homogeneous medium is dominated by brehmsstrahlung originating by multiple scattering processes. In the plane orthogonal to the muon trajectory the direction of the emitted photons is thus largely random, but these photons do not travel very far before depositing their energy in calorimeter cells. Hence the position of additional cells lit up by photons traveling away from the muon trajectory contains a good deal of extra information on the position of the radiating particle. 
        
        We construct a statistical estimator of the muon curvature by determining two separate points in the $xz$ plane, using separately the first 25 and the second 25 layers of crystals in $z$. We compute the following weighted averages:
        
        \begin{center}
        \large
        
        \begin{math}
           \hat{x_1} = \frac{\sum_{i_1} E_i x_i w_i}{\sum_{i_1} E_i w_i}   
        \end{math}
        
        \begin{math}
           \hat{z_1} = \frac{\sum_{i_1} E_i z_i w_i}{\sum_{i_1} E_i w_i} 
        \end{math}
           
        \begin{math}
           \hat{x_2} = \frac{\sum_{i_2} E_i x_i w_i}{\sum_{i_2} E_i w_i}   
        \end{math}
        
        \begin{math}
           \hat{z_2} = \frac{\sum_{i_2} E_i z_i w_i}{\sum_{i_2} E_i w_i}  
        \end{math}
        
        \normalsize
        \end{center}
        
        \noindent where the sums over indices $i_1$ run on calorimeter cells in the first 25 layers, and sums over indices $i_2$ run on calorimeter cells in the second 25 layers along $z$, and where weights $w_i$ are defined as follows:\par
        
        \begin{center}
        \begin{math}
           w_i = exp(-(|y_i-y_\mu|)/50)
        \end{math}
        \end{center}
        
        \noindent with \si{\mm} units, and where $y_\mu$ is the center of the towers in the $y$ plane containing the highest amount of measured energy. In other words, calorimeter cells are assumed to contain information on the $xz$ position of the radiating particle in proportion to their detected energy, and inversely proportional to the distance of the cell to the $y$ coordinate at which the particle track lays.
        
        The two points $(x_1, z_1)$, $(x_2,z_2)$ in the $xz$ plane allow the construction of an estimator for the radius of the muon trajectory: we first specify the equation of a circumference as \par
        
        \begin{center}
        \begin{math}
            (x-x_0)^2 + z^2 = R^2
        \end{math}
        \end{center}
        
        \noindent
        from which we get \par
        
        \begin{center}
            \large
            \begin{math}
                x_0 = \frac{{z_2}^2-{z_1}^2+{x_2}^2-{x_1}^2}{2(x_2-x_1)}
            \end{math}
            \normalsize
        \end{center}
        
        \noindent
        and from it the radius estimator as \par
        
        \begin{center}
            \large
            \begin{math}
                R = \sqrt{{x_1}^2-2 x_1 x_0 +{x_0}^2 + {z_1}^2}
            \end{math}
            \normalsize
        \end{center}
        
        %if we define $\Delta_x=x_2-x_1$ and $\Delta_z = z_2-z_1$, simple trigonometry provide it as \par
        %
        %\begin{center}
        %    \large
        %    \begin{math}
        %        R = \frac{\sqrt{4 {\Delta_z}^4 + {\Delta_x}^4 + 5 {\Delta_x}^2 {\Delta_z}^2 }}{2 %{\Delta_x}}
        %    \end{math}
        %    \normalsize
        %    \end{center}
        
\begin{figure*}[h!]
            \begin{center}
                \begin{subfigure}[t]{0.30\textwidth}
                    \begin{center}
                        \includegraphics[width=\textwidth]{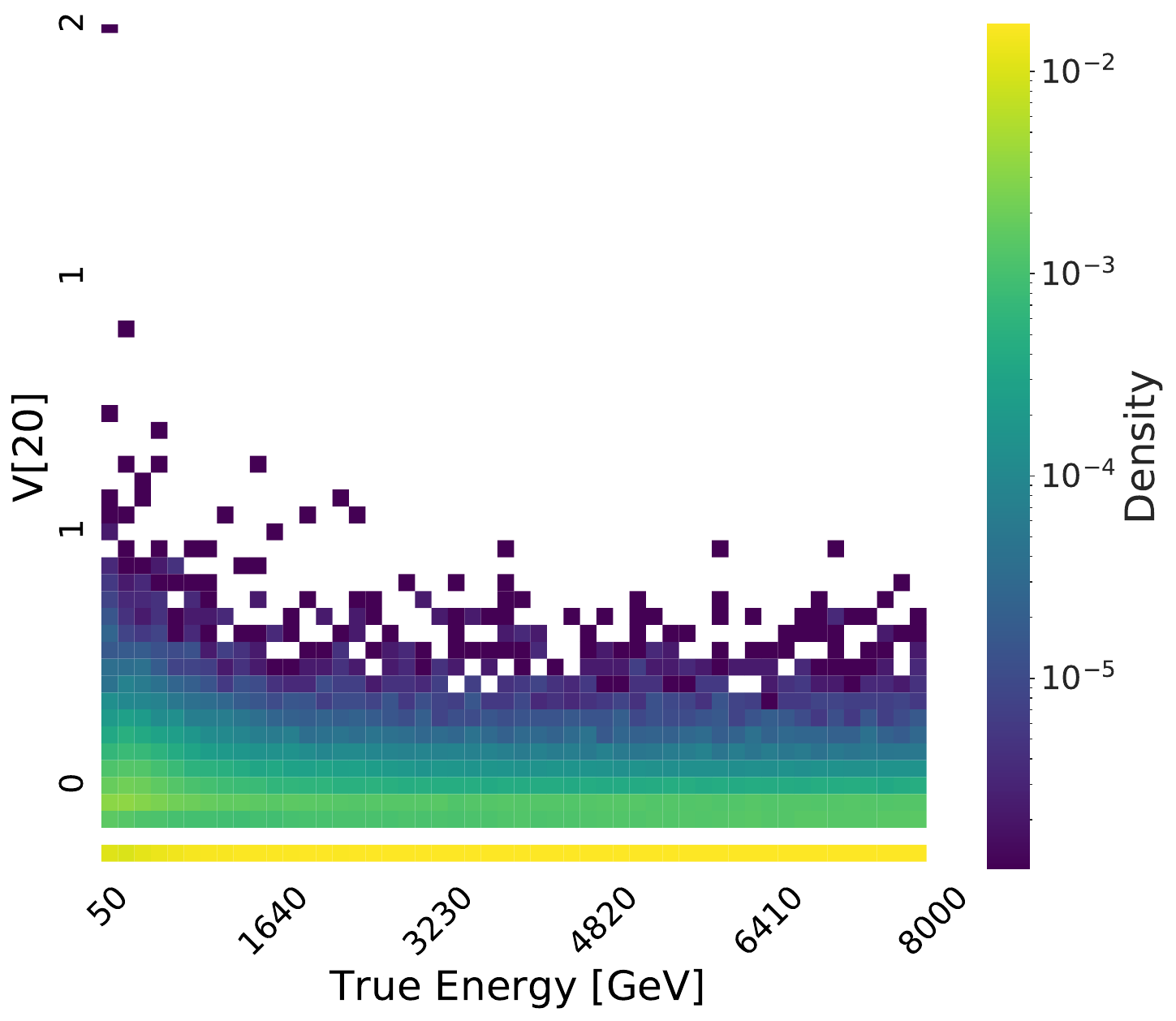}
                    \end{center}
                \end{subfigure}
                \begin{subfigure}[t]{0.30\textwidth}
                    \begin{center}
                        \includegraphics[width=\textwidth]{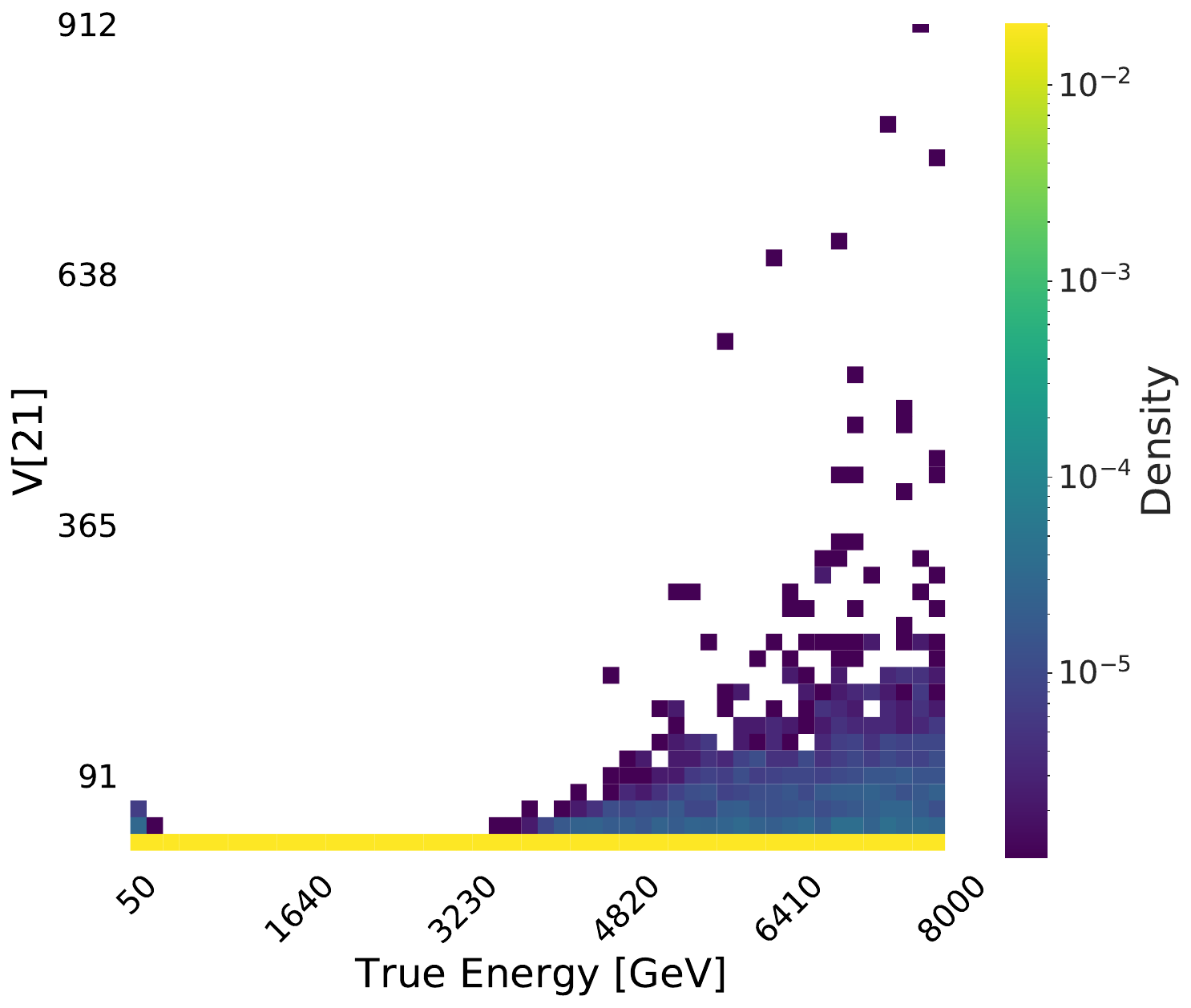}
                    \end{center}
                \end{subfigure}
                \begin{subfigure}[t]{0.30\textwidth}
                    \begin{center}
                        \includegraphics[width=\textwidth]{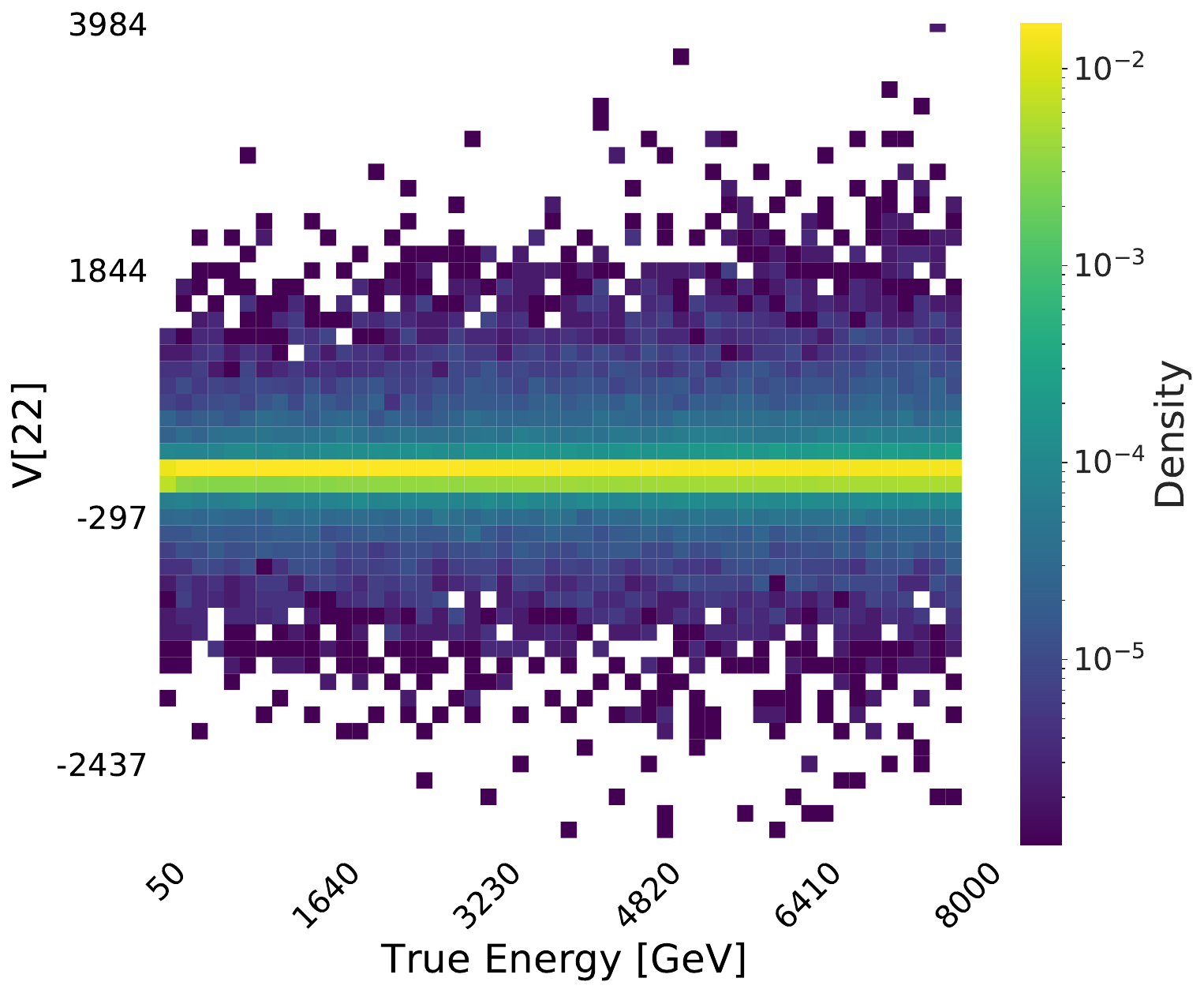}
                    \end{center}
                \end{subfigure}
                \begin{subfigure}[t]{0.30\textwidth}
                    \begin{center}
                        \includegraphics[width=\textwidth]{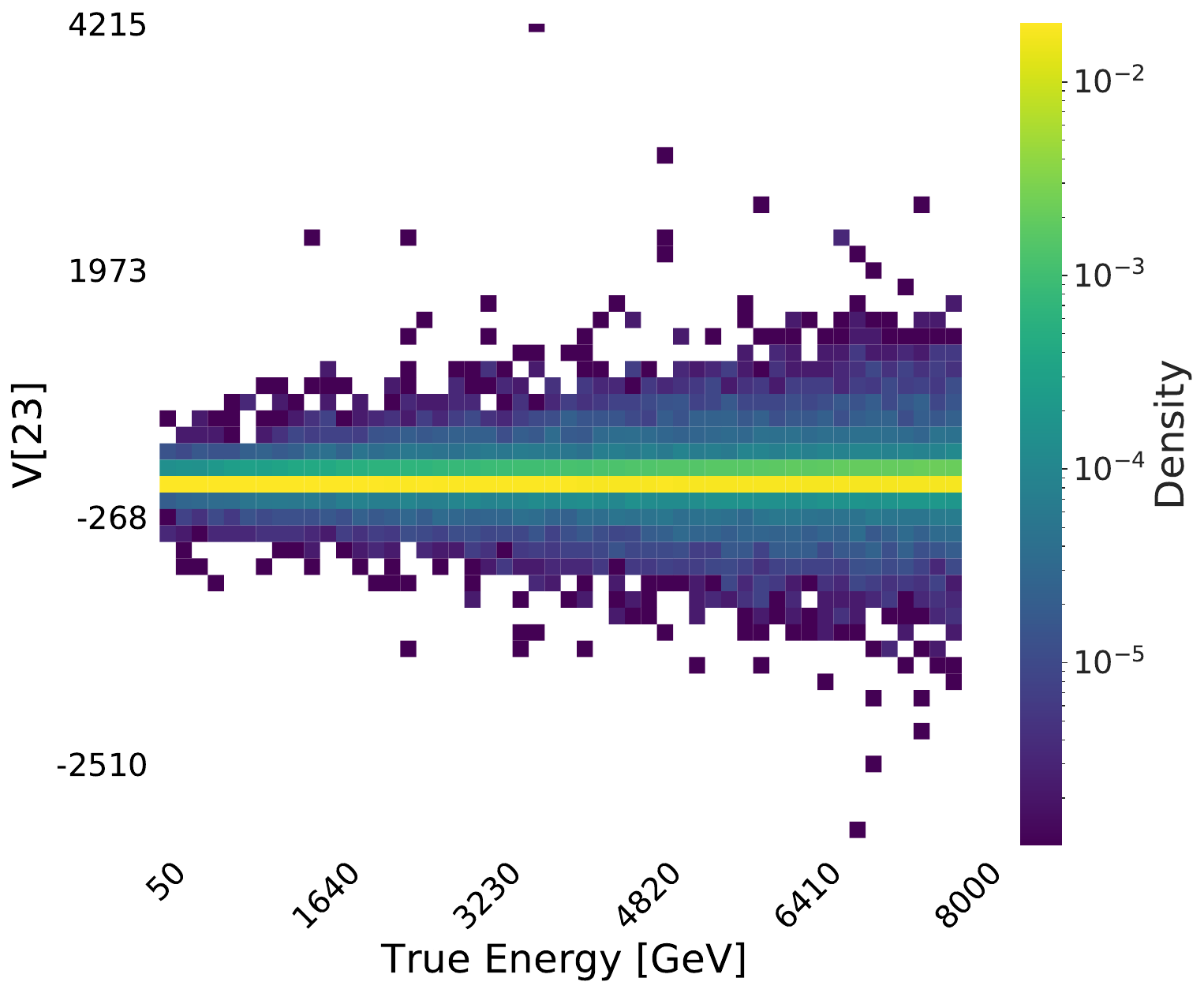}
                    \end{center}
                \end{subfigure}
                \begin{subfigure}[t]{0.30\textwidth}
                    \begin{center}
                        \includegraphics[width=\textwidth]{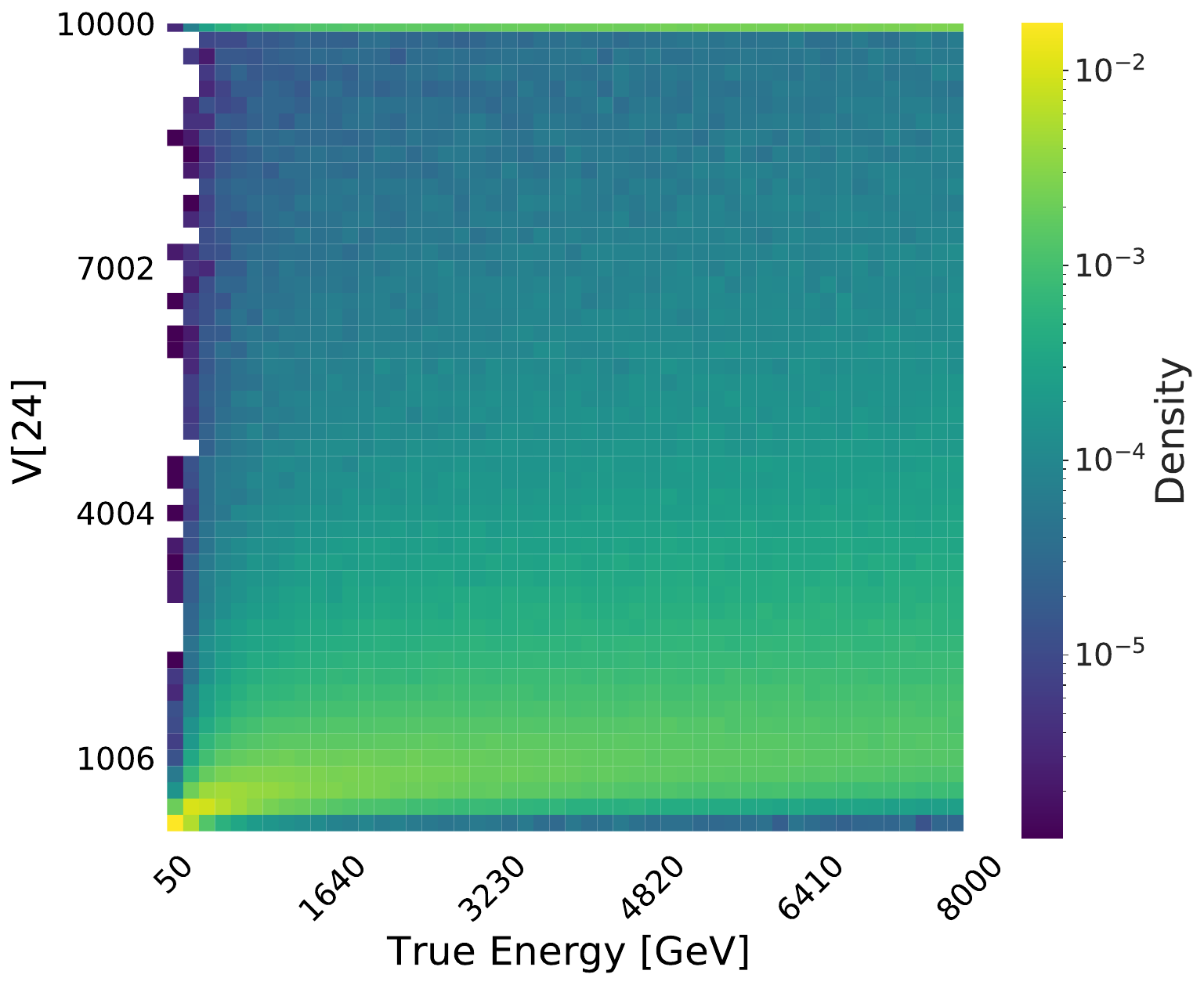}
                    \end{center}
                \end{subfigure}
                \begin{subfigure}[t]{0.30\textwidth}
                    \begin{center}
                        \includegraphics[width=\textwidth]{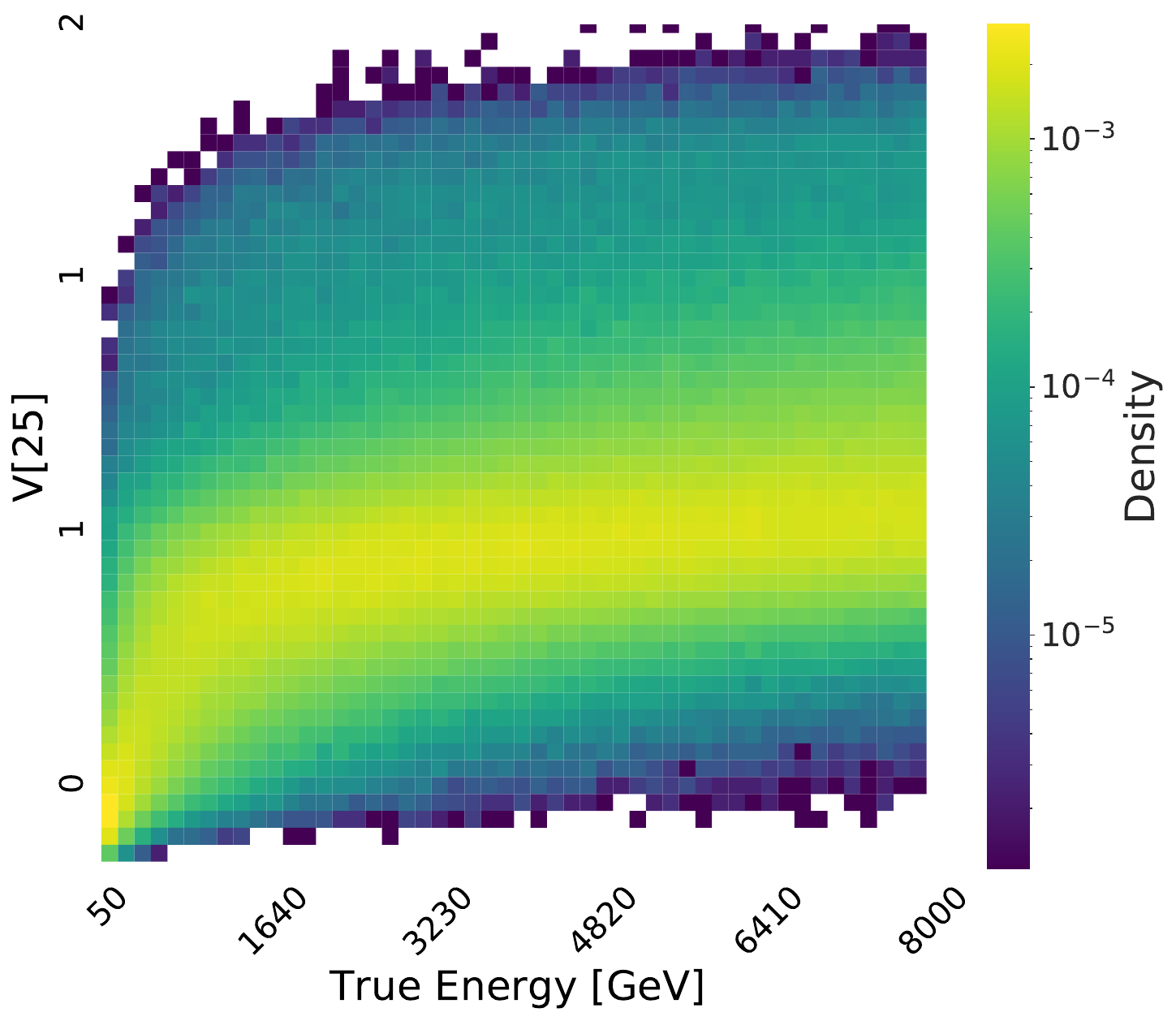}
                    \end{center}
                \end{subfigure}
                \begin{subfigure}[t]{0.30\textwidth}
                    \begin{center}
                        \includegraphics[width=\textwidth]{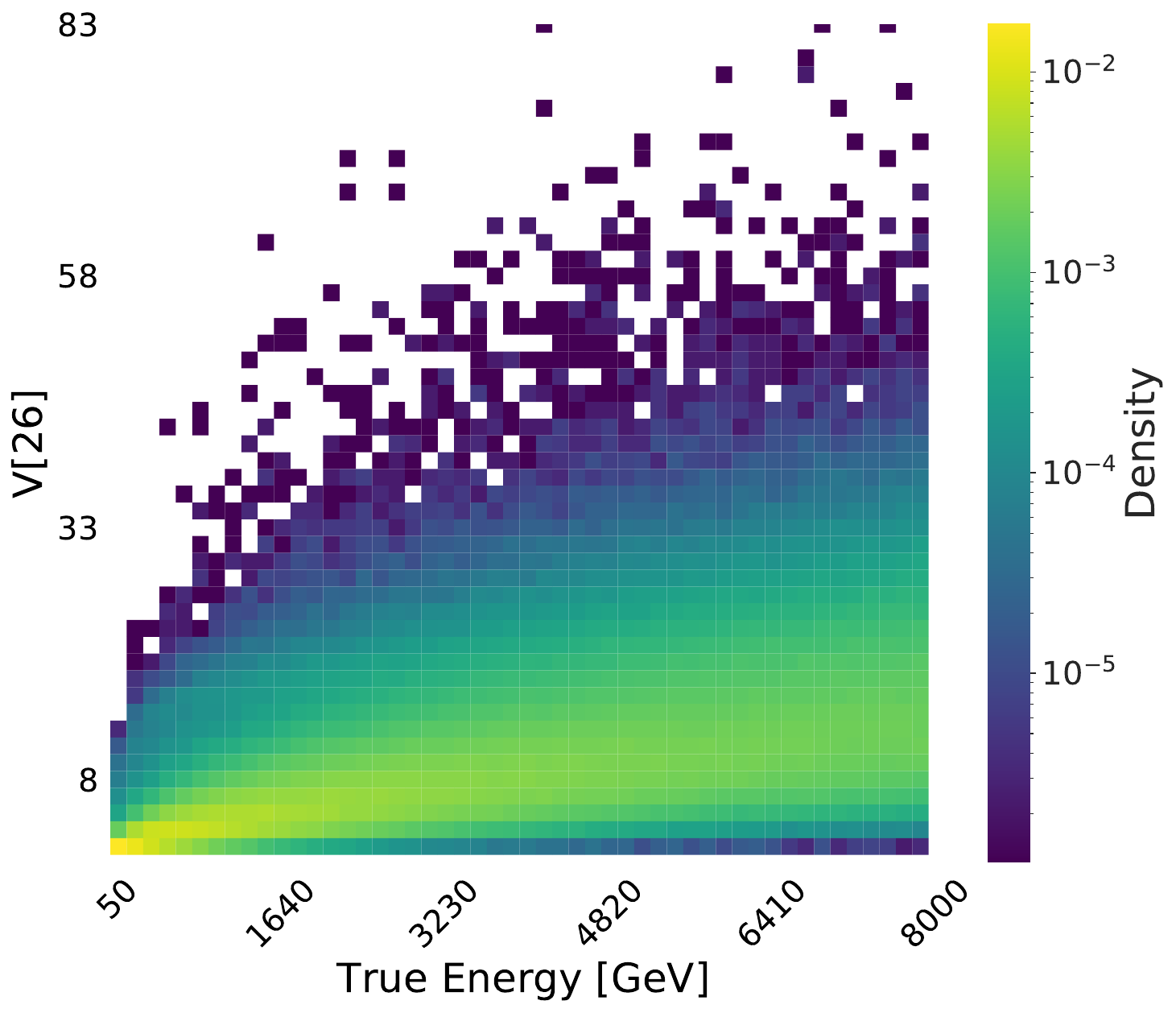}
                    \end{center}
                \end{subfigure}
                \begin{subfigure}[t]{0.30\textwidth}
                    \begin{center}
                        \includegraphics[width=\textwidth]{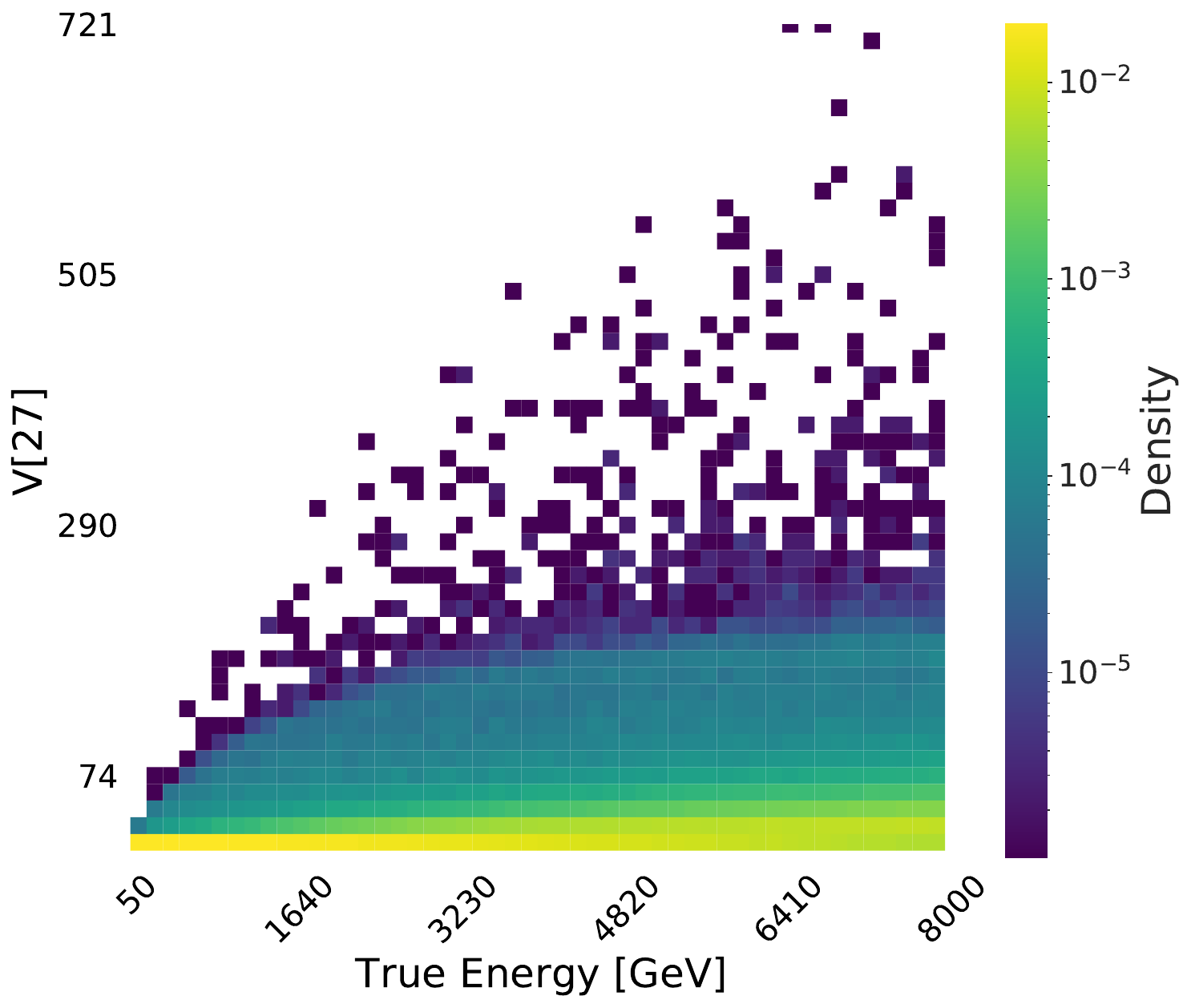}
                    \end{center}
                \end{subfigure}
                \caption{2D histograms showing the dependence of features V[20] to V[27] (on the y axes) on true muon energy (on the x axes). Features are defined in Section~\ref{s:features_description}.}
                \label{f:2d_feats:20-27}
            \end{center}
        \end{figure*}

        \noindent
        Variable $V[24]$ is then defined as $V[24] = 0.3 B R$. It provides useful information for muon momenta below about \SI{500}{\gev}, as can be seen in \autoref{f:2d_feats:20-27}.
        
        %%%%%%%%%%%%%%%%%%%%%%%%%%%%%%%%%%%%%%%%%%%%%%%%%%%%%%%%%%%%%%%
        \subsection{Description of other global features}\label{s:features_description}
        
        \noindent
        We list below the other features we compute for each muon:\par
        
        \begin{itemize}
            \item V[0]: The total energy recorded in the calorimeter in cells above the $E_{\text{thr}}>\SI{0.1}{\gev}$ threshold;
            \item V[1]: We define $H_x = \sum_i E_i\cdot\Delta x_i$ and $H_y = \sum_i E_i\cdot\Delta y_i$, where $\Delta x_i$ and $\Delta y_i$ are the spatial distances in the $x$ and $y$ directions to the centre of the cell which is hit by the muon at the calorimeter front face; from these we derive V[1] = $\sqrt{H_x^2+H_y^2}/\sum_i E_i$. In this calculation, all cells are used;
            \item V[2]: This variable results from the same calculation extracting $V[1]$, but it is performed using in all sums only towers exceeding the $E_{\text{thr}}=\SI{0.1}{\gev}$ threshold;
            \item V[3]: The second moment of the energy distribution around the muon direction in the transverse plane, computed with all towers as 
            $V[3]= \sum_i [E_i (\Delta x_i^2 + \Delta y_i^2)]/ \sum_i E_i$, 
            where indices run on all towers and the distances are computed in the transverse plane, as above;
            \item V[4]: The same as V[3], but computed only using towers located in the first \SI{400}{\mm}-thick longitudinal section of the detector along $z$;
            \item V[5]: The same as V[3], but computed only using towers in the $400<z_i<\SI{800}{\mm}$ region;
            \item V[6]: The same as V[3], but computed only using towers in the $800<z_i<\SI{1200}{\mm}$ region;
            \item V[7]: The same as V[3], but computed only using towers in the     $1200<z_i<\SI{1600}{\mm}$ region;    
            \item V[8]: The same as V[3], but computed only using towers in the $z_i\geq\SI{1600}{\mm}$ region;
            \item V[9]-V[14] and V[16]-V[21]: See {\em supra} (\autoref{s:cluster});
            \item V[15]: Second-highest maximum energy in a $3\times\!3\times\!3$ cubic box from cells not included in type-1 or type-2 clusters;
            \item V[22]: The first moment of the energy distribution along the x axis, relative to the x position of the incoming muon track;
            \item V[23]: The first moment of the energy distribution along the y axis, relative to the y position of the incoming muon track;
            \item V[24]: See {\em supra} (\autoref{s:bendingvar});
            \item V[25]: Maximum energy in a $3\times\!3\times\!3$ cubic box from cells not included in type-1 or type-2 clusters;
            \item V[26]: Sum of energy recorded in cells with energy below \SI{0.01}{\gev};
            \item V[27]: Sum of energy recorded in cells with energy between 0.01 and $E_{thr}=\SI{0.1}{\gev}$.  
        \end{itemize}
        
\begin{figure*}[h!]
            \begin{center}
                \includegraphics[width=16cm]{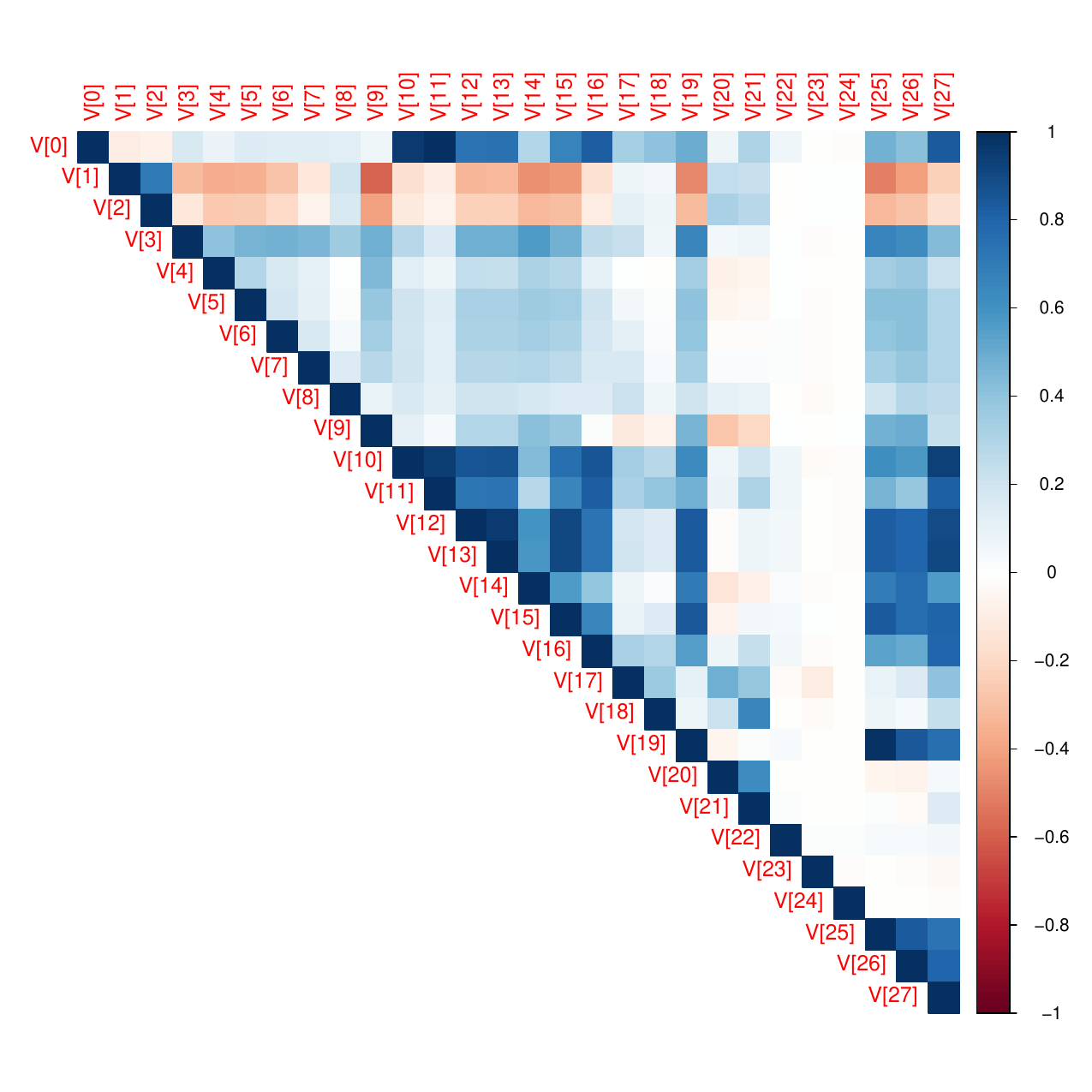}
                \caption{Correlation matrix between the high-level features. See the text for details. }
                \label{f:corrmatrix}
            \end{center}
        \end{figure*}

        \noindent
        The correlation matrix of the 28 variables is shown in \autoref{f:corrmatrix}.
        
        %\FloatBarrier        
        
        %%%%%%%%%%%%%%%%%%%%%%%%%%%%%%%%%%%%%%%%%%%%%%%%%%        
    \section{Ablation study}\label{sec:ablation}
        The architecture of the network described in \autoref{sec:arch}, and training methodology detailed in \autoref{sec:training}, are both reasonably complex, leveraging a range of recently published, or otherwise unusual, techniques, as well as other aspects that are specific to the task at hand. Similar to the study of the input features in \autoref{sec:input_study}, it is worth quantifying the actual benefits of each of these items in the hope of simplifying the model, or to help inform future studies in similar task regimes.
        
        This ablation study takes the strategy of inspecting separately: the loss, the ensembling, the training, the architecture, and the bias correction. In each inspection, particular aspects of the model or method will be removed, or replaced with more standard approaches, individually ({\em i.e.} only one aspect is ever different) to quantify the benefit of each particular aspect in the presence of the others. Finally, the whole section of the model or method will be replaced with a standard approach, to quantify the overall benefit of that part of the model or method. Unless otherwise stated, the results shown are computed on the monitoring-validation dataset and averaged over the five models trained per test via unique-fold training (see \autoref{sec:ensemble_training}).
        
        \subsection{Loss function}
            As a reminder, the regressor is trained to minimise a Huberised version of a domain-informed loss function, which tracks running averages of its thresholds in bins of true energy, and element-wise losses are weighted according to a function of the true energy. We can simplify this loss function by: using just a single bin of true energy; using a fixed threshold per bin computed solely on the first batch of data; using thresholds per bin computed entirely on the current batch of data without averaging; using unweighted loss elements; or using the mean fraction squared error without Huberisation ({\em i.e.} as in \autoref{eq:MFSE}). Finally we can replace the domain-inspired function with a more standard mean squared-error loss, however we will still include the down-weighting of the data. The results of these studies are summarised in \autoref{tab:ablation:loss}.
            
            \begin{table}[ht]
            	\begin{center}
            		\begin{tabular}{lll}
            			\hline
            			Ablation & MI & Change in MI [\%]\\
            			\hline
            			Default & $19.42\pm0.08$ & N/A \\
            			\hline
                        Single bin & $19.14\pm0.08$ & $-1.5\pm0.6$ \\
                        Batchwise thresholds & $19.25\pm0.04$ & $-0.9\pm0.5$ \\
                        Non-Huberised loss & $19.36\pm0.06$ & $-0.4\pm0.5$ \\
                        Fixed thresholds & $19.39\pm0.05$ & $-0.2\pm0.5$ \\
                        \hline
                        MSE loss & $18.43\pm0.06$ & $-5.1\pm0.5$ \\
                        \hline
                        No down-weighting & $16.5\pm0.2$ & $-15.12\pm1.03$ \\
            			\hline
            		\end{tabular}
                \end{center}
                \caption{Ablation study of the loss. The change in MI is computed as the fractional difference with respect to the default model. The ``MSE" test uses the down-weighting of the data, and the ``No down-weighting" test uses the full adaptive Huberised MFSE loss.}
            	\label{tab:ablation:loss}
            \end{table}
            
            From these results we can confirm that the domain-inspired loss we have adopted is beneficial to the training, and that the down-weighting is also very important. The Huberisation of the loss is potentially useful, however when it is used one should compute separate thresholds in bins of true energy, and either fix these for the whole training, or track their running average.
            
        \subsection{Ensembling and dataset size}
            Whilst in this study the focus is mainly on improving performance, in practice one may also be concerned by retraining time, inference time, and dataset-size requirements. Due to these potential concerns it is worth checking the benefits of ensembling and training on larger datasets. Table~\ref{tab:ablation:ensemble} summarises two ensemble trainings, one via full-fold, and the other via unique-fold. Each training is then interpreted in two ways: one assumes that all five models were trained and applied as an ensemble (full/unique ensemble); the other assumes that only one model was trained and computes the average MI across the five models that were actually trained (full/unique singles).
            
            \begin{table*}[ht]
            	\begin{center}
            		\begin{tabular}{lllll}
            			\hline
            			Ablation & Dataset size & \multicolumn{2}{c}{Times} & MI \\
            			 & & Training [\si{\hour}] & Inference [\si{second} per batch] \\
            			\hline
            			Full ensemble & \num{862085} & 113.4 & 0.47 & 20.72 \\
            			Full singles & \num{862085} & 22.7 & 0.091 & $20.29\pm0.04$ \\
            			Unique ensemble & \num{862085} & 23.3 & 0.47 & 19.83 \\
            			Unique singles & \num{197048} & 4.7 & 0.0.091 & $19.37\pm0.08$\\
            			\hline
            		\end{tabular}
                \end{center}
                \caption{Ablation study of the ensembling and dataset size and usage. Inference time is per batch of 256 muons and excludes the disk-to-RAM time. The MI is computed on the holdout-validation data. ``Full" indicates the model was trained on 34/36 folds of data and monitored on one fold of data. ``Unique" indicates the model was trained on a unique set of seven folds of data, and monitored on one fold of data. In the case of the single models, the MI is averaged across five individual models.}
            	\label{tab:ablation:ensemble}
            \end{table*}
            
            From the above results we can see that both ensembling and using a larger dataset provide performance improvements. We also see that if a large dataset is available, then it is better to train a single model on the whole of it, than an ensemble on unique subsamples of it, both in terms of inference time and MI. Since the training time of the full models, the disk-space-size per data point, and the data generation times are all relatively low compared to many other algorithms used in HEP and trainings can be used for an entire data-taking run, our recommendation would be to use as much training data as possible. The choice between single model or ensembling depends mostly on the time-budget available during application (since other reconstruction algorithms will be being run during processing), and whether the regression is performed online during data-taking for triggering, or during offline reconstruction.
            
        \subsection{Training}
            The nominal training scheme involves changing the learning rate and the momentum of the optimiser during training: first via a 1cycle schedule, to quickly train the model; and second via a step decay of the LR. To check the advantage of this schedule, we can retrain keeping the LR and momentum constant. The LR is set to \num{1e-4}, slightly lower than the maximum LR used for nominal training, to account for the fact that it has no possibility to decrease, other than through ADAM's scaling parameters, and that the momentum will not be able to stabilise the higher LR. The number of epochs and early-stopping criteria are kept the same. Such a training results in a $\SI{5.2\pm0.6}{\%}$ decrease in MI, and an increase in the required training time due to the nominal scheme triggering the early-stopping criterion earlier.
            
        \subsection{CNN architecture}
            The CNN, although inspired by established architectures, is by no means standard, and includes a task-specific component in the form of the energy-pass-through connections. The studies performed are: removal of the squeeze-excitation blocks; removal of the max-average pooling layer, instead flattening the hidden state and feeding all inputs to the fully connected layers; replacing the running batchnorm layers with standard BN layers; removal of BN entirely; removal of the identity paths, {\em i.e.} the paths through the trainable convolutional layers are no longer residual (in this case the positions of the BN and activation layers are changed to always be convolution into activation into BN); and removal of the energy-pass-through connections (in this case the number of channels added at each downsampling stage is increased to maintain a similar number of trainable parameters). Finally, we can remove the CNN head entirely and flatten all \num{51200} cell values into a vector to be fed directly to the full connected layers. Table~\ref{tab:ablation:arch} details the results of these studies.
            
            \begin{table*}[ht]
            	\begin{center}
            		\begin{tabular}{lllr}
            			\hline
            			Ablation & MI & Change in MI [\%] & Parameters\\
            			\hline
            			Default & $19.42\pm0.08$ & N/A & \num{636570} \\
            			\hline
            			No BN & $18.5\pm0.3$ & $-5\pm1$ & \num{635404} \\
            			No identity path & $18.72\pm0.08$ & $-3.6\pm0.6$ & \num{634356}\\
            			Nominal BN & $19.2\pm0.2$ & $-1.1\pm0.9$ & \num{636570}\\
            			No E-pass & $19.30\pm0.05$ & $-0.6\pm0.5$ & \num{671076}\\
                        No SE & $19.33\pm0.09$ & $-0.5\pm0.6$ & \num{631259}\\
                        No pooling & $19.4\pm0.1$ & $-0.4\pm0.7$ & \num{805130}\\
                        \hline
                        No CNN & $17.45\pm0.09$ & $-10.2\pm0.6$ & \num{4111361}\\
            			\hline
            		\end{tabular}
                \end{center}
                \caption{Ablation study of the architecture. The change in MI is computed as the fractional difference with respect to the default model. ``Parameters" refers to the number of trainable parameters in the architecture of each model.}
            	\label{tab:ablation:arch}
            \end{table*}
            
            As expected, the CNN head is essential to avoid over-parameterising the model. Additionally, the use of running batchnorm is necessary to avoid instabilities in the validation performance of the network (running without any BN at all also produces instabilities in the training loss). The identity paths also provide a large improvement to the model. It is interesting to note that the energy pass-through connections provide an improvement, since the model should be able to learn this itself, however similar to \textsc{DenseNet}, the fact that we explicitly retain a part of the previous representation of the data throughout the model, allows a slightly more direct flow of gradient update to the trainable layers. Additionally we are implicitly suggesting that the trainable layers act as small corrections to the recorded energy, rather than allowing the model to learn this approach.
            
        \subsection{Bias correction}
            Whilst not strictly part of the architecture, we can also check whether the minor correction to the predictions that we apply post-training is useful in improving the resolution. Without the bias correction, the change in MI is \SI{-2.1}{\%} on the holdout-validation dataset when using the nominal model and all-fold training (see \autoref{sec:ensemble_training}), so the correction is worth applying.
            
        %%%%%%%%%%%%%%%%%%%%%%%%%%%%%%%%%%
    \section{Resource requirements}\label{sec:requirements}
        %%%%%%%%%%%%%%%%%%%%%%%%%
        \subsection{Regressor}
            The models used for this study were trained on Nvidia V100S GPUs. Training the nominal architecture at a batch size of 256 requires \SI{5}{\giga\byte} of VRAM, \SI{23}{\giga\byte} RAM, and \SI{100}{\%} of both a single (virtual) CPU core (Intel~Xeon~Gold~6248 CPU @ \SI{2.5}{\giga\hertz} in our case) and the GPU. The training time per model is about \SI{23}{\hour}, and about five days for the full ensemble when trained serially; however, with sufficient resources, ensemble training would be trivially parallelisable.
            
            Application of the ensemble takes \SI{61}{\second} for a dataset of \num{24631} muons computed in batches of 256, of which \SI{15}{\second} are spent loading the dataset into RAM. Excluding the disk-to-RAM time, inference is about \SI{0.5}{\second} for a batch of 256 muons (including RAM-to-VRAM time) for the ensemble (\SI{0.1}{\second} per batch per model).
            
            Although some steps are taken to reduce data-loading times (LZF-compression and sparse hit-representation), disk-to-RAM loading time is still significant and training/inference time depends highly on the disk read speed and access latency; whilst production and development was mainly performed in the cloud on powerful and expensive GPUs, local runs on a much cheaper Nvidia~1080~Ti GPU with a solid-state hard-drive were actually just as quick.
            
            Whilst the loading time from RAM to GPU is minor compared to the load-time from disk to RAM, further improvements would be to retain the sparse representation of the data, however sparse tensors in \pytorch are still experimental, and sparse CNN implementations are limited in functionality, let alone implemented for 3D convolutions.
    
        %%%%%%%%%%%%%%%%%%%%%%%%%%%%%%%%%%%%%%%%
        \subsection{Datasets and preparation}
            The time to generation the data via \geant~4 heavily depends on the muon energy, however by running the generation as \num{7000} simultaneous jobs on a batch system, the dataset was processed in about one day. The raw ROOT files require \SI{183}{\giga\byte} of storage space.
            
            Computation of the high-level features is performed in C\texttt{++} and is also run as \num{7000} jobs on a batch system, taking a few hours to complete. The resulting uncompressed CSV files require \SI{246}{\mega\byte} of space.
            
            Processing of the raw hits from ROOT into the HDF5 files required by \lumin, and combination with the high-level features is a three-step process:
            \begin{enumerate}
                \item Each ROOT file is processed into an LZF-compressed HDF5 file containing only the raw hits and the muon energy (the ROOT files also contain additional information which is no longer required). This takes about six hours and requires about a further \SI{44}{\giga\byte} of space.
                \item Meta data required to pre-process the HL-features is computed via a loop over the CSV files, which takes a few seconds.
                \item The individual HDF5 files are combined with the CSV files into two LZF-compressed HDF5 files with the training and validation data being split into 36 folds, and the testing data split into 18 folds. At this point the HL-features are pre-processed based on the meta data computed beforehand, and the raw energy deposits are transformed into a sparse format (which reduces loading time). This requires several hours; the final training file has a size of \SI{32}{\giga\byte} and the testing file of \SI{12}{\giga\byte}.
            \end{enumerate}
            
    %\FloatBarrier

    \section{Software}
        The investigation performed in this project depended on many open-source software packages. These are summarised in \autoref{tab:software}. A public version of the research code is available from Ref.~\cite{muon_reg_repo}. The pre-processed datasets are available from Ref.~\cite{muon_reg_dataset}, and are designed to be used directly with the code-base.
        \begin{table*}[hbt]
        	\begin{center}
        		\begin{tabular}{llll}
        			\hline
        			Software & Version & References & Use/Notes\\
        			\hline
        			\lumin & 0.8 & \cite{lumin} & Wrapping \pytorch to implement networks\\
        			\pytorch & 1.8 & \cite{pytorch} & Implementing neural networks\\
        			\textsc{Seaborn} & 0.9 & \cite{Seaborn} & Plot production\\
        			\textsc{Matplotlib} & 3.2 & \cite{MatPlotLib} & Plot production\\
        			\textsc{Pandas} & 1.2 & \cite{Pandas} & Data analysis and computation\\
        			\textsc{NumPy} & 1.21 & \cite{Numpy} & Data analysis and computation\\
        			\textsc{Scikit-Learn} & 0.22.0 & \cite{sklearn} & Data shuffling \& splitting\\
        			\geant & 4 & \cite{GEANT4_0,GEANT4_1} & Detector simulation\\
        			\Root & 6 & \cite{Root} & Processing of data\\
        			\textsc{Uproot} & 3.11 & \cite{uproot} & Processing of data\\
        			\hline
        		\end{tabular}
            \end{center}
            \caption{Software used for the investigation}
        	\label{tab:software}
        \end{table*}
        
%\end{appendices}

\FloatBarrier
% \clearpage

%%%%%%%%%%%%%%%%%%%

\bibliographystyle{utphys_mod.bst}  
\bibliography{main}

\end{document}